\documentclass[usenatbib]{mn2e}

\newcommand{\apjl}{ApJL}
\newcommand{\aap}{A\&A}

\newcommand{\apjs}{ApJS}

\usepackage{epsfig,amsmath,enumerate,amssymb,aas_macros,bm,booktabs,multirow,xcolor,caption,subcaption}
\usepackage{graphics}

\newcommand{\solarmass}{ \rm M_\odot}

\def\gsim { \lower .75ex \hbox{$\sim$} \llap{\raise .27ex \hbox{$>$}}}
\def\lsim { \lower .75ex \hbox{$\sim$} \llap{\raise .27ex \hbox{$<$}}}

\usepackage{natbib}
\usepackage{times}

\title[${\rm H_2}$-based star formation laws in galaxy formation models]{${\rm
H_2}$-based star formation laws in hierarchical models of galaxy  
formation}
\author[Xie et al.]{Lizhi Xie$^{1}$ \thanks{Email:lzxie@oats.inaf.it}, 
  Gabriella De Lucia$^{1}$ ,
  Michaela Hirschmann$^{2}$,
  Fabio Fontanot$^{1}$,
  Anna Zoldan$^{3,1}$
\\    
$^{1}$INAF - Astronomical Observatory of Trieste, via G.B. Tiepolo 11,
I-34143 Trieste, Italy\\
$^{2}$Sorbonne Universit\'es, UPMC-CNRS, UMR7095, Institut d'~Astrophysique de
Paris, F-75014 Paris, France\\
$^{3}$Physics Department, Universit{\'a} degli Studi di Trieste, Via Valerio 2, 34127-Trieste, TS, Italy
}

\begin{document}
\label{first page} \maketitle

\begin{abstract}
We update our recently published model for GAlaxy Evolution and Assembly
(GAEA), to include a self-consistent treatment of the partition of cold gas in
atomic and molecular hydrogen. Our model provides significant improvements with
respect to previous ones used for similar studies. In particular, GAEA (i)
includes a sophisticated chemical enrichment scheme accounting for 
non-instantaneous recycling of gas, metals, and energy; (ii) reproduces the
measured evolution of the galaxy stellar mass function; (iii) reasonably reproduces the
observed correlation between galaxy stellar mass and gas metallicity at
different redshifts. These are important prerequisites for models considering a
metallicity dependent efficiency of molecular gas formation. We also update our
model for disk sizes and show that model predictions are in nice agreement with
observational estimates for the gas, stellar and star forming disks at
different cosmic epochs. We analyse the influence of different star formation
laws including empirical relations based on the hydrostatic pressure of the
disk, analytic models, and prescriptions derived from detailed hydrodynamical
simulations. We find that modifying the star formation law does not affect
significantly the global properties of model galaxies, neither their
distributions. The only quantity showing significant deviations in different
models is the cosmic molecular-to-atomic hydrogen ratio, particularly at high
redshift. Unfortunately, however, this quantity also depends strongly on the
modelling adopted for additional physical processes. Useful constraints on 
the physical processes regulating star formation can be obtained focusing on 
low mass galaxies and/or at higher redshift. In this case, self-regulation 
has not yet washed out differences imprinted at early time.
\end{abstract}


\section{Introduction}
\label{sec:intro}

A proper description of how galaxies form and evolve requires necessarily an
understanding of the physical mechanisms regulating the star formation 
process within dense regions of molecular clouds. At the microscopic 
level, star formation arises from a complex interplay between e.g. turbulence, 
rotation and geometry of the cloud, and magnetic fields, making a 
self-consistent treatment of the process from `first principles' unfeasible in 
theoretical models of galaxy formation and evolution. Fortunately, clear 
and tight correlations are measured between the rate at which stars form 
within a (disc) galaxy and the amount of gas in the disc. Such correlations 
have, for decades now, been a crucial element of theoretical models of 
galaxy formation.

One commonly adopted star formation formulation is based on the so-called
Schmidt-Kennicutt law \citep{schmidt1959,kennicutt1998}, which relates the
surface density of the star formation rate $\Sigma_{\rm SF}$ to that of the 
gas $\Sigma_{\rm gas}$ via a simple power law: $\Sigma_{\rm SF}\propto\Sigma_{\rm gas}^n$, 
with $n=1.4$\footnote{\citet{kennicutt1998} show that a formulation that assumes 
the surface density of star formation rate scales with the ratio of the gas
density to the average orbital time scale, fitted their data equally well.}. 
In many galaxy formation models, a slightly different formulation is used, which
 assumes the star formation rate declines rapidly for surface densities below 
a critical value, often estimated using the disk stability criterion introduced 
by \citet{toomre1964}. For the sample presented in \citet{kennicutt1998}, the 
correlation between $\Sigma_{\rm SF}$ and $\Sigma_{\rm gas}$ (including 
both molecular and atomic hydrogen) was stronger than that with the surface 
density of molecular gas  $\Sigma_{\rm H_2}$. Albeit this and earlier work
pointed out that the larger scatter of the latter relation could be at least in part due to variations
in the CO/H$_2$ conversion factor, most models up to a few years ago 
simply assumed that the star formation rate depends on the amount (and/or
surface density) of `cold gas' (typically all gas below $10^4$~K), with no
attempt to partition it in its molecular and atomic components.

In the last decade, our phenomenological understanding of star formation in
galaxies has improved significantly thanks to the advent of high-quality
spatially resolved observations in HI \citep[e.g. the HI Nearby Galaxy Survey
-][]{walter2008} and CO, (e.g. The BIMA Survey of Nearby Galaxies
- \citealt{helfer2003}, and The HERA CO Line Extragalactic Survey
- \citealt{leroy2009}) and, at the same time, of more reliable estimates of the
star formation at different wavelengths for large samples of nearby galaxies
(e.g. The  Spitzer Infrared Nearby Galaxies Survey - \citealt{kennicutt2007,
calzetti2007}, and the Galaxy Evolution Explorer Nearby Galaxies Survey - 
\citealt{gildepaz2007}). These data clearly demonstrate that 
star formation correlates strongly with the molecular gas in a galaxy, and
poorly or not at all with the atomic gas \citep[e.g.][]{wong2002,kennicutt2007,
leroy2008}. 
In non-barred spiral galaxies, the fraction of molecular gas 
increases towards the centre, where the HI gas surface density remains flat or weakly declines 
\citep{bigiel2008}. The threshold in star formation suggested by
early observations \citep{kennicutt1989,martin2001} can therefore be
interpreted as a transition to a different regime of star formation activity.
 Although it is unclear if molecular gas is a necessary condition for
star formation \citep[see e.g.][and references therein]{gloverclark2012,hu2016},
The observational data provide a detailed characterization of the star formation law 
in terms of molecular hydrogen.

Based on a relatively small sample of nearby galaxies, \citet{blitz2006} argue
that the ratio of molecular-to-atomic hydrogen surface density is determined by
the hydrostatic pressure of the disk. The scatter in the \citet{blitz2006}
relation is relatively large, and alternative interpretations have been provided
for the observations. A different view considers the molecular fraction as 
determined by a balance between the production of molecular hydrogen on the 
surface of dust grains and dissociation of the molecules by radiation from young
stars \citep{krumholz2009b,gnedin2011}.

While the physical processes regulating star formation remain to be
understood, the new rich phenomenology described above has also triggered
significant activity devoted to update and test the influence of H$_2$ based
star formation laws both in hydrodynamical
simulations \citep[e.g.][]{gnedin2011,kuhlen2012}, and in semi-analytic models
of galaxy formation \citep[e.g.][]{fu2010,lagos2011a,somerville2015}. Given
their flexibility and limited computational costs, the latter represents an
ideal interpretative tool for large ongoing surveys of cold gas in nearby and
distant galaxies \citep{fu2012,lagos2011b,popping2014}, as well as future
projects planned on facilities such as the Atacama Large
Millimeter/sub-millimeter Array \citep[ALMA -][]{wootten2009}, the Square 
Kilometre Array \citep[SKA -][]{carilli2004} and its pathfinders 
\citep{booth2009,johnston2008}, and the Five-hundred-meter Aperture Spherical 
radio Telescope \citep[FAST -][]{nan2011}.

In this work, we extend our new and recently published semi-analytic model for 
GAlaxy Evolution and Assembly (GAEA) by including an explicit treatment for 
the partition of cold gas in its atomic and molecular component. As one of its
major features, GAEA includes a sophisticated scheme for chemical enrichment based
on non-instantaneous recycling of gas, energy and metals \citep{delucia2014}. 
\citet{hirschmann2016} show that GAEA also successfully reproduces the 
evolution of the observed correlation between the galaxy stellar mass and 
cold gas metallicity - an important prerequisite for schemes that assume the 
molecular-to-atomic ratio depends on the gas metallicity. 

This paper is organized as follows: in Section~\ref{sec:simsam}, we introduce 
our semi-analytic model and describe in detail the star formation laws that 
we considered. In Section~\ref{sec:growandmf}, we describe how these different 
star formation laws affect the physical properties of galaxies, 
and compare basic statistics on the distribution of stellar masses, HI and 
H$_2$ with available data. In Section~\ref{sec:scaling}, we compare model 
predictions with observed scaling relations between the amount of molecular 
and atomic hydrogen, gas metallicity, size of the star forming discs and 
galaxy stellar mass. In Section~\ref{sec:discussion}, we discuss our results 
also in the framework of previous work. Finally, in 
Section~\ref{sec:conclusion}, we summarize our findings and give our conclusions.

\section{Semi-analytic model}
\label{sec:simsam}

In this work, we take advantage of our recently published model GAEA 
\citep[][hereafter HDLF16]{hirschmann2016}. This model builds on that 
described in \citet{delucia2007}, with modifications introduced to follow 
more accurately processes on the scales of the Milky Way satellites
\citep*{delucia2008,li2010}. The evolution of the baryonic component of dark
matter haloes is traced by following four different reservoirs: stars in
galaxies, cold gas in the galaxy discs, diffuse hot gas associated with dark
matter haloes, and an ejected gas component.  The transfer of mass and energy
between these components is modelled assuming specific prescriptions for: gas
cooling, star formation, stellar feedback (including metal enrichment,
reheating of cold gas, and gas ejection), galaxy mergers (and associated
star-bursts), bulge formation during mergers and driven by disk instability. The
model also includes prescriptions for cold (merger driven) and hot gas
accretion onto super massive black holes, and for the suppression of cooling
flows in massive haloes from radio loud Active Galactic Nuclei (AGN).

Our physical model for the evolution of the baryonic components of galaxies is
coupled to the output of cosmological dark matter simulations, as detailed in
\citet{delucia2007}. In this study, we use dark matter merger trees from
two large-scale cosmological simulations: the Millennium Simulation
\citep[][MS]{springel2005}, and the Millennium II Simulation
\citep[][MSII]{boylankolchin2009}. Both simulations consist of $2160^3$ particles; the box
size is $500\,{\rm Mpc}\,{\rm h}^{-1}$ for the MS and $100\,{\rm Mpc}\,{\rm
  h}^{-1}$ for the MSII, while the particle mass is
  $8.6\times10^8\solarmass\,{\rm h}^{-1}$ for the MS and
  $6.89\times10^6\solarmass\,{\rm h}^{-1}$ for MSII.  Both simulations assume a
  WMAP1 cosmology, with $\Omega_{m}=0.25$, $\Omega_{b}=0.045$,
  $\Omega_{\lambda}=0.75$, $h=0.73$, and $\sigma_8=0.9$.  Recent measurements
from PLANCK \citep{planck2015} and WMAP9 \citep{bennett2013} provide slightly
different cosmological parameters and, in particular, a larger value for
$\Omega_{m}$ and a lower one for $\sigma_8$.  As shown in previous work,
however, these differences are expected to have little influence on model
predictions, once model parameters are returned to reproduce a given set of
observables in the local Universe \citep{wang2008,guo2013}.

In the following, we will briefly summarize the physical prescriptions that are
relevant for this work, and discuss in detail our modifications to include a
modelling of star formation that depends on the amount of molecular
hydrogen.

\subsection{Star formation and stellar feedback in the GAEA model}
\label{subsec:gaea}

In our work, we will assume as a reference `fiducial' model the one presented
in HDLF16 including prescriptions for stellar feedback based on the {\sc FIRE}
simulations, plus the modifications discussed below in
Sections~\ref{subsec:diskmodel} and \ref{subsec:bhmodel}.

In this model, the rate of star formation depends on the amount of `cold gas',
defined as all gas with temperature below $10^4$~K, associated with a model
galaxy. In particular, we assume:  
\begin{equation}
\dot{M}_{\star}= \alpha_{\rm sf}\times M_{\rm sf}/\tau_{\rm dyn},
\end{equation}
where $\alpha_{sf}=0.03$ is the efficiency at which gas is converted into
stars, and $\tau_{\rm dyn}= r_{\rm disk}/V_{\rm vir}$ is the dynamical time of
the galaxy. $r_{\rm disk}$ is the radius of the star forming region.
We assume this is equal to $3$ times the scale length of the disk (assuming an
exponential profile, as in our case, this means that the star forming region
includes $\sim 99.6$ per cent of the total gas mass).  $V_{\rm vir}$ is the
virial velocity of the parent substructure (or the virial velocity at the last
time there was a resolved subhalo for orphan galaxies.)

$M_{\rm sf}$ is the amount of gas available for star formation. Following
\citet{delucia2008}, this is computed by integrating the surface density of 
the cold gas disk, assumed to be exponential, out to the radius ($r_{\rm
crit}$) at which the gas surface density drops below the following critical
value
\citep{kennicutt1989}:

\begin{equation}
\Sigma_{\rm crit}[{\rm M}_{\odot}\,{\rm pc}^{-2}] = 0.59 \times V_{\rm vir}
      [{\rm km}/{\rm s}]/( r_{\rm disk}[\rm kpc]).
\end{equation}
GAEA features a detailed treatment for chemical enrichment that accounts for
the finite lifetime of stars and its dependence on stellar mass, and allows us
to trace individual chemical abundances and non-instantaneous recycling of
metals, gas, and energy. We refer to \citet{delucia2014} for a detailed
description of the relevant prescriptions. Briefly, our model includes separate
sets of chemical yields for Asymptotic Giant Branch stars (AGBs) and both
Supernovae Type Ia (SnIa - the main contributors of iron-peak elements) and
Type II (SnII - that mainly release $\alpha$ elements, including O, Mg, Si, S,
Ca). The assumed delay time distribution for SnIa corresponds to a fraction of
prompt\footnote{`Prompt' is here defined as exploding within $10^8$~yr from the
star formation episode. The fraction increases to about 23 per cent when
considering SnIa events within $4\times10^8$~yr.}  SnIa of about 5 per
cent. When a star formation event takes place, our code stores the information
about the metals, energy and mass of Helium and Hydrogen that will be available
at any time in the future. These information are then included as galaxy
evolution proceeds forward in cosmic time. \citet{delucia2014} argue that this
approach provides a more accurate accounting of the timings and properties of
individual star formation events than alternative methods based on the storage
and binning of the past star formation history of model galaxies. We note that
all previous semi-analytic models that include an explicit treatment of the
partition between atomic and molecular
gas \citep{fu2010,lagos2011a,somerville2015} are based on an
instantaneous recycling approximation.

The energy released by supernovae and stellar winds is assumed to reheat some
of the cold gas in the disk and to drive large-scale galactic winds, ejecting
gas out of the parent halo. Our model for stellar feedback is based on
parametrizations extracted from the FIRE hydrodynamical simulations
\citep{hopkins2014,muratov2015}. Specifically, the reheating rate of the cold
gas depends on the star formation rate and scales both with redshift and with
the potential well of the galaxy: 
\begin{equation}
\dot{M}_{\rm reheat} = \epsilon_{\rm reheat}(1+z)^{1.25}\left(\frac{V_{\rm
    max}}{60\;{\rm km}/{\rm s}}\right)^{\alpha} \times \dot{M}_{\star}.
\end{equation}
$V_{\rm max}$ is the maximum circular velocity of the parent halo. 
When $V_{\rm max} < 60 \,{\rm km}/{\rm s}$, the index $\alpha$ is $-3.2$, while 
for larger values of $V_{\rm max}$, $\alpha = -1.0$. The reheating efficiency,
$\epsilon_{\rm reheat}$ is assumed to be constant and equal to $0.3$. 
The total energy released by massive stars can be expressed as:
\begin{equation}
\dot{E} = \epsilon_{\rm eject}(1+z)^{1.25}\left(\frac{V_{\rm max}}{60\;{\rm km}/{\rm
    s}}\right)^{\alpha} \times  0.5\cdot \dot{M}_{\star}\cdot V_{\rm SN}^2
\end{equation}
where $0.5\,V_{\rm SN}^2$ is the mean kinetic energy of SN ejecta per unit mass
of stars formed, and $\epsilon_{\rm eject}=0.1$ is the ejection efficiency.  An
ejection rate can then be computed as:
\begin{equation}
\dot{M}_{\rm
  eject}=\frac{\dot{E}-0.5\dot{M}_{\rm reheat}V^2_{\rm vir}}{0.5V^2_{\rm vir}} 
\end{equation}
Following the approach by \citet{henriques2013}, we assume that ejected gas can
be re-accreted on a time-scale that depends on the virial mass of the parent
halo. 

As discussed in HDLF16, this stellar feedback scheme allows us to to reproduce
the measured evolution of the galaxy stellar mass function, and the observed
correlation between galaxy stellar mass and its gaseous/stellar metallicity
content. In particular, this model also reproduces the observed evolution of
the mass-cold gas metallicity relation to higher redshift. This is an important
aspect of our reference model since some of the star formation laws we will
discuss below depend explicitly on the metallicity of the cold gas.

\subsection{Disk sizes}
\label{subsec:diskmodel}
As explained above, the rate at which gas is converted into stars depends
sensibly on the size of the gaseous disk. As described below, this is the case
also for the fraction of molecular to atomic hydrogen. In the GAEA model, no
distinction is made between the sizes of gaseous and stellar discs. Both are
assumed to have an exponential surface density profile:
\begin{equation}
\Sigma_{\rm disk} = \Sigma_0 \, {\rm exp}\left(-\frac{r}{r_{\rm disk}}\right)
\label{eqn:diskprofile}
\end{equation}
where $\Sigma_0= M / 2\pi r^2_{\rm disk}$, with $M$ equal to the mass of cold
gas or stars in the disk, and $r_{\rm disc}$ the scale length of the (gaseous
and stellar) disk. Assuming conservation of specific angular momentum, cold gas
is assumed to settle in a rotationally supported disk with scale-length given
by:
\begin{equation}
r_{\rm disk}=\frac{\lambda}{\sqrt{2}} R_{200}
\label{eqn:disksize}
\end{equation}
where $\lambda$ is the spin parameter of the dark matter halo, and $R_{200}$ is
the radius within which the mean mass density is $200$ times of the critical
density of the Universe \citep{mo1998}. At each time-step, the scale-length of
the disk is recomputed by taking the mass-weighted average gas profile of the
existing disk and that of the new material being accreted (cooling).
 
In this study, we use an improved modelling of disk sizes which distinguishes
between gas and stellar discs and allows them to grow continuously in mass and
angular momentum in a physically plausible fashion. Specifically, we follow the
model introduced by \citet{guo2011} that we briefly summarize here. When gas
cools onto galaxies, we assume it carries a specific angular momentum, $j_{\rm
cooling}$, that matches the current value of the parent friend-of-friend
halo. \footnote{Recent hydrodynamical simulations have shown that
cooling gas carries a few times the specific angular momentum of the halo
\citep{danovich2015,stevens2016}. We plan to analyse consequences of these 
findings in our model in future work.}  The gaseous disk gains angular
momentum $\bm{J}_{\rm cooling}=\bm{j}_{\rm cooling}
\times M_{\rm cooling}$ during cooling, where $M_{\rm cooling}$ is the mass of
new cooling gas.  When star formation occurs, we assume that the stars formed
have the same specific angular momentum of the gaseous disk, $\bm{j}_{\rm
SF}$. When gas is recycled to the inter-stellar medium, it carries the same
specific angular momentum of the stellar disk $\bm{j}_{\rm
recycling}$. Finally, during galaxy mergers, the angular momentum of the
accreted gas $\bm{J}_{\rm acc,gas}$ and accreted stars $\bm{J}_{\rm acc,\star}$
are transferred from the merging satellites to the remnant centrals. The
variation of the total angular momentum vector of the gaseous disk, during one
time-step of integration, can then be expressed as:
\begin{equation}
\Delta \bm{J}_{\rm gas}= \bm{J}_{\rm cooling} -\bm{J}_{\rm SF} + \bm{J}_{\rm  recycling} 
+ \bm{J}_{\rm acc,gas}, 
\end{equation}
while for the stellar disk we can write:
\begin{equation}
\Delta \bm{J}_{\star}= \bm{J}_{\rm SF} - \bm{J}_{\rm recycling} + \bm{J}_{\rm acc,\star}.
\end{equation}
Assuming both the stellar and gaseous disks have an exponential profile, their
scale-lengths can be expressed as: 
\begin{equation}
r_{\rm gas,d} = \frac{J_{\rm gas}/M_{\rm gas}}{2V_{\rm max}}, \qquad
r_{\star,d} = \frac{J_{\rm \star,d}/M_{\rm \star,d}}{2V_{\rm max}} 
\end{equation}
where $V_{\rm max}$ is the maximum circular velocity of the host halo. 

In Appendix~\ref{app:disk}, we compare the disk sizes resulting from our
updated model to those from HDLF16. The updated model predicts  significantly
larger gas and stellar disks than HDLF16 at the massive end. Nevertheless, these
difference cause negligible variations for other properties like e.g.  the
stellar mass function, and the mass-metallicity relation.

\subsection{Black hole growth model}
\label{subsec:bhmodel}

In the GAEA model, the growth of super-massive black holes occurs both during
galaxy mergers, by accretion of cold disc gas and by merging with each other
(this is the so-called `quasar-mode'), and through hot gas accretion from
static haloes (the so-called `radio-mode').

Specifically, when a satellite with baryonic mass $M_{\rm sat}$ merges with a
galaxy of mass $M_{\rm cen}$, the black hole accretion rate is modelled
following \citet{kh2000} and \citet{croton2006}:

\begin{equation}
\dot{M}_{\rm BH,qmode} = \frac{f_{\rm BH}\cdot (\frac{M_{\rm sat}}{M_{\rm
cen}})  \cdot
  M_{\rm cold}}{(1 + 280\,{\rm km}\,{\rm s}^{-1}/V_{\rm vir})^2}
\end{equation}
where $f_{BH}=0.03$ is a free parameter, tuned to reproduce the local relation
between the black-hole mass and the bulge mass. $M_{\rm cold}$ 
is the cold gas mass of both central galaxy and satellite galaxy.
and $V_{\rm vir}$ is the virial velocity of the host halo.

For black holes hosted by central galaxies of static haloes:
\begin{equation}
 \dot{M}_{\rm BH,rmode}=\kappa_{\rm radio}\frac{M_{\rm BH}}{10^8
   M_{\odot}\,{\rm h}^{-1}}\frac{f_{\rm hot}}{0.1}\left(\frac{V_{\rm 200}}{200\,{\rm
     km}\,{\rm s}^{-1}}\right)^3 
\end{equation} 
where $f_{\rm hot}=M_{\rm hot}/M_{\rm 200}$ is the hot gas ratio, and
$\kappa_{\rm radio}= 10^{-3}$ is the accretion efficiency. 

In GAEA, as well as in all previous versions of the model adopting the same
formulation, the accretion rates driven by galaxy mergers are not Eddington
limited. So, effectively, black holes are created by the first gas rich galaxy
mergers. We find that this scheme introduces significant resolution problems,
particularly when adopting models where the star formation efficiency depends
on the metallicity of the cold gas component. In this case, star formation is
delayed in low-metallicity galaxies leading to an excess of cold gas that
drives very high accretion rates during later mergers. The net effect is that
of a systematic increase of the black hole masses, and therefore a stronger
effect of the radio-mode feedback. We discuss this issue in detail in
Appendix~\ref{app:resolution}.

To overcome these problems, we introduce a black hole seed at the centre of
haloes with virial temperatures above $10^4$~K (cooling is suppressed below 
this limit). The mass of the black hole seed is assumed to scale with that of
the parent halo according to the following relation:  
\begin{equation}
M_{\rm BH} = \left(\frac{M_{\rm 200}}{10^{10}M_{\odot}\,{\rm
      h}^{-1}}\right)^{1.33} \times \frac{10^{10}\,M_{\odot}{\rm h}^{-1}}{3 \times
  10^6}  
\label{eq:bhvc}
\end{equation}
The power low index $1.33$ is derived assuming $M_{\rm BH}\propto V_c^4$ as
found in \citet[][see also \citealt{dimatteo2003}]{volonteri2011}, and 
using $V_c\propto (1+z)^{1/2}M_{200}^{1/3}$ \citep{mo2002}. We neglect here
the redshift dependence in the last equation. The mass of black hole seeds 
in our model ranges from $1000-10^{5}\solarmass$ in the MS, and 
$10-10^{4}\solarmass$ in the MSII.

Some recent studies
\citep{sabra2015,bogdan2015} argue for a weaker relation between the black
hole mass and circular velocity. We note, however, that we use
Equation~\ref{eq:bhvc} only at high redshift, to generate the black hole
seeds. Later on, black holes grow through accretion and mergers following
the specific modelling discussed above. The normalization in
Equation~\ref{eq:bhvc} is chosen to obtain a good convergence for the black
hole-stellar mass relation at redshift $z=0$ (see Appendix~\ref{app:resolution}).  
Both the quasar and radio mode accretion rates onto black holes are Eddington 
limited in our new model. 

\subsection{Star formation laws}
\label{subsec:sfr}

As described in Section~\ref{subsec:gaea}, our fiducial GAEA model assumes that
stars form from the total reservoir of cold gas, i.e. all gas that has cooled
below a temperature of $10^4$~K. This is inconsistent with the observational
studies referred to in Section~\ref{sec:intro}, showing that the star formation
rate per unit area correlates strongly with the surface density of molecular
gas. In order to account for these observational results, it is necessary to
include an explicit modelling for: (i) the transition from atomic (HI) to
molecular (H$_2$) hydrogen, and (ii) the conversion of H$_2$ into stars. We
refer to these two elements of our updated model as `star formation law', and
consider four different models that are described in detail in the
following. 

In all cases, we assume that the star formation rate per unit area
of the disk is proportional to the surface density of the molecular gas:
\begin{equation}
  \Sigma_{\rm sf} = \nu_{\rm sf}\Sigma_{{\rm H}_2} 
  \label{eqn:sf_law}
\end{equation}
where $\nu_{\rm sf}$ is the efficiency of the conversion of H$_2$ into
stars,  and assumes a different expression for different star formation
laws.  In the following, we also assume that Helium, dust and ionized gas
account for $26$~per cent of the cold gas at all redshifts. The remaining gas
is partitioned in HI and H$_2$ as detailed below. As in previous studies
\citep{fu2010,lagos2011a,somerville2015}, we do not attempt to model
self-consistently the evolution of molecular and atomic hydrogen. Instead, we
simply consider the physical properties of the inter-stellar medium at each
time-step of the evolution, and use them to compute the molecular hydrogen
fraction. This is then adopted to estimate the rate at which H$_2$ is converted
into stars. We only apply the new star formation law to quiescent star
formation events. Merger driven star-bursts (that contribute to a minor
fraction of the cosmic star formation history in our model) are treated
following the same prescriptions adopted in our fiducial GAEA
model \citep{hirschmann2016}.

\begin{figure*}
\includegraphics[width=0.7\textwidth]{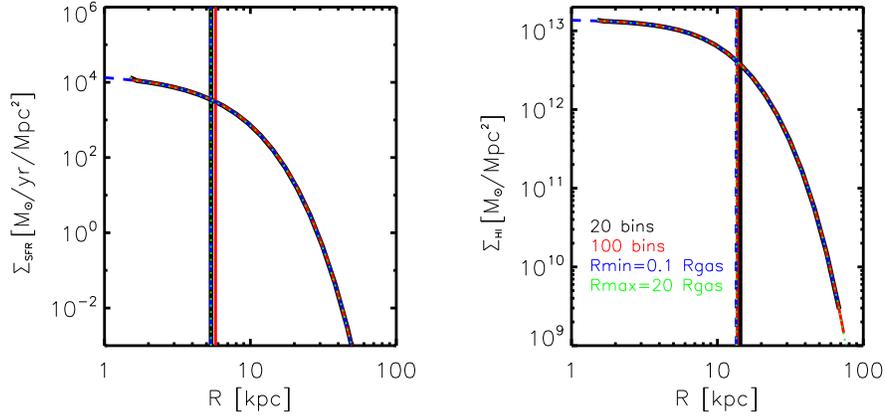}
\caption{Star formation (left panel) and HI (right panel) surface density profiles for one
  particular galaxy at $z=0$ in different runs. These correspond to a number of
  bins larger (red) than our default choice (black), smaller inner radius
  (blue), and larger outer radius (green). This figure refers to the BR06
  model, but results are similar for the other models considered. The vertical
  lines mark the effective radius. }
\label{fig:profile}
\end{figure*}

In all models considered, both the star formation time-scale and molecular
hydrogen ratio depend on the gas surface density. In our calculations, we
divide the gaseous disk in 20 logarithmic annuli from $0.2\, r_{\rm gas,d}$ to
$10\,r_{\rm gas,d}$, where $r_{\rm gas,d}$ is the scale length of the cold gas
disk and is computed as detailed in Section~\ref{subsec:diskmodel}. For each
annulus, we compute the fraction of molecular hydrogen and the corresponding
star formation rate. Equation~\ref{eqn:sf_law} becomes: 
\begin{equation}
  \Sigma_{{\rm sf}, i} = \nu_{\rm sf,i}\Sigma_{{\rm H}_2, i}.
\end{equation}
where $\Sigma_{{\rm sf},i} $, $\nu_{{\rm sf},i}$, and $\Sigma_{{\rm
H}_2, i}$ represent the average SFR density, star formation efficiency, and molecular
surface density in each annulus (with $i$ going from 1 to 20). Then the total
star formation rate is:
\begin{equation}
\dot{M_{\star}}=\sum_{i=1}^{20} \Sigma_{{\rm sf}, i} S_{i}.
\end{equation}
where $S_i $ is the area of each annulus. The annuli are not `fixed' as in
e.g. \citet{fu2010}, but recomputed for each star formation episode. We checked
that results are not significantly affected by the number and size of the rings. 
In particular, we carried out test runs using 100 annuli, a larger  outer 
radius ($[0.2\,r_{\rm rgas,d},20\,r_{\rm gas,d}]$), or a smaller inner radius 
($[0.1\,r_{\rm rgas,d},10\,r_{\rm gas,d}]$), and find little difference in the final properties of galaxies. 
Fig.~\ref{fig:profile} shows the surface 
density profile of the star formation rate and HI for one particular galaxy 
at $z=0$. Only results for one of the models described below (the BR06) are 
shown, but these are similar for all models considered. The vertical lines 
mark the effective radius, defined as the radius that includes half of the 
total SFR or half of the HI mass. 
We find that different choices for the division of the disks in annuli 
cause less than $5$ per cent differences for the sizes of the cold gas disks and 
stellar disks, for all galaxies above the resolution limit of our simulations.
 We verify that also the relations between 
SFR, HI mass, stellar disk sizes and stellar mass are not significantly 
affected by different choices for the number or the size of the annuli. 

In the next subsections, we discuss in detail the four star formation
laws used in our study. Their parameters have been chosen to reproduce the
galaxy stellar mass function, HI mass function, and H$_2$ mass function (less
weight has been given to this observable because of the relevant uncertainties
in the CO to H2 conversion) at $z=0$ using the MS . All parameters entering the
modelling of other physical processes are kept unchanged with respect to our
fiducial model.

\subsubsection{The \citet{blitz2006} star formation law (BR06)}
\label{subsec:br06}

This star formation law is based on the relation observed in local galaxies
between the ratio of molecular to atomic hydrogen ($R_{\rm mol}$) and the
mid-plane pressure acting on the galactic disc ($P_{\rm ext}$) \citep{blitz2006}. Specifically:
\begin{equation}
R_{\rm mol,br} = \frac{\Sigma_{{\rm H}_2}}{\Sigma_{\rm HI}} = \left(\frac{P_{\rm ext}}{P_0}\right)^{\alpha}
\end{equation}
where $P_0$ is the external pressure of molecular clumps. Based on their sample
of 14 nearby galaxies, \citet{blitz2006} find $P_0$ ranging between $0.4
\times 10^{4} {\rm cm}^{-3}$~K and $7.1 \times10^4 {\rm cm}^{-3}$~K, and values
for $\alpha$ varying between $0.58$ and $1.64$. We assume log$(P_0 /k_B
[{\rm cm}^{-3}{\rm K}]) = 4.54 $ and $\alpha=0.92$, that correspond to the mean
values.

The hydrostatic pressure at the mid-plane can be written as follows 
\citep{elmegreen1989}:
\begin{equation}
P_{\rm ext} = \frac{\pi}{2} G\Sigma_{\rm gas}[\Sigma_{\rm gas}+f_{\sigma}\Sigma_{\star}] 
\label{eq:br}
\end{equation}
where $\Sigma_{\rm gas}$ and $\Sigma_{\star}$ are the surface density of the
cold gas and of the stars in each annulus, and $f(\sigma)=\sigma_{\rm
gas}/\sigma_{\star}$ is the ratio between the vertical velocity dispersion of
the gas and that of the stellar disk.  We assume a constant velocity dispersion
for the gaseous disk of $\sigma_{\rm gas}=10 \,{\rm km}\,{\rm s}^{-1}$
\citep{leroy2008}, while for the stellar disk we follow \citet{lagos2011a} 
and assume $\sigma_{\star}=\sqrt{\pi G h_{\star}\Sigma_{\star}}$ and $h_{\star}=r_{\rm \star,d}/7.3$, 
based on observations of nearby disc galaxies
\citep{kregel2002}. For pure gaseous disks, Eq.~\ref{eq:br} is simplified by
setting to zero the stellar surface density. 

Following \citet{lagos2011a}, we assume for this model: 
\begin{equation}
\nu_{\rm sf,br} = \nu_{{\rm br},0}[1+\left(\frac{\Sigma_{\rm gas}}{\Sigma_{0,{\rm
        br}}}\right)^{0.4}]
\end{equation}
where $\Sigma_{0,{\rm br}} = 200 \,{\rm M}_{\odot}\,{\rm pc}^{-2}$ is the
critical density where the slope of the relation between $\Sigma_{\rm SFR}$ and
$\Sigma_{\rm H_2}$ steepens \citep{bigiel2008}. $\nu_{{\rm br}, 0}=0.4 \,{\rm
Gyr}^{-1}$ is chosen to reproduce the observed H$_2$ mass function and galaxy 
stellar mass function at z=0.

\subsubsection{The \citet{krumholz2009b} star formation law
  (KMT09)} 

In a series of papers, \citet{krumholz2008,krumholz2009a,krumholz2009b}
developed an analytic model to determine the fraction of molecular hydrogen,
within a single atomic-molecular complex, resulting from the balance between
dissociation of molecules by interstellar radiation, molecular self-shielding,
and formation of molecules on the surface of dust grains. Accounting for the
fact that the ratio between the intensity of the dissociating radiation field
and the number density of gas in the cold atomic medium that surrounds the
molecular part of a cloud depends (weakly) only on metallicity 
\citep{wolfire2003}, the molecular to total fraction can be written as:
\begin{equation}
 f_{\rm mol,kmt}=\frac{\Sigma_{{\rm H}_2}}{\Sigma_{{\rm H}_2}+\Sigma_{\rm
     HI}}=1-[1+\left(\frac{3}{4}\frac{s_{\rm kmt}}{1+\delta}\right)^{-5}]^{-1/5} ,
\end{equation}
where,
\begin{equation}
 s_{\rm kmt} ={\rm ln}(1+0.6\chi_{\rm kmt})/(0.04\,\Sigma_{{\rm comp},0}\,Z'),
\end{equation}
\begin{equation}
 \chi_{\rm kmt}=0.77(1+3.1\,Z'^{0.365}),
\label{eqn:chikmt09}
\end{equation}
\begin{equation}
 \delta = 0.0712\,(0.1s^{-1}_{\rm kmt}+0.675)^{-2.8},
\end{equation}
and
\begin{equation}
 \Sigma_{{\rm comp},0} = \Sigma_{\rm comp}/(1\,{\rm M}_{\odot} {\rm pc}^{-2}).
\end{equation}
$\Sigma_{\rm comp}$ is the surface density of a giant molecular cloud (GMC) on
a scale of $\sim 100$~pc, and $Z'$ is the metallicity of the gas normalized to
the solar value (we assume $Z_{\odot}=0.02$). Following \citet{krumholz2009b},
we assume $\Sigma_{\rm comp}=f_c \Sigma_{\rm gas}$, where $f_{c}\geq 1$ is a
`clumping factor' that approaches 1 on scales close to 100 pc, and that we
treat as a free parameter of the model. In previous studies, values assumed for
this parameter range from $1.5$ \citep{fu2010} to $5$ \citep{lagos2011a}. In
our case, $f_{c}=3$ provides predictions that are in reasonable agreement with
data, while larger values tend to under-predict the HI content of massive
galaxies. \citet{krumholz2009b} stress that some of the assumptions made in
their model break at gas metallicities below roughly 5 per cent solar
($Z'<0.05$). As discussed e.g. in \citet{somerville2015}, POP III stars will
rapidly enrich the gas to metallicities $\sim 10^{-3}Z_{\odot}$ at high
redshift. Following their approach, when computing the molecular fraction, we
assume this threshold in case the metallicity of the cold gas is lower. We
adopt the same treatment also in the GK11 model and K13 models described below.

As for the efficiency of star formation, we follow \citet{krumholz2009b} and
assume:
\begin{equation}
\nu_{\rm sf,kmt} =\left\{\begin{matrix} 
\nu_{{\rm kmt},0} \times (\frac{\Sigma_{\rm gas}}{\Sigma_{\rm
    kmt}})^{-0.33},  
\Sigma_{\rm gas} < \Sigma_{\rm kmt}
\\
\\
\nu_{{\rm kmt},0} \times (\frac{\Sigma_{\rm gas}}{\Sigma_{\rm kmt}})^{0.33},
\Sigma_{\rm gas} > \Sigma_{\rm kmt}
\end{matrix}\right.
\end{equation}
where $\Sigma_{\rm kmt} = 85\,{\rm M}_{\odot}\,{\rm pc}^{-2}$ is the average
surface density of GMCs in Local Group galaxies \citep{bolatto2008}, and
$\nu_{\rm kmt,0}=0.38\,{\rm Gyr}^{-1}$ is the typical value found in GMCs of
nearby galaxies. We find a better agreement with H$_2$ mass function at
$z=0$ when using a slightly larger values for this model parameter: $\nu_{{\rm
kmt},0} = 0.5\,{\rm Gyr}^{-1}$.

\subsubsection{The \citet{krumholz2013} star formation law (K13)}
\label{sec:modelk13}

\citet{krumholz2013} extend the model described in the previous section to 
the molecule-poor regime (here the typical star formation rate is significantly
lower than that found in molecular-rich regions). KMT09 assumes the cold
neutral medium (CNM) and warm neutral medium (WDM) are in a two-phase
equilibrium. In this case, the ratio between the interstellar radiation field
($G'_0$) and the column density of CNM ($n_{_{\rm CNM}}$) is a weak function of
metallicity.  However the equilibrium breaks down in HI-dominated regions. 
Here, $G'_0$ and $n_{_{\rm CNM}}$ are calculated as summarized below.

The molecular hydrogen fraction can be written as:
\begin{equation}
f_{{\rm mol, k13}}=\left\{\begin{matrix}
1-(3/4)s_{\rm k13}/(1+0.25s_{\rm k13}) ,\; s_{\rm k13}<2
\\
\\ 
0,\qquad\qquad\qquad\qquad\qquad\qquad  s_{\rm k13}\geq2
\end{matrix}\right.
\end{equation}
 where,
 \begin{equation}
  s_{\rm k13}\approx \frac{{\rm ln}(1+0.6\chi_{\rm k13} +0.01 \chi^{2}_{\rm
      k13})}{0.6\tau_{c,{\rm k13}}},
 \end{equation}
\begin{equation}
 \tau_{c,{\rm k13}} = 0.066 f_{c} Z'\Sigma_{0,{\rm k13}},
 \label{eqn:kmt+tau}
\end{equation}
\begin{equation}
\chi_{\rm k13} = 7.2 \frac{G'_0}{n_{_{\rm CNM}}/10\,{\rm cm}^{-3}},
\label{eqn:chik13}
\end{equation}
and $\Sigma_{0,{\rm k13}}=\Sigma_{\rm gas}/ 1\,{\rm M}_{\odot}\,{\rm
pc}^{-2}$. 

As for the KMT09 model, we assume $f_{c}=3$ and use $Z'=0.001\,Z_{\odot}$ to
estimate the molecular fraction when the cold gas metallicity $Z_{\rm
gas}<10^{-3}Z_{\odot}$.  In the above equations, $\chi_{\rm k13}$ represents a
dimensionless radiation field parameter.  Our model adopts a universal initial
mass function (IMF) for star formation, both for quiescent episodes and
star-bursts. UV photons are primarily emitted by OB stars, and the UV
luminosity can be assumed to be proportional to the star formation rate. To
estimate $G'_0$, we use the star formation rate integrated over the entire
gaseous disk, averaged over the time interval between two subsequent snapshots
(this correspond to 20 time-steps of integration) \footnote{ A similar 
modelling has been adopted in \citet{somerville2015}. We note that a more 
physical expression for the intensity of the interstellar radiation field 
would be in terms of the surface density of the star formation rate. We have 
tested, however, that within our semi-analytic framework such alternative 
expression does not affect significantly our model predictions. Results of 
our tests are shown in Appendix~\ref{app:test_g0rhosd}.}. Specifically, we can write:
\begin{equation}
G'_0 \approx \frac{\dot{M}_{\star}}{\dot{M}_{\star,{\rm MW}}},
\end{equation}
and assume $\dot{M}_{\star,{\rm MW}}=1\,{\rm M}_{\odot}{\rm yr}^{-1}$ for the
total SFR of the Milky Way (observational estimates range from $0.68$ to $2.2\,
{\rm M}_{\odot}{\rm yr}^{-1}$, e.g. \citealt{murray2010,robitaille2010}).

$n_{_{\rm CNM}}$ is assumed to be the largest between the minimum CNM 
density in hydrostatic balance and that in two-phase equilibrium:
\begin{equation}
n_{_{\rm CNM}} = {\rm max}(n_{_{\rm CNM, 2p}}, n_{_{\rm CNM,hydro}}).
\end{equation} 
In particular, the column density of the CNM in two-phase equilibrium 
can be written as:
\begin{equation}
n_{_{\rm CNM, 2p}} = 23 G'_{0} \left(\frac{1+3.1Z'^{0.365}}{4.1}\right)^{-1} {\rm
cm}^{-3}, 
\end{equation}
while
\begin{equation}
 n_{_{\rm CNM, hydro}} = \frac{P_{th}}{1.1k_B T_{_{\rm CNM},max}}.
\end{equation}
$k_{B}$ is the Boltzmann constant, $T_{_{\rm CNM},max} = 243K$ is the maximum 
temperature of the CNM \citep{wolfire2003}, and $P_{th}$ is the thermal 
pressure at mid-plane \citep{ostriker2010}:
\begin{equation}
P_{th} = \frac{\pi G\Sigma^2_{\rm HI}}{4\alpha}\{1+2R_{\rm H_2}+\left[(1+2R_{\rm H_2})^2
+\frac{32\zeta_{\rm d} \alpha \sigma^2_{\rm gas}
 \rho_{\rm sd}}{\pi G \Sigma^2_{\rm HI}}\right]^{1/2}\} . 
\label{eqn:k13_pth}
\end{equation}
In the above equation, $R_{\rm H_2}={M_{\rm H_2}}/{M_{\rm gas}-M_{\rm H_2}}$ is
the molecular hydrogen mass after star formation at the last time-step, $M_{\rm
gas}$ is the current total cold gas mass, and $\rho_{\rm sd}$ is the
volume density of stars and dark matter. To compute the latter quantity, we
assume a NFW profile for dark matter haloes and assign to each halo, at a given
redshift and of given mass ($M_{200}$), a concentration using the calculator
provided by \citet{zhao2009}. Once the halo concentration is known, we can
compute the density of dark matter at a given radius. The volume density of
stars is computed assuming an exponential profile for the stellar disk and a
\citet{jaffe1983} profile for the stellar bulge. For the stellar disk height,
we assume $h_{\star}=r_{\rm \star,d}/7.3$.
The other parameters correspond to
the velocity dispersion of gas $\sigma_{\rm gas}=10 \,{\rm km}\,{\rm s}^{-1}$,
and a constant $\zeta_{\rm d} \approx 0.33$.

In the GMC regime, the free fall time of molecular gas is:
\begin{equation}
t_{\rm ff} = 31\Sigma_0^{-1/4} {\rm Myr} .
\end{equation}
Then the star formation efficiency of transforming molecular gas to stars is
given by:
\begin{equation}
\nu_{\rm sf,k13}
 =\frac{0.01}{31\,\Sigma_{0,{\rm k13}}^{-1/4}} {\rm Myr}^{-1}
\end{equation}

\subsubsection{The \citet{gnedin2011} star formation law (GK11)}
\label{subsubsec:gk11}

\citet{gnedin2011} carry out a series of high resolution hydro-simulations
including non-equilibrium chemistry and an on-the-fly treatment for radiative
transfer. Therefore, their simulations are able to follow the formation and
photo-dissociation of molecular hydrogen, and self-shielding in a
self-consistent way. \citeauthor{gnedin2011} provide a fitting function that
parametrizes the fraction of molecular hydrogen as a function of the dust-to-gas
ratio relative to that of the Milky Way ($D_{\rm MW}$), the intensity of the
radiation field ($G'_0$), and the gas surface density ($\Sigma_{\rm
gas}=\Sigma_{{\rm HI}+{\rm H}_2}$). In particular:
\begin{equation}
f_{\rm mol,gk}=\frac{\Sigma_{{\rm H}_2}}{\Sigma_{\rm gas}} = [ 1 +
  \frac{\Sigma_{\rm c}}{\Sigma_{\rm gas}} ] ^{-2},
\end{equation}
where $\Sigma_{\rm c}$ is a characteristic surface density of neutral gas
at which star formation becomes inefficient.
\begin{equation}
\Sigma_{c} = 20 \, {\rm M}_{\odot} {\rm pc^{-2}}\frac{\Lambda^{4/7}}{D_{\rm
    MW}}\frac{1}{\sqrt{1+G'_{0}D_{\rm MW}^2}},
\end{equation}
with:
\begin{equation}
\Lambda = {\rm ln}(1+gD_{\rm MW}^{3/7}(G'_{0}/15)^{4/7}),
\end{equation}
\begin{equation}
g=\frac{1+\alpha_{\rm gk} s_{\rm gk}+s^2_{\rm gk}}{1+s_{\rm gk}},
\end{equation}
\begin{equation}
s_{\rm gk}=\frac{0.04}{D_{\star}+D_{\rm MW}},
\end{equation}
\begin{equation}
\alpha_{\rm gk}=5\frac{G'_{0}/2}{1+(G'_{0}/2)^2},
\end{equation}
\begin{equation}
D_{\star}= 1.5 \times 10^{-3} {\rm ln}(1+(3G'_0)^{1.7}),
\end{equation}

Following GK11, we use the metallicity of cold gas to get the dust ratio:
$D_{\rm MW}\approx Z' = {Z_{\rm gas}}/{Z_{\odot}}$. For $G'_0$, we assume the
same modelling used for the K13 star formation law.  We note that the
simulations by \citet{gnedin2011} were carried out varying $D_{\rm MW}$ from
$10^{-3}$ to $3$, and $G'_0$ from $0.1$ and $100$.  Their fitting formulae
given above are not accurate when $D_{MW}\leq 0.01$. We assume 
$D_{\rm MW}=10^{-3}$ to calculate the molecular fraction when the cold gas 
metallicity $Z_{\rm gas}<10^{-3}Z_{\odot}$.

GK11 also provide the star formation efficiency necessary to fit the
observational results in \citet{bigiel2008} in their simulations:
\begin{equation}
\nu_{\rm sf, gk} 
=\frac{1}{0.8\, {\rm Gyr}}\times\left\{\begin{matrix}
\quad 1 \qquad\qquad \,\; \Sigma_{\rm gas} \geq \Sigma_{\rm gk}
\\
(\frac{\Sigma_{\rm gas}}{\Sigma_{\rm gk}})^{\beta_{\rm
gk}-1}\quad \; \Sigma_{\rm gas} < \Sigma_{{\rm gk}}
\end{matrix}\right.
\end{equation}
where $\Sigma_{\rm gas}$ is the surface density of cold gas, 
$\Sigma_{\rm gk} = 200\,{\rm M}_{\odot}{\rm pc}^{-2}$, and 
$\beta_{\rm gk} = 1.5$.


\section{The influence of different star formation laws on galaxy physical
properties} 
\label{sec:growandmf}

As mentioned in Section~\ref{sec:simsam}, we run our models on two
high-resolution cosmological simulations: the Millennium Simulation (MS), and
the Millennium II (MSII). Our model parameters are calibrated using the MS, and
merger trees from the MSII are used to check resolution convergence. The main
observables that are used to calibrate our models are: the galaxy stellar mass
function, and the HI and ${\rm H}_2$ mass functions at $z=0$.
A comparison between observational data and predictions from one of our models
(BR06) for galaxy clustering in the local Universe has been presented recently
in \citet{zoldan2016}.

In this section, we analyse in more detail the differences between the star
formation laws considered, and discuss how they affect the general properties
of galaxies in our semi-analytic model. Table~\ref{tab:model} lists all star
formation laws considered in this work and the corresponding parameters. 

\begin{table*}
\centering
\begin{tabular}{ | p{1.5cm} || p{5.cm} p{5.cm} p{3.5cm} p{0.001cm} |} 
\centering{Model (color)} & \centering{Molecular fraction} [$R_{\rm mol}=\frac{\Sigma_{{\rm H}_2}}{\Sigma_{\rm HI}}$, $f_{\rm mol}=\frac{\Sigma_{{\rm H}_2}}{\Sigma_{\rm gas}}$]& 
\centering{Star formation efficiency} [$\nu_{\rm SF}$, $\Sigma_{\rm SF}
= \nu_{\rm sf}\Sigma_{{\rm H}_2}$] & 
\centering{Model parameters} &  \\ \hline \hline

\centering{1. Fiducial (black)}  & 
\centering{Fixed molecular fraction  $R_{\rm mol}=0.4$.} &
\centering{$ \dot{M}_{\star}= \alpha_{\rm sf}\times M_{\rm sf}/\tau_{\rm dyn} $,
  	$\alpha_{\rm sf}=0.03$, $\tau_{\rm dyn}= r_{\rm disk}/V_{\rm vir}$} & 
\centering{same as in HDLF16}  & \\ 
\hline

\centering{2. BR06 (red)}  & 
\centering{
	$ R_{\rm mol,br} = (\frac{P_{\rm ext}}{P_0})^{\alpha}$,
	$P_{\rm ext} = \frac{\pi}{2} G\Sigma_{\rm gas}[\Sigma_{\rm
    	gas}+f_{\sigma}\Sigma_{\star}]$, $f(\sigma)\propto 1/\sqrt{r_{\star}\,\sigma_{\star}}$ 
	} &
\centering{$ 
	\nu_{\rm sf,br} = \nu_{{\rm br},0}[1+(\frac{\Sigma_{\rm gas}}{\Sigma_{0,{\rm
        br}}})^{0.4}] $} & 
\centering{ 
	 $\alpha = 0.92$, $P_0 /k_B [{\rm cm}^{-3}{\rm K}] = 10^{4.54} $
	 $\nu_{{\rm br}, 0}=0.4 \,{\rm Gyr}^{-1}$}  & \\
\hline

\centering{3. KMT09 (blue)}  & 
\centering{
	$f_{\rm mol,kmt}=1-[1+(\frac{3}{4}\frac{s_{\rm kmt}}{1+\delta})^{-5}]^{-1/5} $,
 	$\delta = 0.0712(0.1s^{-1}_{\rm kmt}+0.675)^{-2.8}$, 
	$s_{\rm kmt} = \frac{\ln (1+ 0.462(1+3.1Z)^{0.365})}{f_c \Sigma_{\rm gas}Z}$
	} &
\centering{$
	\nu_{\rm sf,kmt} =\nu_{{\rm kmt},0} \times (\frac{\Sigma_{\rm gas}}{\Sigma_{\rm
	    kmt}})^{-0.33}$ \\ 
	if $\Sigma_{\rm gas} < \Sigma_{\rm kmt}$, \\
	$\nu_{\rm sf,kmt} =\nu_{{\rm kmt},0} \times (\frac{\Sigma_{\rm gas}}{\Sigma_{\rm kmt}})^{0.33}$ \\
	if $\Sigma_{\rm gas} > \Sigma_{\rm kmt}$
	} & 
\centering{	
	$f_{c}=3$,\\
	$\nu_{\rm kmt,0} = 0.5\,{\rm Gyr}^{-1}$,\\
	$Z'_{\rm min}=0.001Z_{\odot}$, \\
	$\Sigma_{\rm kmt} = 85\,{\rm M}_{\odot}{\rm pc}^{-2}$
	} & \\ 
\hline

\centering{4. K13 (yellow)}  & 
\centering{
	$f_{{\rm mol, k13}}=1-(3/4)s_{\rm k13}/(1+0.25s_{\rm k13})$ \\ if $ s_{\rm k13}<2$, \\
	$f_{{\rm mol, k13}}=0$ if $s_{\rm k13}\geq2$,
	$s_{\rm k13}\approx \frac{{\rm ln}(1+0.6\chi_{\rm k13} +0.01 \chi^{2}_{\rm
  	k13})}{0.6f_{c} Z \Sigma_{0,{\rm k13}} }$,
	$\chi_{\rm k13} \propto \dot M_{\star}/n_{_{\rm CNM}}$,
	$n_{_{\rm CNM}}=max(n_{_{\rm CNM, 2p}},n_{_{\rm CNM,hydro}})$
	} &
\centering{$
	\nu_{\rm sf,k13}=\frac{\nu_{\rm k13, 0}}{\Sigma_{0,{\rm k13}}^{-1/4}}$,\\
	$\Sigma_{0,{\rm k13}}=\Sigma_{\rm gas}/ 1\,{\rm M}_{\odot}\,{\rm pc}^{-2}$
	} & 
\centering{
	$f_c =3$, \\
	$\nu_{\rm k13, 0} = 0.32\,{\rm Gyr}^{-1}$, \\
	$Z'_{min}=0.001 Z_{\odot}$
	} & \\ 
\hline

\centering{5. GK11 (green)}  & 
\centering{
	$f_{\rm mol,gk}= [ 1 + 	\frac{\Sigma_{\rm c}}{\Sigma_{\rm gas}} ] ^{-2}$, \\
	$\Sigma_{c} \propto \frac{\Lambda^{4/7}
	}{Z}\frac{1}{\sqrt{1+\dot M_{\star} Z^2}}$, \\
	$\Lambda \propto {\rm ln}(1+g\, Z^{3/7}(\dot{M}_{\star}/15)^{4/7})$, \\
	$g=\frac{1+\alpha_{\rm gk}\, s_{\rm gk}+s^2_{\rm gk}}{1+s_{\rm gk}}$, \\
	$s_{\rm gk}\propto\frac{1}{\ln (1+(3\dot{M}_{\star} )^{1.7}) +Z}$, \\
	$\alpha_{\rm gk}\propto \frac{\dot{M}_{\star}}{1+(\dot{M}_{\star}/2)^2}$
	} &
\centering{$
	\nu_{\rm sf, gk} = \nu_{\rm gk, 0} \times 1$ if $\Sigma_{\rm gas} \geq \Sigma_{\rm gk}$, \\
	$\nu_{\rm sf, gk} = \nu_{\rm gk, 0} \times (\frac{\Sigma_{\rm gas}}{\Sigma_{\rm gk}})^{\beta_{\rm gk}-1}$ \\ 
	if $\Sigma_{\rm gas} < \Sigma_{{\rm gk}}$
	} & 
\centering{
	 $\nu_{\rm gk, 0}=1.25\,{\rm Gyr}^{-1}$,\\
	 $\beta_{\rm gk} \approx 1.5$, \\
	 $Z'_{min}=0.001$ } , \\
	$\Sigma_{\rm gk} \approx 200\,{\rm M}_{\odot}pc^{-2}$ &  \\
\\ \hline
\end{tabular}
\caption{A summary of the star formation laws considered in this work,
including a list of the corresponding free parameters. Column 2 gives the
adopted parametrization of the molecular fraction, while column 3 gives the
assumed star formation efficiency. Column 4 lists the values assumed for the
model free parameters.}
\label{tab:model}
\end{table*}

\subsection{Differences between H$_2$ star formation laws}
\label{sec:diffsflaws}

As discussed in the previous section, the star formation laws used in this
study can be separated in a component given by the calculation of the 
molecular fraction $f_{\rm mol}=\Sigma_{\rm H2}/\Sigma_{\rm gas}$ (or $R_{\rm
mol}= \Sigma_{\rm H2}/\Sigma_{\rm HI}$) and one given by the star formation 
efficiency $\nu_{\rm sf}$. 

\begin{figure*}
\includegraphics[width=0.8\textwidth]{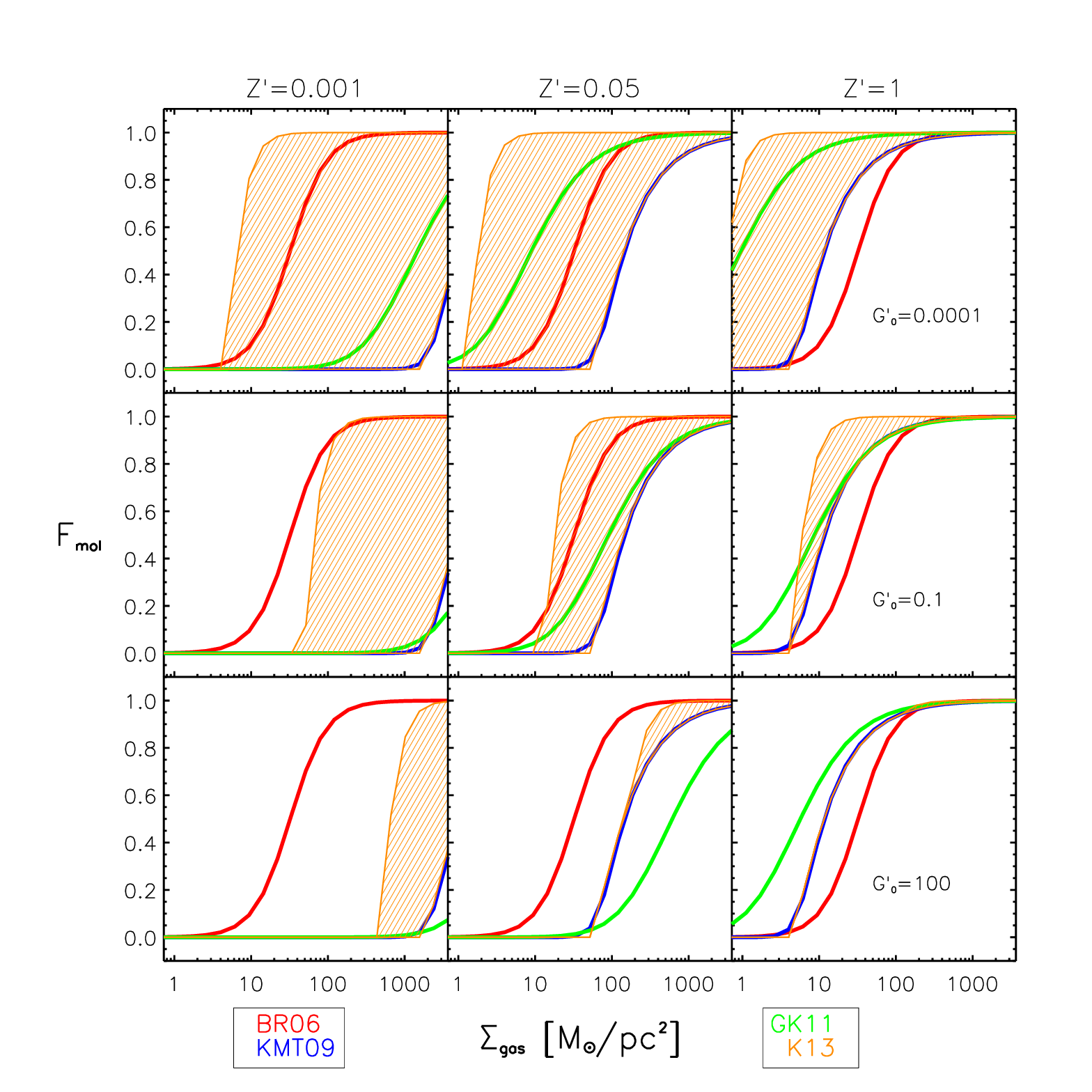}
\caption{The molecular fraction predicted by all models considered in this
study (different colours, as indicated in the legend), as a function of the
cold gas surface density. Different panels show results for different
interstellar radiation intensity ($G'_0 =
{\dot{M}_{\star}}/{\dot{M}_{\star,{\rm MW}}}$, different rows) and gas
metallicities ($Z' = Z_{\rm gas}/Z_{\odot}$, different columns) as
labelled. The stellar disk pressure is assumed to be zero for the BR06
model. The shaded area shows the range of possible values for the molecular
fraction corresponding to the K13 model (see details
in Sec.~\ref{sec:diffsflaws}).}
\label{fig:fmol}
\end{figure*}

\begin{figure}
\includegraphics[width=0.4\textwidth]{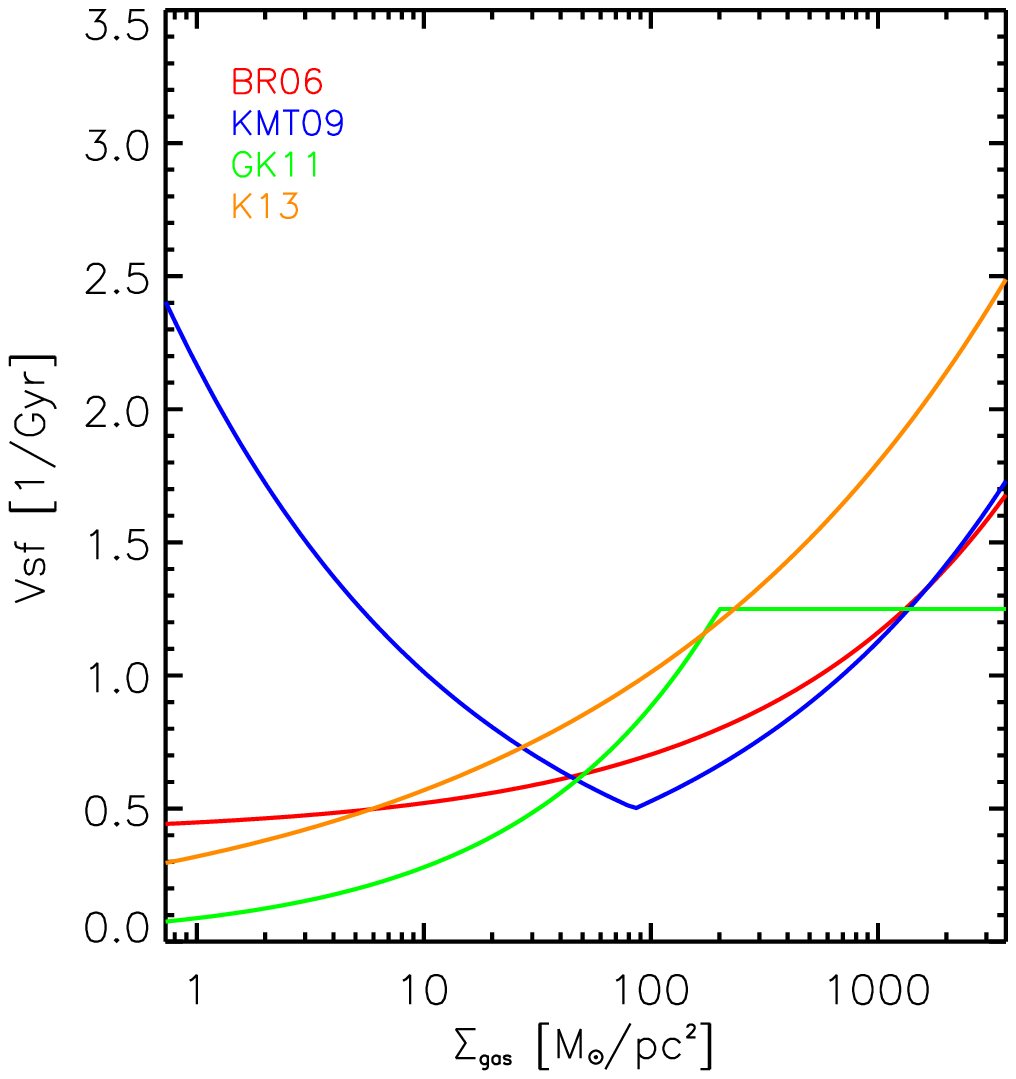}
\caption{Star formation efficiency as a function of cold gas surface density.
Different colours correspond to the different models considered in our study
as indicated in the legend.}
\label{fig:vsf}
\end{figure}

Fig.~\ref{fig:fmol} shows the molecular fraction predicted by the models
considered in this study in three bins of interstellar radiation intensities
and gas metallicity, as a function of the gas surface density. Lines of
different colours correspond to different models, as indicated in the legend.
The molecular fraction in BR06 depends only on the disk pressure, so the red
curve is the same in each panel. The stellar disk pressure is assumed to be
zero for the line shown. Assuming a positive value for the pressure of the
stellar disk, BR06 would predict a slightly higher $f_{\rm mol}$, but this
would not affect our conclusions.  In the K13 model, the molecular fraction
calculation is based on the molecular ratio at last time-step
(Equation~\ref{eqn:k13_pth}). The shaded region shown in the figure highlights
the minimum and maximum value for the molecular fraction, corresponding to the
case its value at the previous time-step is $f_{\rm mol}=1$ (H$_2$-dominated
region) or $f_{\rm mol}=0$ (HI-dominated region) respectively.
Since we do not have halo information for K13, we assume
$\rho_{\rm sd}=2.6\times10^{-5}Q_{\rm g}^2 \frac{\Sigma^2_{\rm
gas}}{1 \solarmass pc^{-2}} \solarmass pc^{-3}$ and $Q_{\rm
g}=2$ \citep[][see equation 35]{krumholz2013}. In
Appendix~\ref{app:test_g0rhosd}, we show that this assumption gives results
that are very similar to those obtained using the approach described in
Section~\ref{sec:modelk13} to compute $\rho_{\rm sd}$.

The predicted molecular fraction differs significantly among the models
considered. For a metal poor galaxy with little star formation and therefore
low interstellar radiation (this would correspond to the initial phases of
galaxy formation), BR06 and K13 predict higher molecular fraction than GK11 and
KMT09 (top left panel). At fixed radiation intensity, an increase of the gas
metallicity corresponds to an increase of the molecular fraction predicted by
the all models but BR06. This is because a higher gas metallicity corresponds to a
larger dust-to-gas ratios, which boosts the formation of hydrogen molecules.
For the highest values of gas metallicity considered (top right panel) the GK11
model produces the highest molecular fraction, BR06 the lowest. When the
interstellar radiation increases (from top to bottom rows) hydrogen molecules
are dissociated more easily and so the molecular fraction, at fixed metallicity
and gas surface density, decreases. In particular, the GK11, KMT09 and K13
models predict a very low molecular fraction for the lowest metallicity and
largest radiation intensity considered (bottom left panel). As metallicity in
cold gas increases, GK11 predicts more molecular gas than the other models.  As
expected by construction, in H$_2$-dominated region, K13 gives similar
molecular fraction to KMT09. For metal-rich galaxies (right column), GK11
predicts more molecular gas than the other models, particularly at low surface
densities. The lowest molecular fractions are instead predicted by the BR06
model. 

Fig.~\ref{fig:vsf} shows the star formation efficiency corresponding to the
four star formation laws implemented, as a function of the gas surface density 
(see third column of Table~\ref{tab:model}).
BR06 and K13 predict an increasing star formation efficiency $\nu_{\rm sf}$ with
increasing surface density of cold gas. GK11 predicts a monotonic increase of
the star formation efficiency up to gas surface density $\sim 100\,{\rm
M}_{\odot}\,{\rm pc}^{-2}$ and then a flattening. Finally, the KMT09 model
predicts a decreasing star formation efficiency up to $\Sigma_{\rm
gas}=85 \solarmass/pc^2$. For higher values of the gas surface density, the
predicted star formation efficiency increases and is very close to that
predicted by the BR06 model. It is interesting to see if these different
predictions translate into a correlation between the star formation rate
surface density and gas surface density that is in agreement with the latest
observations.

\begin{figure*}
\includegraphics[width=1.\textwidth]{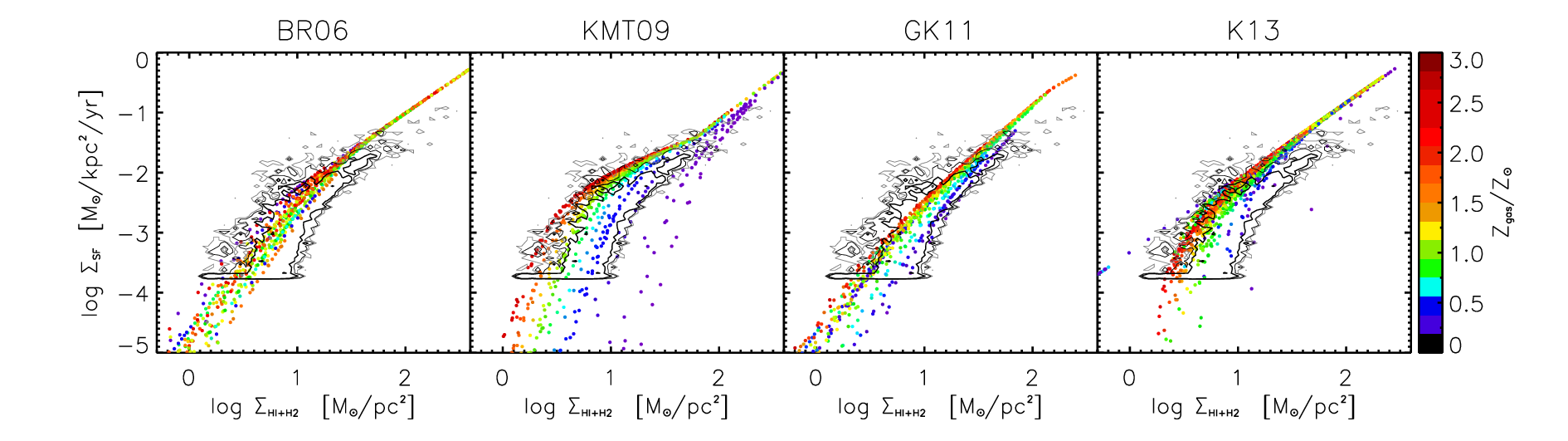}
\caption{The star formation rate surface density against neutral gas surface
density.  Coloured dots are results of model galaxies at redshift $z=0$ with 
different colours responding to different metallicity of the cold gas. Black
contours show the distribution of observed galaxies from \citet{bigiel2010}
(only points in the optical disk are included). Different star formation laws 
are shown in different panels.}
\label{fig:sfdensity}
\end{figure*}

Fig~\ref{fig:sfdensity} shows the surface density of star formation rate
$\Sigma_{\rm sf}$ against the surface density of neutral gas $\Sigma_{\rm HI+H_2}$.
We select galaxies in MSII at redshift $z=0$ and compare with observational
estimates compiled in \citet{bigiel2010}.  Dots
correspond to the surface density of star formation rate and neutral gas in
each annulus of model galaxies. Their colour indicates their cold gas
metallicity. The figure shows that all four star formation laws considered in
our work reproduce observations relatively well. The dependence on
metallicity for the KMT09 model is obvious. In the GK11 and K13 models, the
star formation rate depends also on the radiation intensity and the metallicity
dependence is weaker. \citet{somerville2015} present their predicted 
$\Sigma_{\rm sf}-\Sigma_{\rm HI+H_2}$ relation in their Fig.~6. They find a
clear metallicity dependence also for their prescription where H2 is determined
by the pressure of the interstellar medium, while for our BR06 model we do not
find a clear dependence on metallicity.  We believe that the reason is 
the different chemical enrichment models. \citet{somerville2015} use a 
fixed yield parameter, which naturally leads to a tight relation between 
stellar surface density and cold gas metallicity. In contrast, our model 
includes a detailed recycling and the metallicity of the cold gas and the 
disk pressure are not highly correlated for our simulated galaxies.

\subsection{The growth of galaxies in models with different star formation laws}
\label{subsec:grow}

\begin{figure*}
\includegraphics[width = 0.7\textwidth]{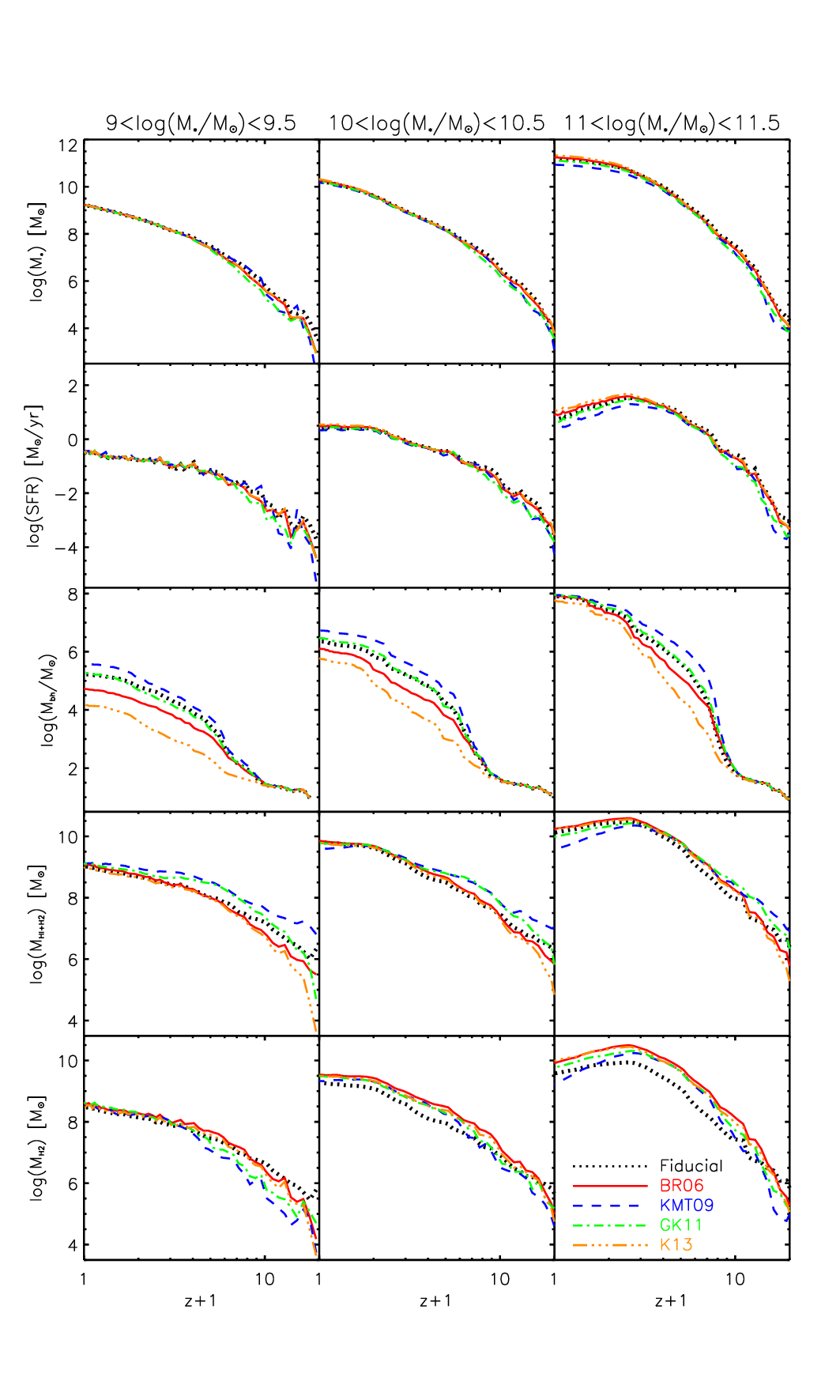}
\caption{The average growth history of $100$ randomly selected central galaxies 
in three stellar mass bins at $z=0$. Galaxies are selected in our fiducial
model and their growth history is compared to the corresponding results based
on runs using the different star formation laws considered in this
study. Different panels (from top to bottom) show the mean evolution
of the stellar mass, the SFR, the central black hole's mass, the mass of
neutral hydrogen, and the H$_2$ mass.}
\label{fig:grow_ms2}
\end{figure*}

To show the influence of different star formation laws on the star formation
history of model galaxies, we select a sample of central model galaxies
in our fiducial model and compare their history to that of the same galaxies
modelled using the different star formation laws considered. In particular, we
randomly select $100$ galaxies in three stellar mass bins in the fiducial
model\footnote{The final stellar masses are not significantly different in the
other models, as shown in Fig.~\ref{fig:grow_ms2}}: $\log (M_{\star}
/M_{\odot})\sim [9,9.5],[10,10.5],[11,11.5]$. For each galaxy, we trace
back in time its main progenitor (the most massive progenitor at each node of
the galaxy merger tree). Fig.~\ref{fig:grow_ms2} compares the average growth
histories of these galaxies. For this analysis, we use our runs based on
the MSII. The HI and H$_2$ masses in the fiducial model are obtained assuming
a constant molecular ratio ${M_{{\rm H}_2}}/{M_{\rm HI}}=0.4$.

Let us focus first on galaxies in the lowest mass bin considered ($9 <
log(M_{\star}) < 9.5\, \solarmass$ at $z=0$, the left column in
Fig.~\ref{fig:grow_ms2}). In all H$_2$-based star formation laws considered,
star formation starts with lower rates than that in our fiducial model. This
happens because the amount of molecular hydrogen at high redshift is
lower than that in the fiducial model (see bottom left panel).  In addition, star
formation in the fiducial model takes place only after the gas surface density
is above a critical value, so most galaxies in this model form stars 
intermittently (this does not show up because Fig.~\ref{fig:grow_ms2} shows 
a mean for a sample of galaxies): once enough gas is accumulated, stars can form at a rate
that is higher than that predicted by our H$_2$-based star formation laws. Then
for one or a few subsequent snapshots, the star formation rate is again
negligible until the gas surface densities again overcomes the critical
value. In contrast, for the H$_2$-based models considered, star formation at
early times is low but continuous for most of the galaxies.
Predictions from the BR06 and K13 models are very close to each other while the
slowest evolution is found for the KMT09 model. The cold gas masses of low mass
galaxies are different between models at early times. KMT09 and GK11 predict
more cold gas than fiducial model, while BR06 and K13 predict the lowest cold
gas mass.  All models converge to very similar values at $z\sim 5$
for stellar mass and SFR, within a factor of $1.5$. The mass of
molecular hydrogen converges only at $z\sim2$. The average mass of cold gas
remains different until present (at $z=0$ the mass of cold gas predicted by
KMT09 model is about $1.3$ times of that predicted by K13 model).

For the other two stellar mass bins considered (middle and right columns in
Fig.~\ref{fig:grow_ms2}), the trends are the same, but there are larger
differences at low redshift. In particular, for the most massive bin
considered, the amount of molecular gas in the fiducial model stays almost
constant at redshift $z < 2$, while it decreases for the other models. This
is particularly evident for the KMT09 model and is due to the fact that the
black hole mass is larger and therefore the AGN feedback is more efficient. 
For the same reason, both the star formation rate and the stellar mass
predicted by this model are below those from the other ones over the same
redshift interval. 

As explained in Section~\ref{subsec:bhmodel}, black holes grow through smooth
accretion of hot gas and accretion of cold gas during galaxy mergers. Galaxies
in the fiducial model have more cold gas than those in BR06 and K13 at early
times, thus the fiducial model predicts more massive black holes. The KMT09 and
GK11 models predict even more massive black holes because, when mergers take 
place there are significant amounts of cold gas available that has not yet 
been used to form stars. 

\subsection{The galaxy stellar mass function}
\label{subsec:smf}

\begin{figure*}
\includegraphics[width=0.8\textwidth]{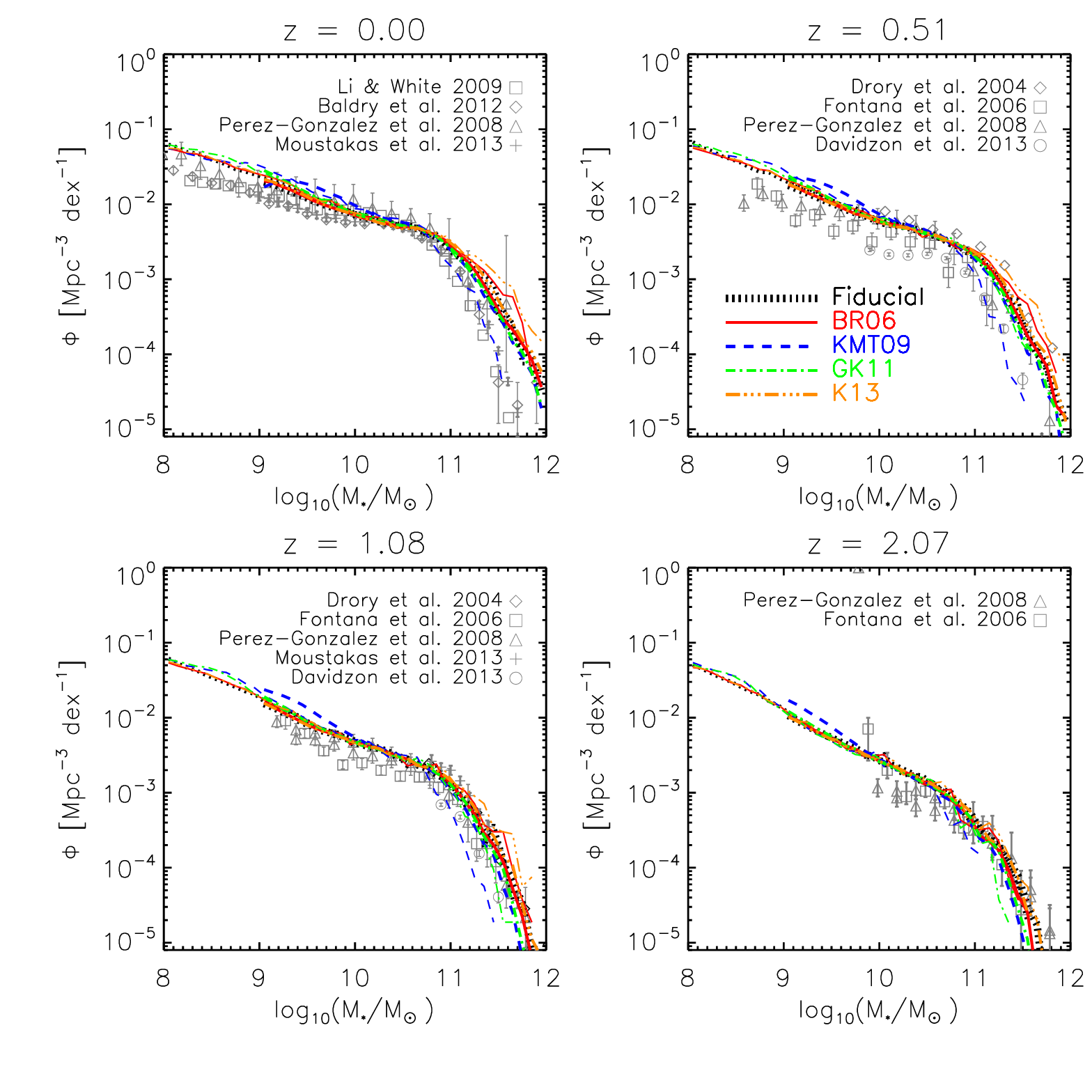}
\caption{Galaxy stellar mass functions at redshift $z=0$, $z\sim0.5$, $z\sim1$,
and $z\sim2$.  Gray symbols show different observational
estimates \citep{li2009,
baldry2012,perezgonz2008,moustakas2013,drory2004,fontana2006,davidzon2013},
while lines of different colours and types correspond to different star
formation laws, as indicated in the legend. Thicker lines are used for the MS,
while thinner lines correspond to the MSII. }
\label{fig:smf}
\end{figure*}

Fig.~\ref{fig:smf} shows the galaxy stellar mass functions predicted by the
different models considered in our study and compare them to observational
measurements at different cosmic epochs. In this figure (and in all the
following), thicker lines are used for the MS (about $1/50$ of the
entire volume) and thinner lines for the MSII (about $1/5$ of the
volume), while different colours correspond to different star formation
laws. We note that the stellar mass function corresponding to our fiducial
model run on the MS at $z=0$, shows a higher number density of massive galaxies
with respect to the results published in HDLF16. We verified that this is due
to our updated black hole model (see Appendix~\ref{app:resolution}).

Predictions from all models are close to those obtained from our fiducial
model, at all redshifts considered. The KMT09 and GK11 models tend to predict
lower number densities for galaxies above the knee of the mass function,
particularly at higher redshift. This is due to the fact that black holes in
KMT09 and GK11 are slightly more massive than in the fiducial model. 
In contrast, black holes in the BR06 and K13 models are less massive 
than those in the fiducial model for the MSII. As a consequence, the BR06 and 
K13 models predict more massive galaxies above the knee of the mass function 
with respect to the fiducial model. We have not been
 able to find one unique parametrization for the black hole seeds, or
 modification of the black hole model, that are able to provide a good
 convergence between the MS and MSII for all four star formation laws in our
 study.  Below the knee of the mass function, model predictions are very close
 to each other with only the KMT09 model run on the MS predicting slightly
 larger number densities. The predictions from the same model based on the MSII
 are very close to those obtained from the other models, showing this is
 largely a resolution effect.

\subsection{The HI and H$_2$ mass functions}
\label{subsec:ISM1}

\begin{figure*}
\includegraphics[width=0.8\textwidth]{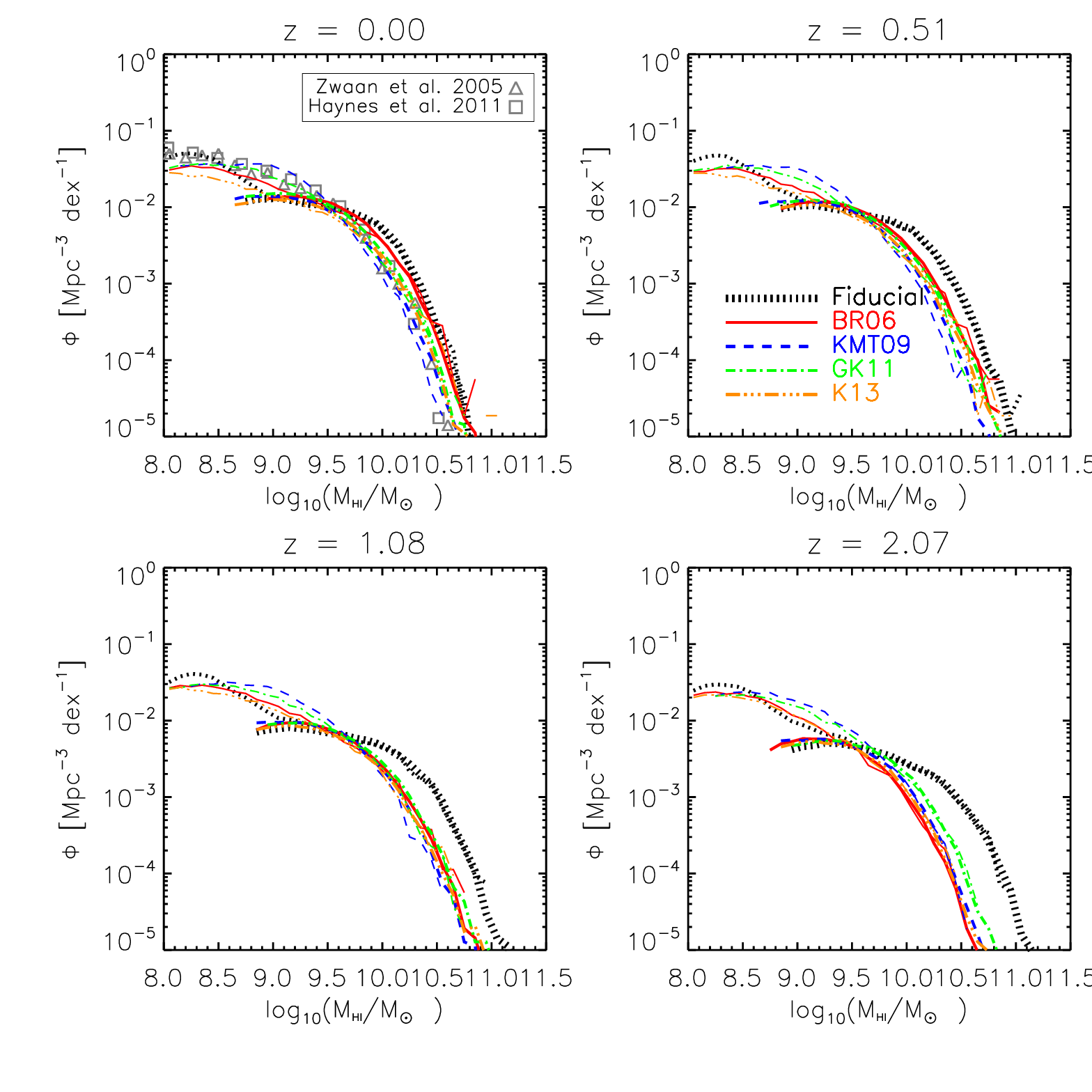}
\caption{HI mass function at redshift $z=0$, $z\sim0.5$, $z\sim1$, $z\sim 2$. 
  Gray triangles and squares show the observational measurements by 
  \citet{zwaan2005} and \citet{haynes2011}, respectively. 
  Thicker lines are used for the MS, while thinner lines correspond to the MSII.}
\label{fig:himf}
\end{figure*}

Fig.~\ref{fig:smf} shows that the galaxy stellar mass function is 
complete down to $\sim 10^9 \solarmass$ for the MS and $\sim 10^8 \solarmass$
for the MSII. Only galaxies above these limits are considered in this section.

Fig.~\ref{fig:himf} shows the predicted HI mass function from all models used
in this study. For our fiducial model, we assume a constant molecular fraction
of $M_{{\rm H}_2} /M_{\rm HI}=0.4$ to estimate the amount of HI from the total
cold gas associated with model galaxies. The grey symbols correspond to
observational data by \citet[][triangles]{zwaan2005}
and \citet[][squares]{haynes2011}.

All models agree relatively well with observations at $z=0$, by construction
(we tune the free parameters listed in Table~\ref{tab:model} so as to
obtain a good agreement with the HI and H$_2$ mass function at $z=0$). Comparing
results based on the MS and MSII, the figure shows that resolution does not
affect significantly the number densities of galaxies with HI mass above $\sim
10^{9.5}\solarmass$ at all redshifts. Below this limit, the number density
predicted from all models run on the MS are significantly below those obtained
using the higher resolution simulation. The fiducial model tends to predict
higher number densities of HI rich galaxies, particularly at higher
redshift. This is due to the fact that all H$_2$-based star formation laws
predict increasing molecular fractions with increasing redshift, in
qualitative agreement with what inferred from observational
data \citep[e.g.][]{popping2015}.

While it is true that predictions from the other models are relatively close to
each other, the figure shows that there are some non negligible differences
between them. In particular, the KMT09 model tend to predict the lowest number
densities for galaxies above the knee, and the highest number densities for HI
masses in the range $\sim 10^{8.5} - 10^{9.5}\solarmass$. This is because
massive galaxies in the KMT09 model tend to have more massive black holes than
in other models so that radio mode AGN feedback is stronger.  In the same
model, low mass galaxies tend to have lower star formation rates at high
redshift and are therefore left with more cold gas at low redshift (see
Fig.~\ref{fig:grow_ms2}).  The BR06 model has the opposite behaviour, predicting the largest
number densities for galaxies above the knee (if we exclude the fiducial model)
and the lowest below. The differences between the models tend to decrease with
increasing redshift: at $z\sim 2$ all models are very close to each other with
only the GK11 model being offset towards slightly higher number densities.

\begin{figure*}
\includegraphics[width=0.8\textwidth]{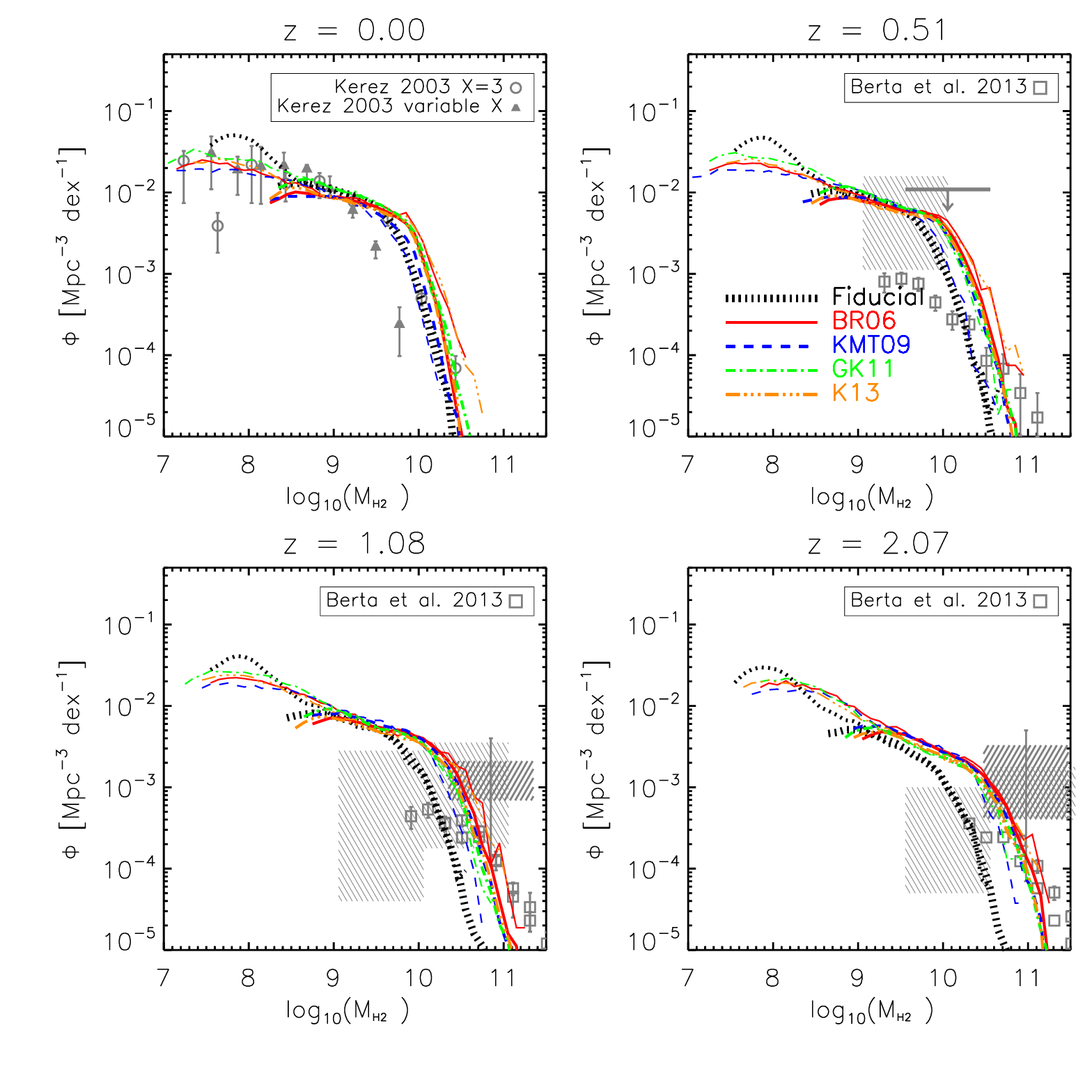}
\caption{H$_2$ mass function at redshift $z=0$, $z\sim0.5$, $z\sim1$, $z\sim
2$. Thicker lines are used for the MS, while thinner lines correspond to the
  MSII.  The observational measurements at $z=0$ are
  from \citet{keres2003}. Open circles correspond to the case $X_{\rm CO}=3$,
  while triangles correspond to a variable $X_{\rm CO}$. Open squares at
  higher redshift are from \citet{berta2013}. They are based on indirect
  estimates of the molecular mass, and include only normal star forming
  galaxies. Gray shaded regions are based on blind CO detections
  by \citet{walter2014}, and \citet{decarli2016} and assume $\alpha_{\rm
  CO}=3.6 \solarmass{\rm (K\,km\,s^{-1}\,pc^{-2})^{-1}}$.}
\label{fig:h2mf}
\end{figure*}

Fig.~\ref{fig:h2mf} shows the H$_2$ mass function from redshift $z\sim2$ to
$z=0$. The observational measurements at $z=0$ are based on the CO luminosity
function by \citet{keres2003}, and assume a constant CO/H$_2$ conversion factor
$X_{\rm CO}=3$ or a variable one \citep{obreschkow2009a}. All models
over-predict the number density of galaxies with log$(M_{\rm H2}) \gtrsim 9$
when considering a variable CO/H$_2$ conversion factor. Results based on the
fiducial and KMT09 model are consistent with measurements based on a constant
conversion factor. The other models tend to predict more H$_2$ at the high mass
end. The trend is the same at higher redshift. Here, we compare our model
predictions with estimates by \citet{berta2013}. These include only main
sequence galaxies and are based on a combination of PACS far-infrared and
GOODS-HERSCHEL data. The molecular mass is estimated from the star formation
rate, measured by using both far-infrared and ultra-violet photometry.  All
models tend to over-predict significantly the number densities of galaxies with
H$_2$ below $\sim 10^{10.5}\solarmass$. This comparison should, however, be
considered with caution as measurements are based on an incomplete sample and
an indirect estimate of the molecular gas mass. We also include, for 
comparisons, results of blind CO surveys \citep{walter2014,decarli2016}. These 
are shown as shaded regions in Fig.~\ref{fig:h2mf}.

For the H$_2$ mass function, resolution starts playing a role at $\sim
10^{8.6}\solarmass$ at $z=0$, but the resolution limit increases significantly
with redshift: at $z\sim 2$ the runs based on the MS become incomplete at
H$_2$ masses $\sim 10^{9.3}\solarmass$. Resolution also has an effect for the
H$_2$ richest galaxies for the KMT09, BR06, and K13 models. We find that this
is due to the fact that black holes start forming earlier in higher resolution
runs, which affects the AGN feedback and therefore the amount of gas in the
most massive galaxies.

To summarize, all star formation laws we consider are able to reproduce the
observed stellar mass function, HI mass function, and H$_2$ mass function.  We
obtain a good convergence between MS and MSII at $M_{\star}>10^9 \solarmass$
for the galaxy stellar mass function, $M_{HI}>10^{9.5}\solarmass$ for the HI
mass function, and $M_{\rm H_2}>10^{8.5}-10^{9.5}\solarmass$ from $z=0$ to
$z=2$ for the H$_2$ mass function. As explained above, model
predictions do not converge for the massive end of the galaxy stellar mass
function and H$_2$ mass function, and this is due to a different effect of AGN feedback (see
Appendix~\ref{app:resolution}).  We do not find significant differences
between predictions based on different star formation laws. Based on these
results, we argue that it is difficult to discriminate among different star
formation laws using only these statistics, even when pushing the
redshift range up to $z\sim2$, and including HI and H$_2$ mass as low as
$M_{\star}\sim10^8 \solarmass$.  Indeed, the systematic differences we find
between different models are very small. Our results also indicate that there
are significant differences between results obtained by post-processing model
outputs and those based on the same physical model but adopting an implicit
molecular based star formation law.

\section{Scaling relations}
\label{sec:scaling}
In this section, we show scaling relations between the galaxy stellar mass and
other physical properties related directly or indirectly to the amount of gas
associated with galaxies, at different cosmic epochs. In order to increase the
dynamic range in stellar mass considered and the statistics, we take advantage
of both the MS and the MSII. In particular, unless otherwise stated, we use all
galaxies with $M_{\star}>10^{10}\solarmass$ from the former simulation, and all
galaxies with $M_{\star}>10^8 \solarmass$ from the latter. As shown in the
previous section, and discussed in detail in Appendix~\ref{app:resolution}, the
convergence between the two simulations is good, and we checked that this
is the case also for the scaling relations as discussed below.

\subsection{Atomic and molecular hydrogen content}

\begin{figure*}
\includegraphics[width = 1.\textwidth]{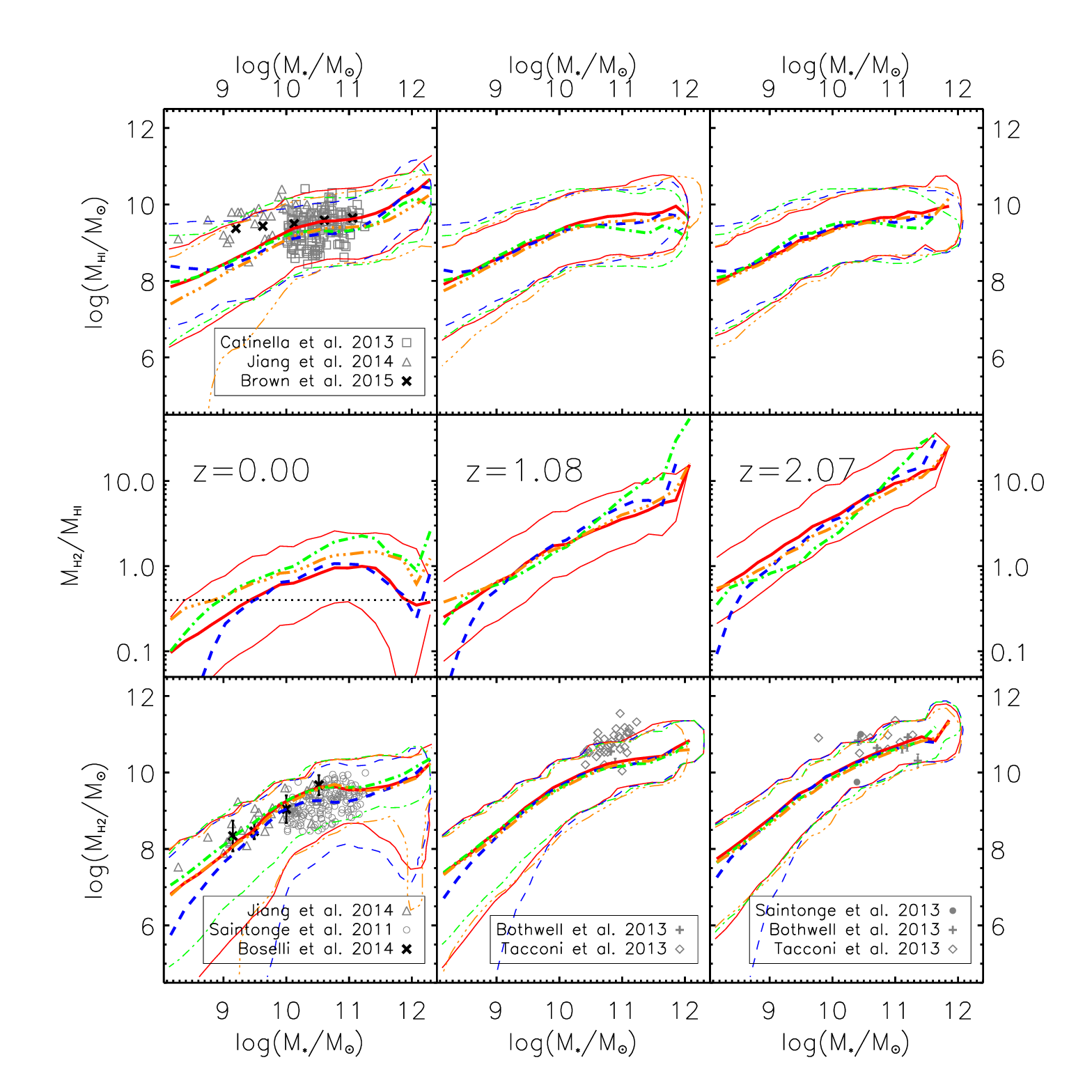}
\caption{From top to bottom panels: HI content of galaxies, ratio between H$_2$
and HI, and H$_2$ mass as a function of the galaxy stellar mass. Different
columns correspond to different redshifts, as indicated in the legend. Symbols
correspond to observational measurements
from \citet{catinella2013}, \citet{saintonge2011,saintonge2013}, \citet{boselli2014}
, \citet{jiang2015}, \citet{bothwell2013}, and \citet{tacconi2013}.  Colored
curves show results from the different models considered in this study,
combining the MS and MSII as described in the text. Thin lines
correspond to contours enclosing $95$ per cent of the galaxies in each stellar
mass bin, while thicker lines correspond to the median of the
distributions. The thin red lines in the middle panel show the 16th and 84th
percentiles for the BR06 model. The other models have a similar scatter.}
\label{fig:scalings}
\end{figure*}

We begin with a comparison between model predictions and observational data for
the amount of atomic and molecular hydrogen associated with galaxies of
different stellar mass, and at different cosmic epochs. This is shown in
Fig.~\ref{fig:scalings}, for all models used in our study. The top panels show
the predicted relation between the HI mass and the galaxy stellar mass, and
compare model predictions with observational estimates of local galaxies from
the GASS survey \citep[][squares]{catinella2013} and from a smaller sample (32
galaxies) with HI measured from ALFALFA \citep[][triangles]{jiang2015}. The
former survey is based on a mass-selected sample of galaxies with $M_{\star} >
10^{10}\solarmass$, while the sample by \citet{jiang2015} includes only star
forming nearby galaxies, and is therefore biased towards larger HI
masses. \citet[][black multiplication sign]{brown2015} provide average results
of NUV-detected galaxies from ALFALFA.  Contours show the distribution of model
galaxies indicating the region that encloses $95$ per cent of the galaxies in
each galaxy stellar mass bin considered. All models predict a similar and
rather large scatter, with results consistent with observational measurements
at $z=0$ for galaxies with stellar mass between $10^{10}$ and
$10^{11}\solarmass$. For lower mass galaxies, all models tend to predict lower
HI masses than observational estimates. This is in part due to the fact that
observed galaxies in this mass range are star forming. If we select star
forming galaxies (${\rm sSFR}>0.1/{\rm Gyr}$) from the BR06 model, the median
mass of HI is $0.3$ dex higher (but still lower than data) than that obtained by
considering all model galaxies.  The relation between HI and stellar mass (as
well as the amplitude of the scatter) evolves very little as a function of
cosmic time.

The middle panels of Fig.~\ref{fig:scalings} show the molecular-to-atomic
ratio as a function of the galaxy stellar mass at different redshifts. At
$z=0$, the ratio tends to flatten for galaxy masses larger than $\sim
10^{10}\solarmass$ and its median value is not much larger than the canonical
$0.4$ that is typically adopted to post-process models (shown as the dotted line in the
left-middle panel) that do not include an explicit partition of the cold gas
into its atomic and molecular components.  For lower galaxy stellar masses, the
molecular-to-atomic ratio tends to decrease with decreasing galaxy mass due to
their decreasing gas surface density.
The BR06 and KMT09 models predict the lowest molecular-to-atomic ratios at
$z=0$, while the GK11 model the highest. At higher redshifts, the relation
becomes steeper also at the most massive end, differences between the different
models become less significant, and the overall molecular-to-atomic ratio tends
to increase at any value of the galaxy stellar mass. Specifically, galaxies
with stellar mass $\sim 10^9\,\solarmass$ have a molecular-to-atomic ratio of
about $0.24$ at $z=0$, $\sim 0.53$ at $z\sim 1$, and $\sim 0.9$ at
$z\sim2$. For galaxies with stellar mass $\sim 10^{11}\,\solarmass$, the
molecular-to-atomic gas ratio varies from $\sim 1.4$ at $z=0$ to $\sim 11.6$ at
$z\sim 2$.  The evolution of the molecular ratio is caused by the evolution of
the size-mass relation: galaxies at high redshift have smaller size and higher
surface density than their counterparts at low redshift. The relations shown
in the middle panel clarify that a simple post-processing adopting a constant
molecular-to-atomic ratio is a poor description of what is expected on the
basis of more sophisticated models. One could improve the calculations by 
assuming a molecular-to-atomic ratio that varies as a function of redshift 
and galaxy stellar mass. We note, however, that there is a relatively large 
scatter in the predicted relations that would not be accounted for.

The bottom panels of Fig.~\ref{fig:scalings} shows, the molecular hydrogen mass
as a function of galaxy stellar mass. Symbols correspond to different
observational measurements. At $z=0$, filled circles are used for data from the
COLDGAS survey \citep{saintonge2011}. These are based on CO(1-0) line
measurements and assume $\alpha_{\rm CO}=3.2\solarmass/(K\,km/s\,pc^2)$ to
convert CO luminosities in H$_2$ masses. Data from \citet[][open
triangles]{jiang2015} include only main sequence star forming galaxies, are
based on CO(2-1) lines, and assume $\alpha_{\rm CO}=4.35 \solarmass/({\rm
K\,kms^{-1}\,pc^{-2}})$. \citet{boselli2014} provide mean values and 
standard deviations of late-type galaxies, classified by morphology and selected 
from the Herschel Reference Survey, with a constant conversion factor 
$\alpha_{\rm CO}=3.6\solarmass/(K\,km/s\,pc^2)$. 
The samples observed at higher redshift are less
homogeneous and likely biased. Measurements by \citet[][dots]{saintonge2013}
are for a sample of 17 lensed galaxies with measurements based on CO(3-2) lines
and metallicity-dependent conversion factors. Data
from \citet[][diamonds]{tacconi2013} are for a sample of 52 star forming
galaxies with measurements based on CO(3-2) lines and assuming $\alpha_{\rm
CO}=4.36 \solarmass/({\rm K\,km\,s^{-1}\,pc^{-2}})$. Galaxies from their sample cover the redshift
range from $0.7$ to $2.3$; we plot all those below $z\sim 1.3$ in the middle
panel and all those above $z\sim1.7$ in the right panel. \citet{bothwell2013}
give data for 32 sub-millimetre galaxies and assume $\alpha_{\rm CO}=1 \solarmass/({\rm K\,km\,s^{-1}\,pc^{-2}})$. As for the
top panels, thick lines show the median relations predicted from the different
star formation laws considered in our paper, while the thin contours mark the
region that encloses $68$ per cent of the galaxies in each stellar mass bin.
At $z=0$, observational data are close to the median relations obtained for the
different models. The data by \citet{jiang2015}, as well as most of
those considered at higher redshift, tend to be above the median relations
although all within the predicted scatter. We verify that this is still
the case even when considering only main sequence star forming galaxies at
$z\sim 1$. Similar results were found by \citet{popping2014}.

\subsection{Galaxy stellar mass - cold gas metallicity relation}
\label{subsec:gzm}

\begin{figure*}
\includegraphics[width=1.\textwidth]{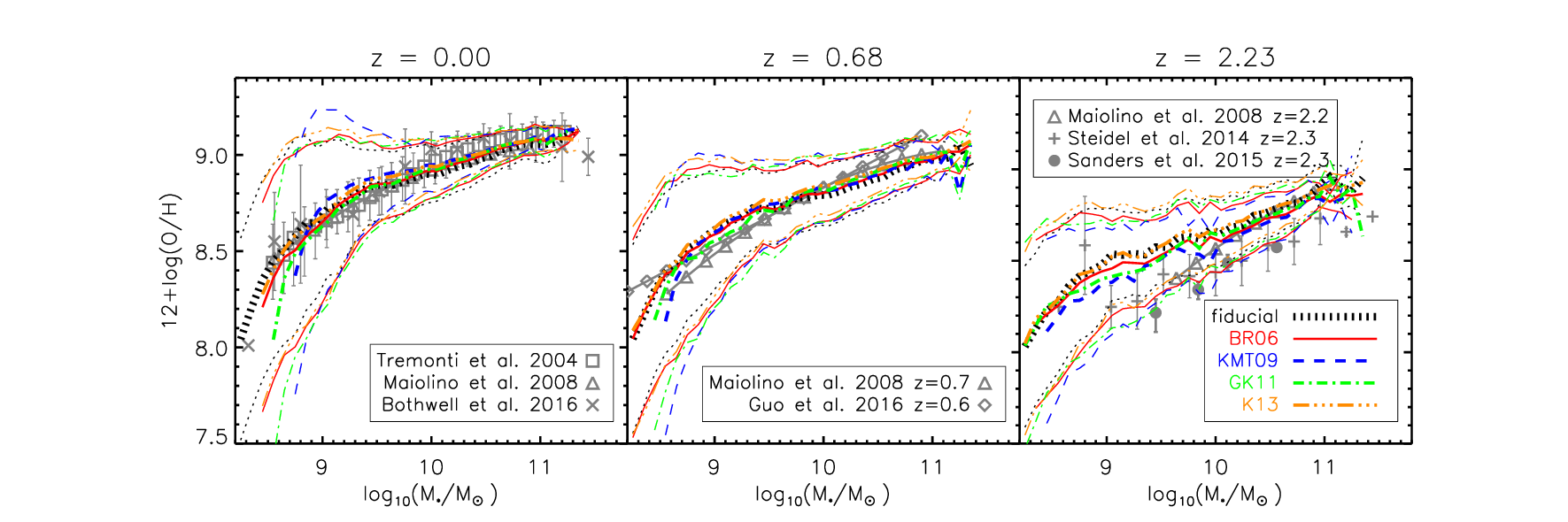}
\caption{The relation between the cold gas metallicity and galaxy stellar mass. 
 Gray symbols with error bars show observational measurements, while colored
lines correspond to the different star formation laws considered in this
study. We only select star forming galaxies ($sSFR > 0.3/t_H$),
without a significant AGN ($M_{\rm BH}<10^6\,\solarmass$), and with cold gas
fraction $M_{\rm gas}/(M_{\rm gas}+M_{\star})>0.1$. In \citet{tremonti2004},
they used a \citet{kroupa2001} IMF to calculate stellar mass. We shift it to a 
Chabrier IMF by dividing the observed masses by a factor $1.06$. Thin
 lines in each panel show the scatter predicted for the BR06 model (the
scatter has similar amplitude for the other star formation laws).}
\label{fig:gzm}
\end{figure*}

Three of the star formation laws used in this study include an explicit
dependence on the metallicity of the cold gas component. Therefore, it is
important to verify that the observed correlation between the galaxy stellar
mass and the gas metallicity is reproduced. Fig.~\ref{fig:gzm} shows the
oxygen abundance of cold gas\footnote{We remove helium ($26\%$) from 
cold gas to get the abundance of Hydrogen, whereas HDLF16 did not. Therefore 
our results for the fiducial model are different from those of the FIRE 
model in Fig.~6 of HDLF16.} from redshift $z=0$ to $z\sim 2$ predicted by all
models considered in this study, and compares model predictions with different
observational measurements. For this figure, we select star forming galaxies 
($\dot{M}_{\star}/M_{\star} >0.3/t_H $, where $t_H$ is the Hubble
time), with no significant AGN ($M_{\rm BH} <10^6\, \solarmass$), and with gas
fraction $M_{\rm gas}/(M_{\rm gas}+M_{\star}) >0.1$. We used this selection in
an attempt to mimic that of the observational samples, that mainly include
star forming galaxies. 

Model results are in quite good agreement with data and predictions from the
different models are relatively close to each other. At $z\sim 2$, all models
tend to over-predict the estimated metallicities compared to
observational measurements by \citet{steidel2014} and \citet{sanders2015}. Our
model predictions are, instead, very close to the measurements for galaxies 
more massive than $\sim 10^{10}\solarmass$ by \citet{maiolino2008}. 
Fig.~\ref{fig:gzm} shows that the GK11 and KMT09
model predict slightly lower gas metallicities for low mass galaxies at the
highest redshift shown. The mass-metallicity relation shown in
Fig.~\ref{fig:gzm} extends the dynamic range in stellar mass shown in HDLF16,
where we also used a slightly different selection for model galaxies. While we
defer to a future study a more detailed comparison with observational data at
the low-mass end, we note that our model is the only published one that
reproduces the estimated evolution of the mass-metallicity relation up
to $z\sim0.7$ (and up to $z\sim 2$ for the most massive galaxies). As
discussed in \citet[e.g.][]{somerville2015}, this is an important prerequisite
for models that are based on metallicity dependent star formation laws.

\subsection{Star forming sequence}
\label{subsec:sfsequence}
\begin{figure*}
\includegraphics[width = 1.\textwidth]{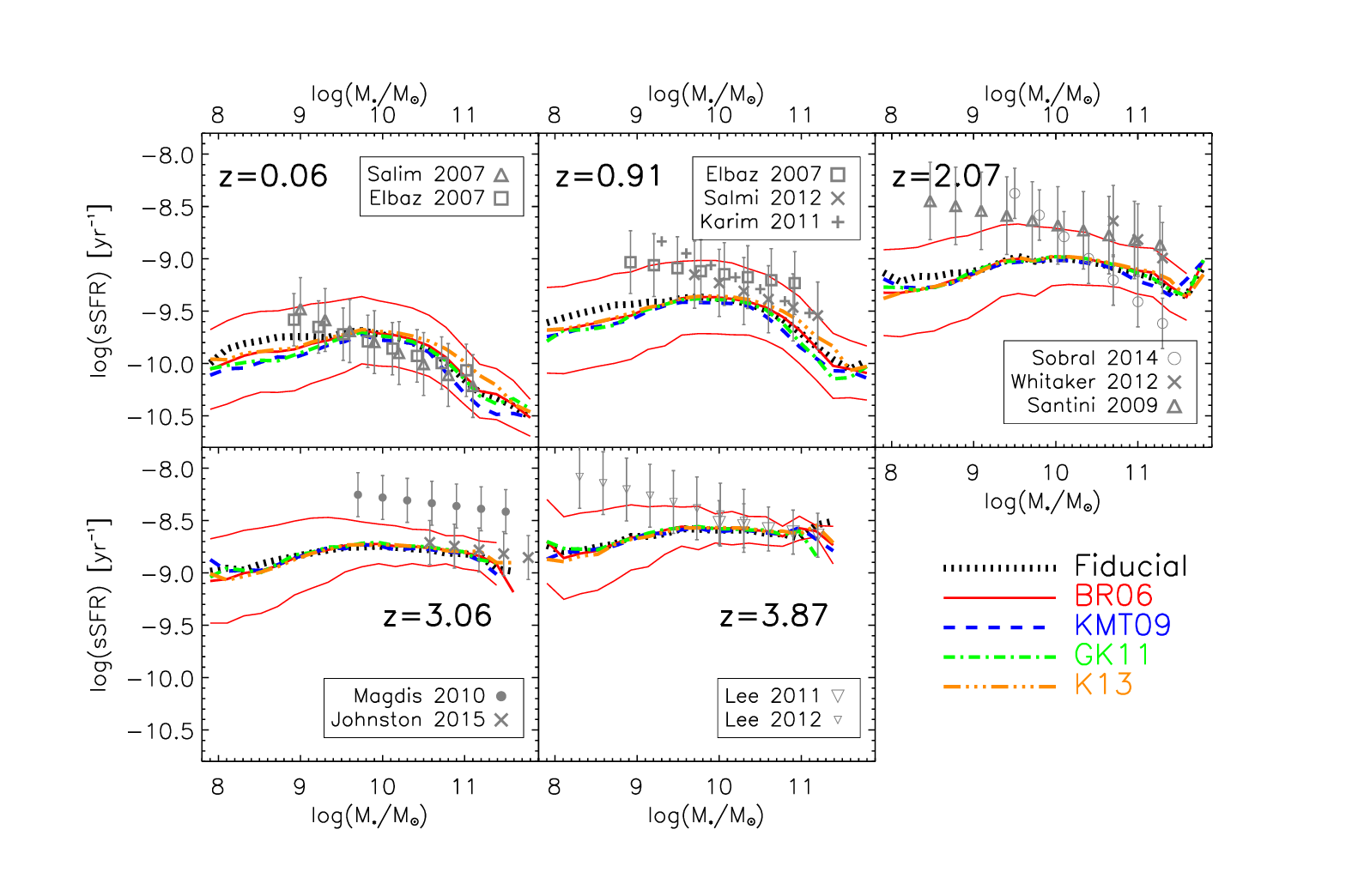}
\caption{Specific star formation rate as a function of galaxy stellar mass at  
different redshifts, as labeled. Gray symbols show different observational
estimates \citep{elbaz2007,salim2007,salmi2012,santini2009,sobral2014,magdis2010,johnston2015,lee2011,lee2012}. All
SFR and stellar mass estimates are converted to a Chabrier IMF, to be
consistent with our model assumptions. Thick lines show the mean relation
obtained for all star formation laws considered in our work, while the 
thinner lines in all panels show the scatter (standard deviation) predicted for
the BR06 model (the other models exhibit a similar scatter).}
\label{fig:sfr}
\end{figure*}

Fig.~\ref{fig:sfr} shows the specific star formation rate (sSFR) as a function
of galaxy stellar mass, from redshift $z=0$ to $z\sim4$. Only model galaxies
with sSFR$>0.3/t_{\rm H}$ are used for this analysis. Gray symbols correspond
to different observational measurements based on
H$_\alpha$ \citep{elbaz2007,sobral2014}, UV \citep{salim2007,johnston2015},
UV+IR \citep{salmi2012,santini2009}, and
FUV \citep{magdis2010,lee2011,lee2012}.  Symbols and error bars correspond to
the best fitting and standard deviation given
in \citet{speagle2014}. All derived stellar masses are converted to a
Chabrier IMF (dividing by $1.06$ in the case of a Kroupa IMF, and $1.7$ in case
of a Salpeter IMF). We have also converted the different estimates of
the star formation rates to a Chabrier IMF using the population synthesis model
by \citet{bruzual2003}.

All models predict decreasing sSFRs with decreasing redshift at fixed stellar
mass, a trend that is consistent with that observed.  Model predictions agree
relatively well with observational measurements up to $z\sim 1$  for
galaxies more massive than $\sim 10^{10}\solarmass$. At lower masses, data
suggests a monotonic increase of the sSFR with decreasing galaxy stellar mass
while the predicted relation are relatively flat. This trend is driven 
by central galaxies whose sSFR decreases slightly with decreasing stellar mass, 
while satellite galaxies are characterized by a flat sSFR - stellar relation. 
For galaxies at $z > 1$, star formation rates are under-estimated in
models, especially for low mass galaxies. The same problem was pointed out in
HDLF16 and is shared by other published galaxy formation models
\citep{fu2012,weinmann2012,mitchell2014,somerville2015,henriques2015}. Although there are
still large uncertainties on the measured sSFRs, particularly at high redshift,
the lack of actively star forming galaxies (or, in other words, the excess of
passive galaxies) at high redshift still represents an important challenge for
theoretical models of galaxy formation. Previous studies argued that
suppressing the star formation efficiency at early times (by using some form of
pre-heating or ad hoc tuned ejection and re-incorporation rates of gas) so as
to post-pone it to lower redshift could alleviate the problem \citep[see
e.g.][]{white2015,hirschmann2016}. A metallicity dependent star formation law
is expected to work in the same direction. However, surprisingly, all different
star formation laws considered in our study predict a very similar relation
between sSFR and galaxy stellar mass, at all redshifts considered.  This is
because different star formation laws predict similar star formation rates for
'high' surface density $\Sigma_{gas}> 20 \solarmass/pc^2$: the majority of
galaxies in our model have gas surface density above this value.  Previous
studies \citep{lagos2011a,somerville2015} also find that the different star
formation laws have little effect for active galaxies.

\subsection{Disk sizes}
\label{subsec:disksize}

In this section, we show model predictions for the radii of the HI and stellar
components, as well as for the star forming radius. We define as effective 
radius the radius that encloses half of the total SFR, HI, or stellar mass,
and assume exponential surface density profiles for both the stellar and the
gaseous disks (see equation~\ref{eqn:diskprofile}). We also assume that the
bulge density profile is well described by a Jaffe law \citep{jaffe1983}. As
discussed in Section~\ref{subsec:diskmodel}, the scale lengths of the gaseous 
and stellar disks are determined assuming conservation of the specific 
angular momentum. The star forming radius is instead measured by integrating
star formation over 20 annuli (see Section~\ref{subsec:sfr}).

\begin{figure*}
\includegraphics[width = 1.\textwidth]{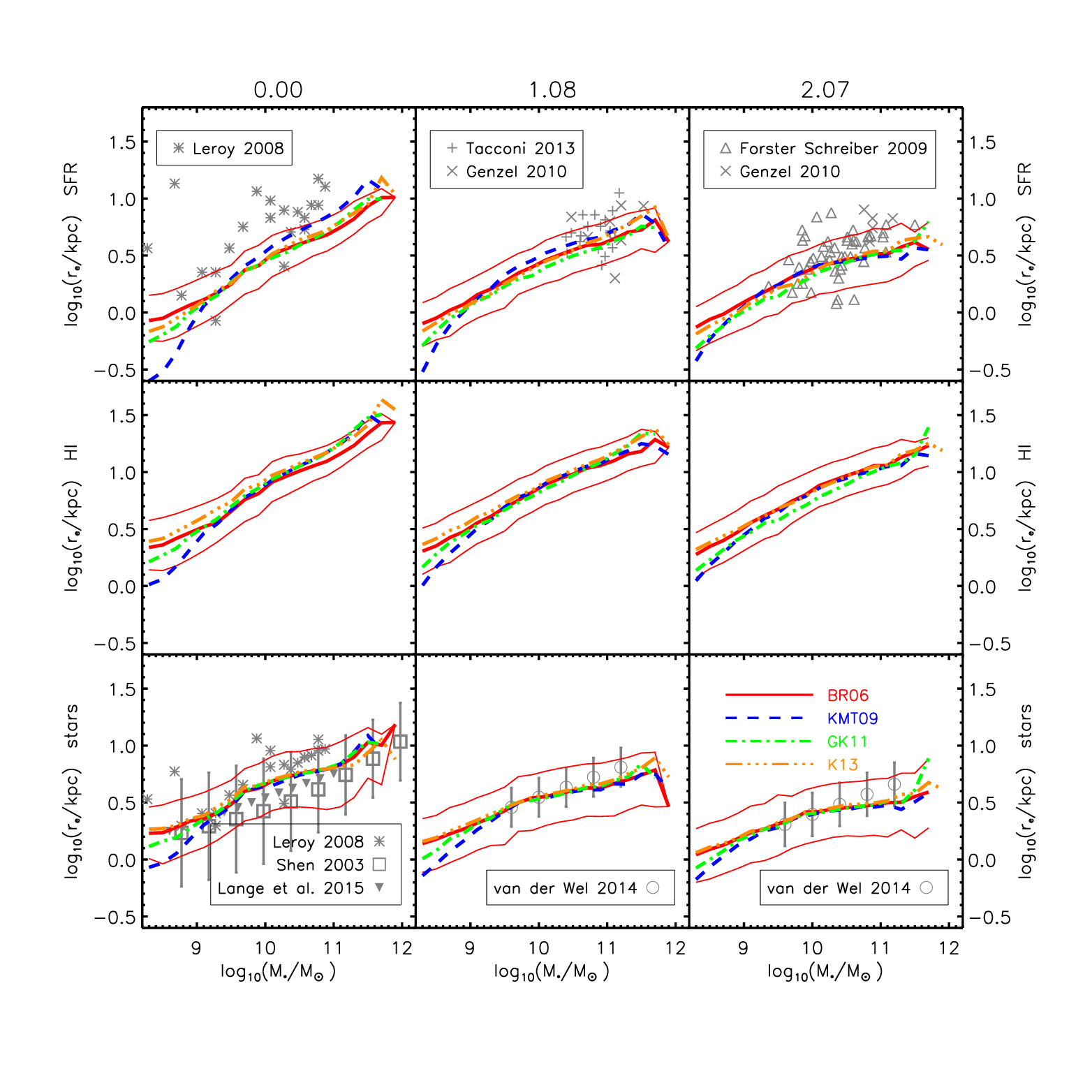}
\caption{The size-mass relation at redshift $z=0$, $z\sim1$, and $z\sim2$, 
from left to right. From top to bottom, the y-axis corresponds to the 
effective radius of the star forming disk, the HI component, and the stars. 
Gray symbols show different observational estimates, as indicated in the 
legend. The squares, upside-down triangles, and open circles correspond to 
the half-light radius in the r- , r-, and K-band 
\citep{shen2003,lange2015,vanderwel2014}. Coloured lines show
the median size-mass relation predicted by the different star formation laws
considered in our study. Thin red lines show the 16th and 84th percentiles of
the distribution for the BR06 model.}
\label{fig:size}
\end{figure*}

Fig.~\ref{fig:size} compares model predictions with observational data at
different redshifts. We only select disk dominated galaxies ($M_{\rm
bulge}/M_{\star} < 0.5$), with gas fraction $M_{\rm gas}/(M_{\rm
gas}+M_{\star})>0.1$, and specific star formation rate s$SFR > 0.3/t_{\rm
H}$ to make fair comparisons with observations.
The data shown in the top panels of Fig.~\ref{fig:size}
correspond to the half-light radii estimates from the PHIBSS survey 
\citep[][based on CO(3-2) lines]{tacconi2013}, from SINS \citep[][based on
H$_\alpha$]{forsterschreiber2009}, and \citet{genzel2010} (a combination of 
\citet{davis2007,noeske2007,erb2006} based on a combination of H$_\alpha$, UV, and CO maps).
 The sizes from \citet{leroy2008} correspond to the
scale lengths of exponential fits to the stellar and star formation surface
density, and are derived from K-band and FUV+24$\mu$m, respectively. The estimated 
scale lengths are multiplied by a factor $1.68$ to convert them in a
half mass radius. The stellar radii shown in the bottom panels correspond to
the half-light radii measured from CANDLES and 3D-HST \citep{vanderwel2014},
 from GAMA \citet{lange2015}, and from SDSS galaxies \citep{shen2003}.

For galaxies with fixed stellar mass, the effective HI and SFR radii evolve
little from redshift $z\sim2$ to present. The ratio between the SFR radius and
the HI radius of a typical galaxy with $M_{\star}=10^{10}\solarmass$ at $z=0$
is $\sim 1.2$ times that of a galaxy with the same stellar mass at $z\sim2$.
In contrast, the stellar size of the same galaxy at $z=0$ is $1.8$ times of
that at $z\sim 2$.  At redshift $z\sim2$, the SFR and stellar effective radii
are similar, while at $z=0$, the stellar radii are nearly $2$ times the star
forming radii. Available data, however, suggest that the star forming radii are
larger than the stellar radii at $z=0$. 
At all redshift, HI size is $2.5$ times of SFR size. Note that the stellar
size-mass relation of \citet{leroy2008} differs from that by \citet{shen2003} and \citet{lange2015}
because of the different selection criteria and different measurements of the
half mass radius. \citet{leroy2008} select star forming galaxies and measured the
half mass radius by fitting exponential profiles to the stellar surface density, 
as we do. \citet{shen2003,lange2015} measured half mass radius of S\'{e}rsic fits and
selected late-type galaxies with S\'{e}rsic index $n<2.5$. 

The predicted stellar radii are comparable with observational estimates at all
redshifts considered. The star forming radii are under-estimated in the models
by about $0.4$ dex at $z=0$, but in relatively good agreement with data at
higher redshift.  The four star formation laws used in our study predict very
similar size-mass relation, at all redshifts considered. This is expected: in
our model, disk sizes are calculated using the angular momentum of the accreted
cold gas. As we already discussed, different star formation laws predict very
similar star formation histories. So the consumption and accretion histories of
cold gas are also very similar. Our results are consistent with those
by \citet{popping2014} who compared star forming radii with a model including
prescriptions similar to our BR06 and GK11 models.

\section{Cosmic evolution of neutral hydrogen}
\label{sec:globaldensity}

\begin{figure}
\includegraphics[width = 0.4\textwidth]{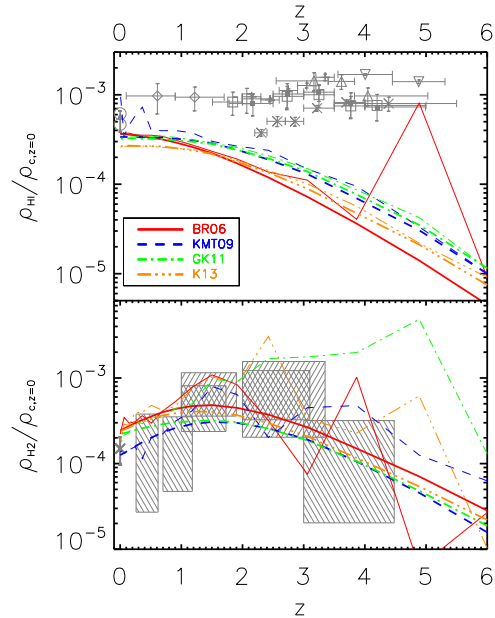}
\caption{The cosmic density evolution of HI (top panel) and H$_2$ (bottom
panel).  Different colours and line styles correspond to the different star
formation laws considered in our study, as indicated in the legend. Thick lines
correspond to densities estimated considering all galaxies down to the
completeness limit of the MSII. Thin lines have been obtained by fitting the HI
and H$_2$ mass functions at different redshifts and extrapolating them towards
lower masses (see text for details).  Gray symbols and shaded regions show
observational estimates.}
\label{fig:globaldensity}
\end{figure}

Fig.~\ref{fig:globaldensity} shows the evolution of the cosmic density of
HI (top panel) and H$_2$ (bottom panel). As shown in Fig.~\ref{fig:smf}, our
galaxy stellar mass functions are complete down to $M_{\star}\sim
10^8 \solarmass$ when run on MSII. The thick lines shown in
Fig.~\ref{fig:globaldensity} correspond to the density of HI and H$_2$ obtained
by summing up all galaxies above the completeness limit of the MSII in the
simulation box. Thin lines correspond to densities estimated by
fitting\footnote{We perform the fit considering the mass range between the peak
of the mass function and the maximum mass.} the predicted HI and H$_2$ mass functions with a \citet{schechter1976} distribution:
\begin{equation}
\phi (M_{\rm HI,H_2}) = \ln10\, \phi_0 \left( \frac{M_{\rm HI,H_2}}{M_0}\right)^{\alpha +1} e^{-\frac{M_{\rm HI,H_2}}{M_0}}
\end{equation} 
, and extrapolating model predictions towards infinite low mass. The resulting cosmic density is:
\begin{equation}
\rho_{\rm HI,H_2} = \Gamma (\alpha + 2) \phi_0 M_0 .
\end{equation} 
The relatively small size of the box and limited dynamic range in masses lead to a
very noisy behaviour of model predictions, particularly for the cosmic density
of molecular hydrogen.

We find that all the star formation laws considered in our work predict a
monotonic decrease of the HI cosmic density with increasing redshift. The BR06 model predicts the most
rapid evolution of the HI density while the GK11 and KMT09 the slowest. A
similar trend was found by \citet{popping2014}. This work, as well
as \citet{lagos2011b}, predict however a mild increase of the HI density
between present and $z\sim 1$, and then a decrease towards higher redshift. We
believe this is due to an excess of galaxies in the HI mass range
$10^8-10^9\,\solarmass$ combined with a faster evolution of the HI mass
function at higher redshift in these models with respect to our predictions
(compare e.g. Fig. 7 in \citealt{popping2014} and Fig. 8
in \citealt{lagos2011b} with our Fig. 7). In the top panel of
Fig.~\ref{fig:globaldensity}, we add observational measurements
by \citet{zwaan2005} and \citet{martin2010} at $z=0$, and measurements inferred
from damped Ly$\alpha$ systems (DLAs) at higher
redshifts \citep{peroux2005,rao2006,guimaraes2009,prochaska2009,zafar2013,noterdaeme2012,crighton2015}. While
our extrapolated estimates using KMT09 are closer to local estimates
(these are also based on fitting the observed HI mass function and
extrapolating it to lower masses), all models under-predict the cosmic density
of HI at higher redshift. The comparison with DLAs should, however, be
interpreted with caution. In fact, HI is attached to galaxies in our model
 while the nature of DLAs and their relationship with galaxies
remains unclear. In addition, low mass galaxies, which are not well resolved in 
our simulation, are gas rich and their contribution could be important at high redshift\citep{lagos2011b}.

In the bottom panel of Fig.~\ref{fig:globaldensity}, our predicted cosmic
density evolution of molecular hydrogen is compared with measurements
by \citet{keres2003} at $z=0$ and estimates based on blind CO surveys at higher
redshifts \citep{walter2014,decarli2016}. The local estimate of the cosmic
density of molecular hydrogen is obtained by fitting the observed mass
distribution and extrapolating towards lower masses, as we do for the thin
lines. A constant conversion factor ($\alpha_{\rm CO}= 4.75 \solarmass
(K\,km\,s^{-1}\,pc^{-2})^{-1}$) is assumed in this case. The higher redshift 
estimates are obtained by summing all observed galaxies and assuming $\alpha_{\rm CO}=3.6 \solarmass
(K\,km\,s^{-1}\,pc^{-2})^{-1}$. 
All models predict a mild increase of the H$_2$ cosmic density between $z=0$
and $z\sim 1-2$, followed by a somewhat more rapid decrease of the molecular
hydrogen density towards higher redshift. These trends are in qualitative
agreement with the estimated behaviour although uncertainties are still very
large. Our model predictions are in qualitative agreement with those
by \citet{popping2014} and \citet{lagos2011b}. The latter study, however, predicts a much
higher peak for the H$_2$ cosmic density at $1<z<2$ and a larger difference
between prediction based on different star formation laws. 

\section{Discussion}
\label{sec:discussion}

\subsection{Comparison with previous work}
\label{subsec:compareprevious}

In the last years, a number of semi-analytic models have been improved to
account for H$_2$-based star formation laws. In particular, \citet{fu2010}, 
\citet{lagos2011a}, and \citet{somerville2015} implement prescriptions 
for molecular gas formation processes in three independently developed 
semi-analytic models, and test the influence of different star formation 
laws. All groups discuss scenarios where the H$_2$ 
is determined by the pressure of the inter-stellar medium (our BR06 model), 
or by the analytic calculations by \citet[][our KMT09 model]{krumholz2008,
krumholz2009a,krumholz2009b}. In addition, \citet{somerville2015} include 
a star formation law based on the simulations presented in \citet{gnedin2011} 
as we do for our GK11 model.

These groups use different approaches for the calibration of models: 
\citet{fu2010} re-tune their AGN and stellar feedback parameters, as well as 
the free parameters entering the adopted star formation laws, to reproduce 
the galaxy stellar mass function, HI, and H$_2$ mass functions at
$z=0$. \citet{lagos2011a} choose the parameters in the modified star formation
laws to fit the observed relation between the surface density of star formation
and surface density of gas in nearby galaxies. All other parameters are left
unchanged. \citet{somerville2015} use an approach closer to that adopted
by \citet{fu2010}, and re-tune both the parameters entering the star formation
laws and those related to other physical processes to reproduce the galaxy
stellar mass function, the total gas fractions as a function of galaxy stellar
mass, and the normalization of the relation between stellar metallicity and
galaxy mass, all at $z=0$. 

In our case, we only modify the parameters entering the star formation laws and
leave unchanged all parameters entering additional prescriptions (e.g. stellar
and/or AGN feedback). As we discuss in Sections~\ref{subsec:diskmodel}
and \ref{subsec:bhmodel}, we update some prescriptions with respect to the
original model presented in HDLF16, but these updates have only a marginal
effect on the physical properties of our model galaxies. Some of the previous
studies
\citep{fu2012,somerville2015} consider separately the effect of the
prescriptions adopted to partition cold gas into its atomic and molecular
components, and those for the conversion of molecular gas in stars. In this
study we do not attempt to separate the effect of these two ingredients.

Our model belongs to the same family of models used by Fu and collaborators,
but differs from the model used in their study in a number of important
aspects. In particular, as discussed in Section~\ref{subsec:gaea}, our model
includes a sophisticated chemical enrichment scheme that allows us to follow
the non instantaneous recycling of gas, energy and different metal species into
the inter-stellar medium. This is the first time H$_2$-based star formation
laws are implemented in a model accounting for non-instantaneous
recycling. This is particularly relevant for prescriptions that depend
explicitly on the gas metallicity (e.g. KMT09, GK11, and K13 models), because
an instantaneous recycling approximation could lead to a too efficient
enrichment of the galaxies interstellar medium.  Another important success of
our model lies in the  relatively good agreement we find between model
predictions and the observed evolution of the relation between galaxy stellar
mass and gas metallicity (see  discussion in
Section~\ref{subsec:gzm}). This is of course an important prerequisite for the
prescriptions that use metallicity of the interstellar medium to estimate the
H$_2$ molecular fractions.  None of the previous models satisfy this
requirement: \citet[][see their Fig.~3]{fu2012} show that, at least in some of
their models, there is significant evolution of the gaseous phase metallicity,
at fixed galaxy mass.  The relation between galaxy stellar mass and
metallicity, however, tends to be too flat compared to observational estimates,
and only one of their models (i.e. that based on the Krumholz et
al. calculations) is in relatively good agreement with measurements at
$z=0$. In contrast, all models considered in \citet{somerville2015} predict a
mass-metallicity relation that is steeper than observed, with very little
evolution as a function of redshift. \citet{lagos2012} show predictions for the
relationship between gas metallicity and B-band luminosity, but only at
$z=0$. The evolution of the mass-metallicity relation based on the model
discussed in \citet[][this is essentially an update of the Lagos et al. model
to the WMAP7 cosmology]{gonzalez2014} is shown in \citet[][see their
Fig.~12]{guo2016}. Also in this case, very little evolution is found as a
function of redshift, and the relation is steeper than observational
estimates. Our Fig.~\ref{fig:gzm} shows that all models considered in this
paper predict a mass-metallicity relation that is in very good agreement with
observational estimates at $z=0$, all the way down to the resolution limit of
the Millennium II simulation. The predicted evolution as a function of redshift
is also in good agreement with data  at $z\sim 0.7$, and up to $z\sim 2$ for
galaxies more massive than $\sim 10^{10}\,{\rm M}_{\sun}$. Less massive
galaxies tend to have higher cold phase metallicities in the models than in the
data at the highest redshift considered ($z\sim 2.2$). We note, however, that
observational samples at this redshift are still sparse and likely strongly
biased.

The implementation of H$_2$ based star formation laws generally includes an
explicit dependence on the sizes of the galaxies (in particular of the disk,
and of its star forming region). Therefore, an additional important requirement
is that the adopted model reproduces observational measurements for the star
forming disks. We show in Section~\ref{subsec:disksize} that our model
satisfies this requirement too. Similar agreement with observational estimates
of disk sizes has been discussed in \citet{popping2014} for two of the models
considered in \citet{somerville2015}. \citet{lagos2011a} fail to reproduce 
the measured relation between the optical size and the luminosity of galaxies 
in the local Universe (see their Fig.~D3). \citet{fu2010} do not discuss the 
sizes of their model galaxies with respect to observational constraints. Finally, we note that our reference model does 
reproduce the observed evolution of the galaxy stellar mass function. As discussed
in HDLF16, this is due to the implementation of an updated
stellar feedback scheme in which large amounts of the baryons are `ejected' and
unavailable for cooling at high redshift, and the gas ejection rate decrease
significantly with cosmic time. \citet{lagos2011a} also reproduce the stellar 
mass function up to $z\sim3$ \citep[][Fig.~A7]{gonzalez2014}. Both the models 
discussed in \citet[][their Fig.~7]{fu2012} and in \citet[][their 
Fig.~7]{somerville2015} exhibit the well known excess of galaxies with 
intermediate to low mass galaxies at high redshift.

\subsection{Can we discriminate among different star formation laws?}
\label{subsec:globalratio}

In agreement with previous studies, we find that modifying the star formation
laws does not have significant impact on the global properties of model
galaxies and their distributions. As discussed in \citet{lagos2011a,somerville2015}
, as well as works based on hydro-simulations \citep{schaye2010,haas2013},
 this can be understood as a result of
self-regulation of star formation: if less stars are formed, stellar feedback
is less efficient in depleting the galaxy inter-stellar medium and more gas is
then available for subsequent star formation. Vice versa, if star formation is
more efficient, significant amounts of gas are ejected and subsequent star
formation is less efficient. The net result of this self-regulation is that the
average star formation histories (as well as the mass accretion histories and
other physical properties of galaxies) are not significantly altered when
different star formation laws are considered. 

In agreement with previous papers, we find that the number densities of
galaxies below the knee of the mass function are insensitive to the adopted star formation laws in
the redshift range $0<z<2$. At variance with previous models, we find significant
differences for the number densities of the most massive galaxies in models
with different star formation laws. We find this is caused by the fact that
differences in the amount of gas available at high redshift lead to a different
growth history for the black holes, and therefore to a different importance of
radio mode AGN feedback. The effect is particularly strong for metallicity
dependent star formation laws that lead to higher accretion rates onto the
central black holes at higher redshift (see Section~\ref{subsec:grow}). 
\citet{somerville2015} use a black hole model that limits the black hole 
mass to the observed BH-bulge relation \citep{hirschmann2012}. The available 
excess cold gas in their GKfid model will, therefore, not lead to excessively 
massive black holes. \citet{fu2010} assume that star formation rates depend
on the surface density of total cold gas, instead of the molecular gas, when $f_{\rm
H_2}<0.5$. In this way, their model based on the Krumholz et al. calculations
predicts star formation rates comparable to those obtained using the alternative
prescriptions based on pressure of the ISM at early times. This leads to very
similar black hole masses at late times when adopting different star formation 
laws. The model of \citet{lagos2011a} with BR06 and KMT09 also leads to a 
large amount of cold gas reservoir in galaxies at high redshift. But this
cold gas contributes to star bursts rather than to black hole growth in their model.  

In agreement with \citet{fu2010} and \citet{somerville2015}, we find that
different prescriptions can be tuned to reproduce the estimated HI and H$_2$
mass functions in the local Universe. The high mass end of the H$_2$ mass
function diverges for the same reasons illustrated above. Similarly, we find
that models based on different star formation laws predict very similar scaling
relations, and very similar evolution for these relations. This is in contrast
with \citet{lagos2011a} who rule out their metallicity dependent prescription
arguing that it does not reproduce well the observed HI mass function and
scaling relations at $z=0$.

\begin{figure}
\includegraphics[width = 0.4\textwidth]{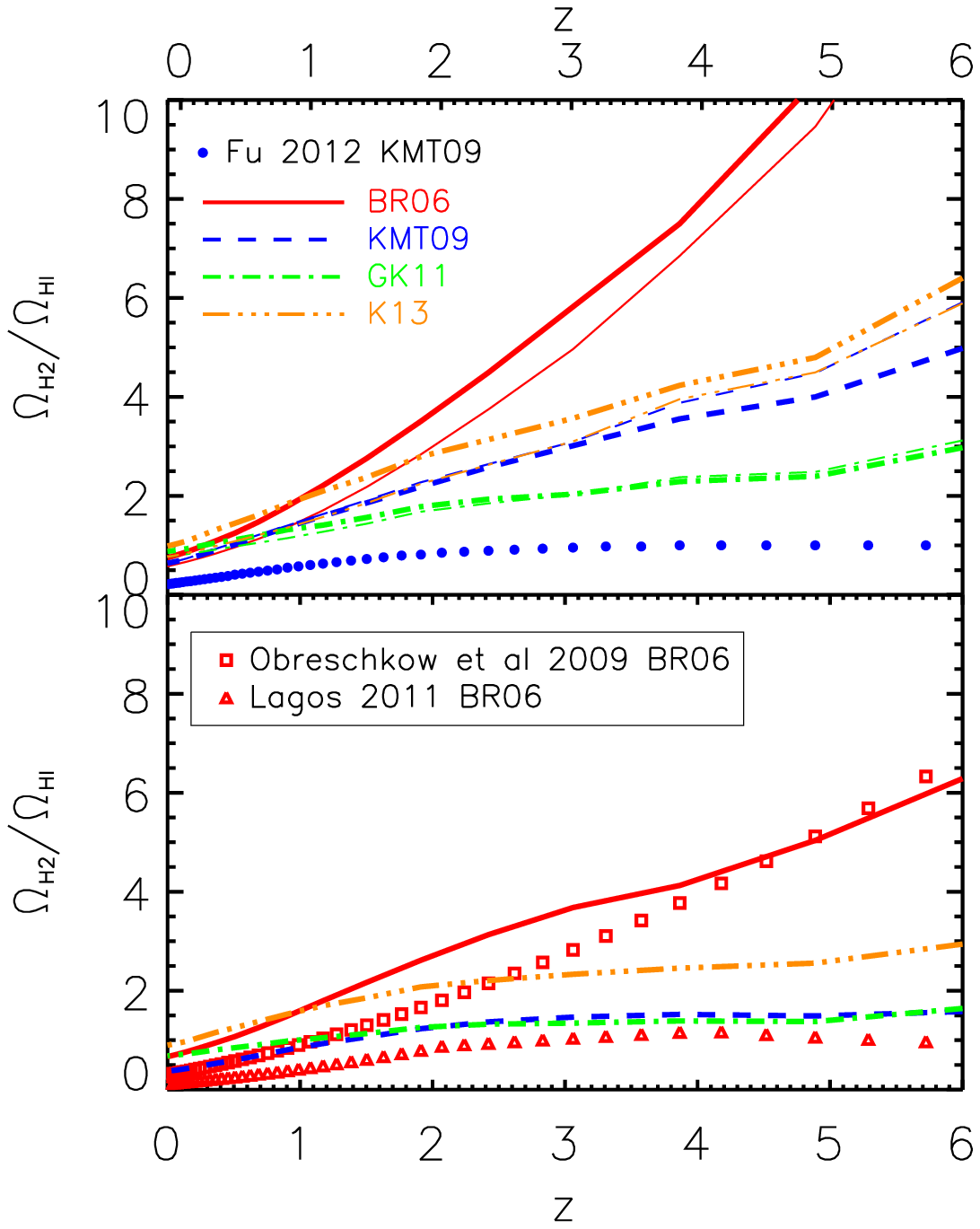}
\caption{The cosmic molecular ratio ${\Omega_{H_2}}/{\Omega_{HI}}$ as a
function of redshift. Thick curves correspond to our runs based on the MSII, while
thin curves correspond to the MS. The red squares, red triangles, and blue dots
 are predictions from \citet[][with BR06]{obreschkow2009b}, 
\citet[][with BR06]{lagos2011b}, and \citet[][with KMT09]{fu2012}, respectively.
The top panel shows results for galaxies above $M_{\star}>10^9 \solarmass$. 
The bottom panel shows results for galaxies with $M_{\star}>10^8 \solarmass$.}
\label{fig:globalratio}
\end{figure}

The only quantity we find to differ significantly between models based on
different star formation laws is the cosmic molecular-to-atomic hydrogen
ratio. The redshift evolution of this quantity is shown in 
Fig.~\ref{fig:globalratio}, for all models used in this study. 
In the top panel, we only consider galaxies with stellar mass larger than
$M_{\star}>10^9 \solarmass$.  Thick lines are based on the MSII, while thin
lines correspond to the MS. All models predict a monotonically increasing
ratio with increasing redshift. Predictions from the different models are very
close up to $z\sim 1$, and diverge significantly at higher redshift. In
particular, the BR06 model predicts the steeper evolution, with an increase of
about a factor $20$ between $z=0$ and $z\sim 6$. Among the models considered in
our study, the milder evolution is predicted by the GK11 model. In this case,
the molecular-to-atomic ratio increases by only a factor $\sim 8$ from $z=0$ to
$z\sim 6$. The other two models, KMT09 and K13, predict similar evolution. 
 The convergence between the two simulations is good although slightly higher 
values are found when using the MSII instead of the MS for the BR06 and K13 
model. This is expected as the resolution limit in this case is higher than 
the mass threshold adopted (see Section~\ref{subsec:ISM1}). The blue dots
 shows prediction by \citet{fu2012} based on prescriptions similar to those 
of our KMT09 model and on the MS. Their model predictions differ from ours 
both in normalization and in evolution as a function of redshift 

In the bottom panel of Fig.~\ref{fig:globalratio}, we show predictions based on
the MSII only and on a lower mass threshold ($M_{\star}>10^8 \solarmass$), and
compare them with predictions from previous work. Since the molecular-to-atomic
ratio decreases for lower mass galaxies, the overall cosmic value also
decreases. The trends described above remain the same: the BR06 model predicts
the strongest evolution and the highest value at $z>1$ among all models used in
our study. The weakest evolution and the lowest values are instead predicted by
the GK11 and KMT09 models. This is due to the overall decrease of the stellar
metallicity at higher redshift, which turns in lower values of the
molecular-to-atomic ratio. The K13 model is in between the BR06 and the other
two models. The red squares and triangles shown in the bottom panel of
Fig.~\ref{fig:globalratio} correspond to predictions from \citet{obreschkow2009b} 
and \citet{lagos2011b}, respectively. Both assume
prescriptions similar to those adopted in our BR06 model to partition the cold
gas in its atomic and molecular component. The former study, however, is based
on a post-processing of the model published in \citet{delucia2007} run on the MS, 
while the latter is applied on Monte-Carlo merger trees. \citet{obreschkow2009b} 
consider all halos more massive than $10^{10} \solarmass$, while \citet{lagos2011b}
use a lower halo mass limit of $5 \times 10^{8} \solarmass$. We have verified that
our model predictions do not differ significantly when using $M_h >10^{10}\solarmass$ 
instead of $M_{\star} > 10^{8}\solarmass$.

Fig.~\ref{fig:globalratio} shows that the cosmic evolution of
${\Omega_{H_2}}/{\Omega_{HI}}$ is the only quantity for which predictions from
different star formation laws are significantly different (although models
start to differentiate only at higher redshift where important systematics in
the data start playing an important role). However, the figure shows that large
variations can be obtained adopting the same star formation law but
 different prescriptions for other physical processes,
 i.e. in independently developed models. This again makes it difficult to use these particular
observations to put direct and strong constraints on the star formation
law.

\vspace{2cm}

\section{Summary and conclusions}
\label{sec:conclusion}
We present an update of our recently published model for GAlaxy
Evolution and Assembly \citep[GAEA]{hirschmann2016}, aimed at
including a self-consistent treatment of the partition of cold gas in
its atomic and molecular components. Our approach is similar to that
followed in previous work based on independently developed
models \citep{fu2010,lagos2011a,somerville2015}, but our model
provides significant improvements over those previously used for
similar studies. In particular, GAEA: (i) includes a sophisticated
chemical enrichment treatment that accounts for the non-instantaneous
recycling of gas, metals, and energy; (ii) reproduces the measured
relation between the metallicity of the cold gas and the galaxy
stellar mass, as well as its evolution as a function of cosmic time;
(iii) includes an updated modelling for stellar feedback, based partly
on results from hydrodynamical simulations, and that allows us to
reproduce the observed evolution of the galaxy stellar mass function.

These represent important prerequisites for our study, particularly
when considering prescriptions to compute molecular-to-atomic fraction
including an explicit dependence on the gas
metallicity \citep{krumholz2009b,gnedin2011,krumholz2013}. We also
consider the empirical relation by \citet{blitz2006}, based on the
 hydrostatic pressure of the disk. We find that
modifying the star formation law does not translate in appreciable
differences for the physical properties of galaxies or their
statistical distributions. In particular, neither the number densities
nor the physical properties of low-mass galaxies are significantly
affected by the adoption of a molecular formation efficiency that
depends on the cold gas metallicity, in contrast with previous
claims \citep[e.g.][]{kd2012}. As discussed in previous studies 
\citep[e.g.][]{lagos2011a,somerville2015}, this behaviour arises from a
self-regulation of the star formation: if less stars form
(because e.g. of low molecular fractions due to low metallicities),
less gas is reheated/ejected due to stellar feedback. As a result,
more gas is available for star formation at later times.

All star formation laws we consider are tuned (modifying only the free
parameters entering these prescriptions) in order to reproduce the local HI and
H$_2$ galaxy mass functions. For all models, we find a remarkable agreement
between model predictions and the observed scaling relations  between HI
and H$_2$ masses and galaxy stellar mass, distributions in optical and star
forming sizes, and the relation between the cold gas phase metallicity and the
galaxy stellar mass. The only quantity that exhibits significant variations
depending on the different H$_2$-based star formation laws is the cosmic
molecular-to-atomic hydrogen ratio. Unfortunately, we find that similar
deviations are obtained when implementing the same H$_2$-based star formation
law into independent semi-analytic models.  These results suggest that it is
very difficult to use available data on the gas content of galaxies to
discriminate between different models. The difficulties will remain also with
larger statistical samples as the scatter in most of the scaling relations is
significant. A more promising avenue to put constraints on the physical
processes affecting star formation laws is that of focusing on smaller galaxies
 and/or on galaxies at earlier cosmic epochs, as these are the regimes
where self-regulation of star formation has not yet effectively washed out
differences by imprinted by different star formation laws \citep[see
also][]{somerville2015}.

\section*{Acknowledgements}
LX and GDL acknowledge financial support from the MERAC foundation. MH
acknowledges financial support from the European Research Council via an
Advanced Grant under grant agreement no. 321323 NEOGAL.

\bibliographystyle{mn2e}
\setlength{\bibhang}{2.0em}
\setlength\labelwidth{0.0em}
\bibliography{h2sam}

\begin{thebibliography}{140}
\expandafter\ifx\csname natexlab\endcsname\relax\def\natexlab#1{#1}\fi

\bibitem[{{Baldry} {et~al}\mbox{.}(2012){Baldry}, {Driver}, {Loveday},
  {Taylor}, {Kelvin}, {Liske}, {Norberg}, {Robotham}, {Brough}, {Hopkins},
  {Bamford}, {Peacock}, {Bland-Hawthorn}, {Conselice}, {Croom}, {Jones},
  {Parkinson}, {Popescu}, {Prescott}, {Sharp}, \& {Tuffs}}]{baldry2012}
{Baldry} I.~K. {et~al.}, 2012, \mnras, 421, 621

\bibitem[{{Bennett} {et~al}\mbox{.}(2013){Bennett}, {Larson}, {Weiland},
  {Jarosik}, {Hinshaw}, {Odegard}, {Smith}, {Hill}, {Gold}, {Halpern},
  {Komatsu}, {Nolta}, {Page}, {Spergel}, {Wollack}, {Dunkley}, {Kogut},
  {Limon}, {Meyer}, {Tucker}, \& {Wright}}]{bennett2013}
{Bennett} C.~L. {et~al.}, 2013, \apjs, 208, 20

\bibitem[{{Berta} {et~al}\mbox{.}(2013){Berta}, {Lutz}, {Nordon}, {Genzel},
  {Magnelli}, {Popesso}, {Rosario}, {Saintonge}, {Wuyts}, \&
  {Tacconi}}]{berta2013}
{Berta} S. {et~al.}, 2013, \aap, 555, L8

\bibitem[{{Bigiel} {et~al}\mbox{.}(2010){Bigiel}, {Leroy}, {Walter}, {Blitz},
  {Brinks}, {de Blok}, \& {Madore}}]{bigiel2010}
{Bigiel} F., {Leroy} A., {Walter} F., {Blitz} L., {Brinks} E., {de Blok}
  W.~J.~G., {Madore} B., 2010, \aj, 140, 1194

\bibitem[{{Bigiel} {et~al}\mbox{.}(2008){Bigiel}, {Leroy}, {Walter}, {Brinks},
  {de Blok}, {Madore}, \& {Thornley}}]{bigiel2008}
{Bigiel} F., {Leroy} A., {Walter} F., {Brinks} E., {de Blok} W.~J.~G., {Madore}
  B., {Thornley} M.~D., 2008, \aj, 136, 2846

\bibitem[{{Blitz} \& {Rosolowsky}(2006)}]{blitz2006}
{Blitz} L., {Rosolowsky} E., 2006, \apj, 650, 933

\bibitem[{{Bogd{\'a}n} \& {Goulding}(2015)}]{bogdan2015}
{Bogd{\'a}n} {\'A}., {Goulding} A.~D., 2015, \apj, 800, 124

\bibitem[{{Bolatto} {et~al}\mbox{.}(2008){Bolatto}, {Leroy}, {Rosolowsky},
  {Walter}, \& {Blitz}}]{bolatto2008}
{Bolatto} A.~D., {Leroy} A.~K., {Rosolowsky} E., {Walter} F., {Blitz} L., 2008,
  \apj, 686, 948

\bibitem[{{Booth} {et~al}\mbox{.}(2009){Booth}, {de Blok}, {Jonas}, \&
  {Fanaroff}}]{booth2009}
{Booth} R.~S., {de Blok} W.~J.~G., {Jonas} J.~L., {Fanaroff} B., 2009, ArXiv
  0910.2935

\bibitem[{{Boselli} {et~al}\mbox{.}(2014){Boselli}, {Cortese}, {Boquien},
  {Boissier}, {Catinella}, {Lagos}, \& {Saintonge}}]{boselli2014}
{Boselli} A., {Cortese} L., {Boquien} M., {Boissier} S., {Catinella} B.,
  {Lagos} C., {Saintonge} A., 2014, \aap, 564, A66

\bibitem[{{Bothwell} {et~al}\mbox{.}(2013){Bothwell}, {Smail}, {Chapman},
  {Genzel}, {Ivison}, {Tacconi}, {Alaghband-Zadeh}, {Bertoldi}, {Blain},
  {Casey}, {Cox}, {Greve}, {Lutz}, {Neri}, {Omont}, \&
  {Swinbank}}]{bothwell2013}
{Bothwell} M.~S. {et~al.}, 2013, \mnras, 429, 3047

\bibitem[{{Boylan-Kolchin} {et~al}\mbox{.}(2009){Boylan-Kolchin}, {Springel},
  {White}, {Jenkins}, \& {Lemson}}]{boylankolchin2009}
{Boylan-Kolchin} M., {Springel} V., {White} S.~D.~M., {Jenkins} A., {Lemson}
  G., 2009, \mnras, 398, 1150

\bibitem[{{Brown} {et~al}\mbox{.}(2015){Brown}, {Catinella}, {Cortese},
  {Kilborn}, {Haynes}, \& {Giovanelli}}]{brown2015}
{Brown} T., {Catinella} B., {Cortese} L., {Kilborn} V., {Haynes} M.~P.,
  {Giovanelli} R., 2015, \mnras, 452, 2479

\bibitem[{{Bruzual} \& {Charlot}(2003)}]{bruzual2003}
{Bruzual} G., {Charlot} S., 2003, \mnras, 344, 1000

\bibitem[{{Calzetti} {et~al}\mbox{.}(2007){Calzetti}, {Kennicutt},
  {Engelbracht}, {Leitherer}, {Draine}, {Kewley}, {Moustakas}, {Sosey}, {Dale},
  {Gordon}, {Helou}, {Hollenbach}, {Armus}, {Bendo}, {Bot}, {Buckalew},
  {Jarrett}, {Li}, {Meyer}, {Murphy}, {Prescott}, {Regan}, {Rieke}, {Roussel},
  {Sheth}, {Smith}, {Thornley}, \& {Walter}}]{calzetti2007}
{Calzetti} D. {et~al.}, 2007, \apj, 666, 870

\bibitem[{{Carilli} \& {Rawlings}(2004)}]{carilli2004}
{Carilli} C.~L., {Rawlings} S., 2004, \nar, 48, 979

\bibitem[{{Catinella} {et~al}\mbox{.}(2013){Catinella}, {Schiminovich},
  {Cortese}, {Fabello}, {Hummels}, {Moran}, {Lemonias}, {Cooper}, {Wu},
  {Heckman}, \& {Wang}}]{catinella2013}
{Catinella} B. {et~al.}, 2013, \mnras, 436, 34

\bibitem[{{Crighton} {et~al}\mbox{.}(2015){Crighton}, {Murphy}, {Prochaska},
  {Worseck}, {Rafelski}, {Becker}, {Ellison}, {Fumagalli}, {Lopez}, {Meiksin},
  \& {O'Meara}}]{crighton2015}
{Crighton} N.~H.~M. {et~al.}, 2015, \mnras, 452, 217

\bibitem[{{Croton} {et~al}\mbox{.}(2006){Croton}, {Springel}, {White}, {De
  Lucia}, {Frenk}, {Gao}, {Jenkins}, {Kauffmann}, {Navarro}, \&
  {Yoshida}}]{croton2006}
{Croton} D.~J. {et~al.}, 2006, \mnras, 365, 11

\bibitem[{{Danovich} {et~al}\mbox{.}(2015){Danovich}, {Dekel}, {Hahn},
  {Ceverino}, \& {Primack}}]{danovich2015}
{Danovich} M., {Dekel} A., {Hahn} O., {Ceverino} D., {Primack} J., 2015,
  \mnras, 449, 2087

\bibitem[{{Davidzon} {et~al}\mbox{.}(2013){Davidzon}, {Bolzonella}, {Coupon},
  {Ilbert}, {Arnouts}, {de la Torre}, {Fritz}, {De Lucia}, {Iovino}, {Granett},
  {Zamorani}, {Guzzo}, {Abbas}, {Adami}, {Bel}, {Bottini}, {Branchini},
  {Cappi}, {Cucciati}, {Franzetti}, {Fumana}, {Garilli}, {Krywult}, {Le Brun},
  {Le F{\`e}vre}, {Maccagni}, {Ma{\l}ek}, {Marulli}, {McCracken}, {Paioro},
  {Peacock}, {Polletta}, {Pollo}, {Schlagenhaufer}, {Scodeggio}, {Tasca},
  {Tojeiro}, {Vergani}, {Zanichelli}, {Burden}, {Di Porto}, {Marchetti},
  {Marinoni}, {Mellier}, {Moscardini}, {Moutard}, {Nichol}, {Percival},
  {Phleps}, \& {Wolk}}]{davidzon2013}
{Davidzon} I. {et~al.}, 2013, \aap, 558, A23

\bibitem[{{Davis} {et~al}\mbox{.}(2007){Davis}, {Guhathakurta}, {Konidaris},
  {Newman}, {Ashby}, {Biggs}, {Barmby}, {Bundy}, {Chapman}, {Coil},
  {Conselice}, {Cooper}, {Croton}, {Eisenhardt}, {Ellis}, {Faber}, {Fang},
  {Fazio}, {Georgakakis}, {Gerke}, {Goss}, {Gwyn}, {Harker}, {Hopkins},
  {Huang}, {Ivison}, {Kassin}, {Kirby}, {Koekemoer}, {Koo}, {Laird}, {Le
  Floc'h}, {Lin}, {Lotz}, {Marshall}, {Martin}, {Metevier}, {Moustakas},
  {Nandra}, {Noeske}, {Papovich}, {Phillips}, {Rich}, {Rieke}, {Rigopoulou},
  {Salim}, {Schiminovich}, {Simard}, {Smail}, {Small}, {Weiner}, {Willmer},
  {Willner}, {Wilson}, {Wright}, \& {Yan}}]{davis2007}
{Davis} M. {et~al.}, 2007, \apjl, 660, L1

\bibitem[{{De Lucia} \& {Blaizot}(2007)}]{delucia2007}
{De Lucia} G., {Blaizot} J., 2007, \mnras, 375, 2

\bibitem[{{De Lucia} \& {Helmi}(2008)}]{delucia2008}
{De Lucia} G., {Helmi} A., 2008, \mnras, 391, 14

\bibitem[{{De Lucia} {et~al}\mbox{.}(2014){De Lucia}, {Tornatore}, {Frenk},
  {Helmi}, {Navarro}, \& {White}}]{delucia2014}
{De Lucia} G., {Tornatore} L., {Frenk} C.~S., {Helmi} A., {Navarro} J.~F.,
  {White} S.~D.~M., 2014, \mnras, 445, 970

\bibitem[{{Decarli} {et~al}\mbox{.}(2016){Decarli}, {Walter}, {Aravena},
  {Carilli}, {Bouwens}, {da Cunha}, {Daddi}, {Ivison}, {Popping}, {Riechers},
  {Smail}, {Swinbank}, {Weiss}, {Anguita}, {Assef}, {Bauer}, {Bell},
  {Bertoldi}, {Chapman}, {Colina}, {Cortes}, {Cox}, {Dickinson}, {Elbaz},
  {G{\'o}nzalez-L{\'o}pez}, {Ibar}, {Infante}, {Hodge}, {Karim}, {Le Fevre},
  {Magnelli}, {Neri}, {Oesch}, {Ota}, {Rix}, {Sargent}, {Sheth}, {van der Wel},
  {van der Werf}, \& {Wagg}}]{decarli2016}
{Decarli} R. {et~al.}, 2016, \apj, 833, 69

\bibitem[{{Di Matteo} {et~al}\mbox{.}(2003){Di Matteo}, {Croft}, {Springel}, \&
  {Hernquist}}]{dimatteo2003}
{Di Matteo} T., {Croft} R.~A.~C., {Springel} V., {Hernquist} L., 2003, \apj,
  593, 56

\bibitem[{{Drory} {et~al}\mbox{.}(2004){Drory}, {Bender}, {Feulner}, {Hopp},
  {Maraston}, {Snigula}, \& {Hill}}]{drory2004}
{Drory} N., {Bender} R., {Feulner} G., {Hopp} U., {Maraston} C., {Snigula} J.,
  {Hill} G.~J., 2004, \apj, 608, 742

\bibitem[{{Dutton} {et~al}\mbox{.}(2011){Dutton}, {van den Bosch}, {Faber},
  {Simard}, {Kassin}, {Koo}, {Bundy}, {Huang}, {Weiner}, {Cooper}, {Newman},
  {Mozena}, \& {Koekemoer}}]{dutton2011}
{Dutton} A.~A. {et~al.}, 2011, \mnras, 410, 1660

\bibitem[{{Elbaz} {et~al}\mbox{.}(2007){Elbaz}, {Daddi}, {Le Borgne},
  {Dickinson}, {Alexander}, {Chary}, {Starck}, {Brandt}, {Kitzbichler},
  {MacDonald}, {Nonino}, {Popesso}, {Stern}, \& {Vanzella}}]{elbaz2007}
{Elbaz} D. {et~al.}, 2007, \aap, 468, 33

\bibitem[{{Elmegreen}(1989)}]{elmegreen1989}
{Elmegreen} B.~G., 1989, \apj, 338, 178

\bibitem[{{Erb} {et~al}\mbox{.}(2006){Erb}, {Steidel}, {Shapley}, {Pettini},
  {Reddy}, \& {Adelberger}}]{erb2006}
{Erb} D.~K., {Steidel} C.~C., {Shapley} A.~E., {Pettini} M., {Reddy} N.~A.,
  {Adelberger} K.~L., 2006, \apj, 647, 128

\bibitem[{{Fontana} {et~al}\mbox{.}(2006){Fontana}, {Salimbeni}, {Grazian},
  {Giallongo}, {Pentericci}, {Nonino}, {Fontanot}, {Menci}, {Monaco},
  {Cristiani}, {Vanzella}, {de Santis}, \& {Gallozzi}}]{fontana2006}
{Fontana} A. {et~al.}, 2006, \aap, 459, 745

\bibitem[{{F{\"o}rster Schreiber} {et~al}\mbox{.}(2009){F{\"o}rster Schreiber},
  {Genzel}, {Bouch{\'e}}, {Cresci}, {Davies}, {Buschkamp}, {Shapiro},
  {Tacconi}, {Hicks}, {Genel}, {Shapley}, {Erb}, {Steidel}, {Lutz},
  {Eisenhauer}, {Gillessen}, {Sternberg}, {Renzini}, {Cimatti}, {Daddi},
  {Kurk}, {Lilly}, {Kong}, {Lehnert}, {Nesvadba}, {Verma}, {McCracken},
  {Arimoto}, {Mignoli}, \& {Onodera}}]{forsterschreiber2009}
{F{\"o}rster Schreiber} N.~M. {et~al.}, 2009, \apj, 706, 1364

\bibitem[{{Fu} {et~al}\mbox{.}(2010){Fu}, {Guo}, {Kauffmann}, \&
  {Krumholz}}]{fu2010}
{Fu} J., {Guo} Q., {Kauffmann} G., {Krumholz} M.~R., 2010, \mnras, 409, 515

\bibitem[{{Fu} {et~al}\mbox{.}(2012){Fu}, {Kauffmann}, {Li}, \& {Guo}}]{fu2012}
{Fu} J., {Kauffmann} G., {Li} C., {Guo} Q., 2012, \mnras, 424, 2701

\bibitem[{{Genzel} {et~al}\mbox{.}(2010){Genzel}, {Tacconi}, {Gracia-Carpio},
  {Sternberg}, {Cooper}, {Shapiro}, {Bolatto}, {Bouch{\'e}}, {Bournaud},
  {Burkert}, {Combes}, {Comerford}, {Cox}, {Davis}, {Schreiber},
  {Garcia-Burillo}, {Lutz}, {Naab}, {Neri}, {Omont}, {Shapley}, \&
  {Weiner}}]{genzel2010}
{Genzel} R. {et~al.}, 2010, \mnras, 407, 2091

\bibitem[{{Gil de Paz} {et~al}\mbox{.}(2007){Gil de Paz}, {Boissier}, {Madore},
  {Seibert}, {Joe}, {Boselli}, {Wyder}, {Thilker}, {Bianchi}, {Rey}, {Rich},
  {Barlow}, {Conrow}, {Forster}, {Friedman}, {Martin}, {Morrissey}, {Neff},
  {Schiminovich}, {Small}, {Donas}, {Heckman}, {Lee}, {Milliard}, {Szalay}, \&
  {Yi}}]{gildepaz2007}
{Gil de Paz} A. {et~al.}, 2007, \apjs, 173, 185

\bibitem[{{Glover} \& {Clark}(2012)}]{gloverclark2012}
{Glover} S.~C.~O., {Clark} P.~C., 2012, \mnras, 421, 9

\bibitem[{{Gnedin} \& {Kravtsov}(2011)}]{gnedin2011}
{Gnedin} N.~Y., {Kravtsov} A.~V., 2011, \apj, 728, 88

\bibitem[{{Gonzalez-Perez} {et~al}\mbox{.}(2014){Gonzalez-Perez}, {Lacey},
  {Baugh}, {Lagos}, {Helly}, {Campbell}, \& {Mitchell}}]{gonzalez2014}
{Gonzalez-Perez} V., {Lacey} C.~G., {Baugh} C.~M., {Lagos} C.~D.~P., {Helly}
  J., {Campbell} D.~J.~R., {Mitchell} P.~D., 2014, \mnras, 439, 264

\bibitem[{{Guimar{\~a}es} {et~al}\mbox{.}(2009){Guimar{\~a}es}, {Petitjean},
  {de Carvalho}, {Djorgovski}, {Noterdaeme}, {Castro}, {Poppe}, \&
  {Aghaee}}]{guimaraes2009}
{Guimar{\~a}es} R., {Petitjean} P., {de Carvalho} R.~R., {Djorgovski} S.~G.,
  {Noterdaeme} P., {Castro} S., {Poppe} P.~C.~D.~R., {Aghaee} A., 2009, \aap,
  508, 133

\bibitem[{{Guo} {et~al}\mbox{.}(2016){Guo}, {Gonzalez-Perez}, {Guo},
  {Schaller}, {Furlong}, {Bower}, {Cole}, {Crain}, {Frenk}, {Helly}, {Lacey},
  {Lagos}, {Mitchell}, {Schaye}, \& {Theuns}}]{guo2016}
{Guo} Q. {et~al.}, 2016, \mnras, 461, 3457

\bibitem[{{Guo} {et~al}\mbox{.}(2013){Guo}, {White}, {Angulo}, {Henriques},
  {Lemson}, {Boylan-Kolchin}, {Thomas}, \& {Short}}]{guo2013}
{Guo} Q., {White} S., {Angulo} R.~E., {Henriques} B., {Lemson} G.,
  {Boylan-Kolchin} M., {Thomas} P., {Short} C., 2013, \mnras, 428, 1351

\bibitem[{{Guo} {et~al}\mbox{.}(2011){Guo}, {White}, {Boylan-Kolchin}, {De
  Lucia}, {Kauffmann}, {Lemson}, {Li}, {Springel}, \& {Weinmann}}]{guo2011}
{Guo} Q. {et~al.}, 2011, \mnras, 413, 101

\bibitem[{{Haas} {et~al}\mbox{.}(2013){Haas}, {Schaye}, {Booth}, {Dalla
  Vecchia}, {Springel}, {Theuns}, \& {Wiersma}}]{haas2013}
{Haas} M.~R., {Schaye} J., {Booth} C.~M., {Dalla Vecchia} C., {Springel} V.,
  {Theuns} T., {Wiersma} R.~P.~C., 2013, \mnras, 435, 2955

\bibitem[{{Haynes} {et~al}\mbox{.}(2011){Haynes}, {Giovanelli}, {Martin},
  {Hess}, {Saintonge}, {Adams}, {Hallenbeck}, {Hoffman}, {Huang}, {Kent},
  {Koopmann}, {Papastergis}, {Stierwalt}, {Balonek}, {Craig}, {Higdon},
  {Kornreich}, {Miller}, {O'Donoghue}, {Olowin}, {Rosenberg}, {Spekkens},
  {Troischt}, \& {Wilcots}}]{haynes2011}
{Haynes} M.~P. {et~al.}, 2011, \aj, 142, 170

\bibitem[{{Helfer} {et~al}\mbox{.}(2003){Helfer}, {Thornley}, {Regan}, {Wong},
  {Sheth}, {Vogel}, {Blitz}, \& {Bock}}]{helfer2003}
{Helfer} T.~T., {Thornley} M.~D., {Regan} M.~W., {Wong} T., {Sheth} K., {Vogel}
  S.~N., {Blitz} L., {Bock} D.~C.-J., 2003, \apjs, 145, 259

\bibitem[{{Henriques} {et~al}\mbox{.}(2015){Henriques}, {White}, {Thomas},
  {Angulo}, {Guo}, {Lemson}, {Springel}, \& {Overzier}}]{henriques2015}
{Henriques} B.~M.~B., {White} S.~D.~M., {Thomas} P.~A., {Angulo} R., {Guo} Q.,
  {Lemson} G., {Springel} V., {Overzier} R., 2015, \mnras, 451, 2663

\bibitem[{{Henriques} {et~al}\mbox{.}(2013){Henriques}, {White}, {Thomas},
  {Angulo}, {Guo}, {Lemson}, \& {Springel}}]{henriques2013}
{Henriques} B.~M.~B., {White} S.~D.~M., {Thomas} P.~A., {Angulo} R.~E., {Guo}
  Q., {Lemson} G., {Springel} V., 2013, \mnras, 431, 3373

\bibitem[{{Hirschmann}, {De Lucia} \& {Fontanot}(2016){Hirschmann}, {De Lucia},
  \& {Fontanot}}]{hirschmann2016}
{Hirschmann} M., {De Lucia} G., {Fontanot} F., 2016, \mnras, 461, 1760

\bibitem[{{Hirschmann} {et~al}\mbox{.}(2012){Hirschmann}, {Somerville}, {Naab},
  \& {Burkert}}]{hirschmann2012}
{Hirschmann} M., {Somerville} R.~S., {Naab} T., {Burkert} A., 2012, \mnras,
  426, 237

\bibitem[{{Hopkins} {et~al}\mbox{.}(2014){Hopkins}, {Kere{\v s}}, {O{\~n}orbe},
  {Faucher-Gigu{\`e}re}, {Quataert}, {Murray}, \& {Bullock}}]{hopkins2014}
{Hopkins} P.~F., {Kere{\v s}} D., {O{\~n}orbe} J., {Faucher-Gigu{\`e}re} C.-A.,
  {Quataert} E., {Murray} N., {Bullock} J.~S., 2014, \mnras, 445, 581

\bibitem[{{Hu} {et~al}\mbox{.}(2016){Hu}, {Naab}, {Walch}, {Glover}, \&
  {Clark}}]{hu2016}
{Hu} C.-Y., {Naab} T., {Walch} S., {Glover} S.~C.~O., {Clark} P.~C., 2016,
  \mnras, 458, 3528

\bibitem[{{Jaffe}(1983)}]{jaffe1983}
{Jaffe} W., 1983, \mnras, 202, 995

\bibitem[{{Jiang} {et~al}\mbox{.}(2015){Jiang}, {Wang}, {Gu}, {Wang}, \&
  {Zhang}}]{jiang2015}
{Jiang} X.-J., {Wang} Z., {Gu} Q., {Wang} J., {Zhang} Z.-Y., 2015, \apj, 799,
  92

\bibitem[{{Johnston} {et~al}\mbox{.}(2015){Johnston}, {Vaccari}, {Jarvis},
  {Smith}, {Giovannoli}, {H{\"a}u{\ss}ler}, \& {Prescott}}]{johnston2015}
{Johnston} R., {Vaccari} M., {Jarvis} M., {Smith} M., {Giovannoli} E.,
  {H{\"a}u{\ss}ler} B., {Prescott} M., 2015, \mnras, 453, 2540

\bibitem[{{Johnston} {et~al}\mbox{.}(2008){Johnston}, {Taylor}, {Bailes},
  {Bartel}, {Baugh}, {Bietenholz}, {Blake}, {Braun}, {Brown}, {Chatterjee},
  {Darling}, {Deller}, {Dodson}, {Edwards}, {Ekers}, {Ellingsen}, {Feain},
  {Gaensler}, {Haverkorn}, {Hobbs}, {Hopkins}, {Jackson}, {James}, {Joncas},
  {Kaspi}, {Kilborn}, {Koribalski}, {Kothes}, {Landecker}, {Lenc}, {Lovell},
  {Macquart}, {Manchester}, {Matthews}, {McClure-Griffiths}, {Norris}, {Pen},
  {Phillips}, {Power}, {Protheroe}, {Sadler}, {Schmidt}, {Stairs},
  {Staveley-Smith}, {Stil}, {Tingay}, {Tzioumis}, {Walker}, {Wall}, \&
  {Wolleben}}]{johnston2008}
{Johnston} S. {et~al.}, 2008, Experimental Astronomy, 22, 151

\bibitem[{{Kauffmann} \& {Haehnelt}(2000)}]{kh2000}
{Kauffmann} G., {Haehnelt} M., 2000, \mnras, 311, 576

\bibitem[{{Kennicutt}(1989)}]{kennicutt1989}
{Kennicutt}, Jr. R.~C., 1989, \apj, 344, 685

\bibitem[{{Kennicutt}(1998)}]{kennicutt1998}
{Kennicutt}, Jr. R.~C., 1998, \apj, 498, 541

\bibitem[{{Kennicutt} {et~al}\mbox{.}(2007){Kennicutt}, {Calzetti}, {Walter},
  {Helou}, {Hollenbach}, {Armus}, {Bendo}, {Dale}, {Draine}, {Engelbracht},
  {Gordon}, {Prescott}, {Regan}, {Thornley}, {Bot}, {Brinks}, {de Blok}, {de
  Mello}, {Meyer}, {Moustakas}, {Murphy}, {Sheth}, \& {Smith}}]{kennicutt2007}
{Kennicutt}, Jr. R.~C. {et~al.}, 2007, \apj, 671, 333

\bibitem[{{Keres}, {Yun} \& {Young}(2003){Keres}, {Yun}, \&
  {Young}}]{keres2003}
{Keres} D., {Yun} M.~S., {Young} J.~S., 2003, \apj, 582, 659

\bibitem[{{Kregel}, {van der Kruit} \& {de Grijs}(2002){Kregel}, {van der
  Kruit}, \& {de Grijs}}]{kregel2002}
{Kregel} M., {van der Kruit} P.~C., {de Grijs} R., 2002, \mnras, 334, 646

\bibitem[{{Kroupa}(2001)}]{kroupa2001}
{Kroupa} P., 2001, \mnras, 322, 231

\bibitem[{{Krumholz}(2013)}]{krumholz2013}
{Krumholz} M.~R., 2013, \mnras, 436, 2747

\bibitem[{{Krumholz} \& {Dekel}(2012)}]{kd2012}
{Krumholz} M.~R., {Dekel} A., 2012, \apj, 753, 16

\bibitem[{{Krumholz}, {McKee} \& {Tumlinson}(2008){Krumholz}, {McKee}, \&
  {Tumlinson}}]{krumholz2008}
{Krumholz} M.~R., {McKee} C.~F., {Tumlinson} J., 2008, \apj, 689, 865

\bibitem[{{Krumholz}, {McKee} \& {Tumlinson}(2009{\natexlab{a}}){Krumholz},
  {McKee}, \& {Tumlinson}}]{krumholz2009a}
{Krumholz} M.~R., {McKee} C.~F., {Tumlinson} J., 2009{\natexlab{a}}, \apj, 693,
  216

\bibitem[{{Krumholz}, {McKee} \& {Tumlinson}(2009{\natexlab{b}}){Krumholz},
  {McKee}, \& {Tumlinson}}]{krumholz2009b}
{Krumholz} M.~R., {McKee} C.~F., {Tumlinson} J., 2009{\natexlab{b}}, \apj, 699,
  850

\bibitem[{{Kuhlen} {et~al}\mbox{.}(2012){Kuhlen}, {Krumholz}, {Madau}, {Smith},
  \& {Wise}}]{kuhlen2012}
{Kuhlen} M., {Krumholz} M.~R., {Madau} P., {Smith} B.~D., {Wise} J., 2012,
  \apj, 749, 36

\bibitem[{{Lagos} {et~al}\mbox{.}(2011{\natexlab{a}}){Lagos}, {Baugh}, {Lacey},
  {Benson}, {Kim}, \& {Power}}]{lagos2011b}
{Lagos} C.~D.~P., {Baugh} C.~M., {Lacey} C.~G., {Benson} A.~J., {Kim} H.-S.,
  {Power} C., 2011{\natexlab{a}}, \mnras, 418, 1649

\bibitem[{{Lagos} {et~al}\mbox{.}(2012){Lagos}, {Bayet}, {Baugh}, {Lacey},
  {Bell}, {Fanidakis}, \& {Geach}}]{lagos2012}
{Lagos} C.~d.~P., {Bayet} E., {Baugh} C.~M., {Lacey} C.~G., {Bell} T.~A.,
  {Fanidakis} N., {Geach} J.~E., 2012, \mnras, 426, 2142

\bibitem[{{Lagos} {et~al}\mbox{.}(2011{\natexlab{b}}){Lagos}, {Lacey}, {Baugh},
  {Bower}, \& {Benson}}]{lagos2011a}
{Lagos} C.~D.~P., {Lacey} C.~G., {Baugh} C.~M., {Bower} R.~G., {Benson} A.~J.,
  2011{\natexlab{b}}, \mnras, 416, 1566

\bibitem[{{Lange} {et~al}\mbox{.}(2015){Lange}, {Driver}, {Robotham}, {Kelvin},
  {Graham}, {Alpaslan}, {Andrews}, {Baldry}, {Bamford}, {Bland-Hawthorn},
  {Brough}, {Cluver}, {Conselice}, {Davies}, {Haeussler}, {Konstantopoulos},
  {Loveday}, {Moffett}, {Norberg}, {Phillipps}, {Taylor},
  {L{\'o}pez-S{\'a}nchez}, \& {Wilkins}}]{lange2015}
{Lange} R. {et~al.}, 2015, \mnras, 447, 2603

\bibitem[{{Lange} {et~al}\mbox{.}(2016){Lange}, {Moffett}, {Driver},
  {Robotham}, {Lagos}, {Kelvin}, {Conselice}, {Margalef-Bentabol}, {Alpaslan},
  {Baldry}, {Bland-Hawthorn}, {Bremer}, {Brough}, {Cluver}, {Colless},
  {Davies}, {H{\"a}u{\ss}ler}, {Holwerda}, {Hopkins}, {Kafle}, {Kennedy},
  {Liske}, {Phillipps}, {Popescu}, {Taylor}, {Tuffs}, {van Kampen}, \&
  {Wright}}]{lange2016}
{Lange} R. {et~al.}, 2016, \mnras, 462, 1470

\bibitem[{{Lee} {et~al}\mbox{.}(2011){Lee}, {Dey}, {Reddy}, {Brown},
  {Gonzalez}, {Jannuzi}, {Cooper}, {Fan}, {Bian}, {Glikman}, {Stern},
  {Brodwin}, \& {Cooray}}]{lee2011}
{Lee} K.-S. {et~al.}, 2011, \apj, 733, 99

\bibitem[{{Lee} {et~al}\mbox{.}(2012){Lee}, {Ferguson}, {Wiklind}, {Dahlen},
  {Dickinson}, {Giavalisco}, {Grogin}, {Papovich}, {Messias}, {Guo}, \&
  {Lin}}]{lee2012}
{Lee} K.-S. {et~al.}, 2012, \apj, 752, 66

\bibitem[{{Leroy} {et~al}\mbox{.}(2009){Leroy}, {Walter}, {Bigiel}, {Usero},
  {Weiss}, {Brinks}, {de Blok}, {Kennicutt}, {Schuster}, {Kramer},
  {Wiesemeyer}, \& {Roussel}}]{leroy2009}
{Leroy} A.~K. {et~al.}, 2009, \aj, 137, 4670

\bibitem[{{Leroy} {et~al}\mbox{.}(2008){Leroy}, {Walter}, {Brinks}, {Bigiel},
  {de Blok}, {Madore}, \& {Thornley}}]{leroy2008}
{Leroy} A.~K., {Walter} F., {Brinks} E., {Bigiel} F., {de Blok} W.~J.~G.,
  {Madore} B., {Thornley} M.~D., 2008, \aj, 136, 2782

\bibitem[{{Li} \& {White}(2009)}]{li2009}
{Li} C., {White} S.~D.~M., 2009, \mnras, 398, 2177

\bibitem[{{Li}, {De Lucia} \& {Helmi}(2010){Li}, {De Lucia}, \&
  {Helmi}}]{li2010}
{Li} Y.-S., {De Lucia} G., {Helmi} A., 2010, \mnras, 401, 2036

\bibitem[{{Magdis} {et~al}\mbox{.}(2010){Magdis}, {Rigopoulou}, {Huang}, \&
  {Fazio}}]{magdis2010}
{Magdis} G.~E., {Rigopoulou} D., {Huang} J.-S., {Fazio} G.~G., 2010, \mnras,
  401, 1521

\bibitem[{{Maiolino} {et~al}\mbox{.}(2008){Maiolino}, {Nagao}, {Grazian},
  {Cocchia}, {Marconi}, {Mannucci}, {Cimatti}, {Pipino}, {Ballero}, {Calura},
  {Chiappini}, {Fontana}, {Granato}, {Matteucci}, {Pastorini}, {Pentericci},
  {Risaliti}, {Salvati}, \& {Silva}}]{maiolino2008}
{Maiolino} R. {et~al.}, 2008, \aap, 488, 463

\bibitem[{{Martin} {et~al}\mbox{.}(2010){Martin}, {Papastergis}, {Giovanelli},
  {Haynes}, {Springob}, \& {Stierwalt}}]{martin2010}
{Martin} A.~M., {Papastergis} E., {Giovanelli} R., {Haynes} M.~P., {Springob}
  C.~M., {Stierwalt} S., 2010, \apj, 723, 1359

\bibitem[{{Martin} \& {Kennicutt}(2001)}]{martin2001}
{Martin} C.~L., {Kennicutt}, Jr. R.~C., 2001, \apj, 555, 301

\bibitem[{{Mitchell} {et~al}\mbox{.}(2014){Mitchell}, {Lacey}, {Cole}, \&
  {Baugh}}]{mitchell2014}
{Mitchell} P.~D., {Lacey} C.~G., {Cole} S., {Baugh} C.~M., 2014, \mnras, 444,
  2637

\bibitem[{{Mo}, {Mao} \& {White}(1998){Mo}, {Mao}, \& {White}}]{mo1998}
{Mo} H.~J., {Mao} S., {White} S.~D.~M., 1998, \mnras, 295, 319

\bibitem[{{Mo} \& {White}(2002)}]{mo2002}
{Mo} H.~J., {White} S.~D.~M., 2002, \mnras, 336, 112

\bibitem[{{Moustakas} {et~al}\mbox{.}(2013){Moustakas}, {Coil}, {Aird},
  {Blanton}, {Cool}, {Eisenstein}, {Mendez}, {Wong}, {Zhu}, \&
  {Arnouts}}]{moustakas2013}
{Moustakas} J. {et~al.}, 2013, \apj, 767, 50

\bibitem[{{Muratov} {et~al}\mbox{.}(2015){Muratov}, {Kere{\v s}},
  {Faucher-Gigu{\`e}re}, {Hopkins}, {Quataert}, \& {Murray}}]{muratov2015}
{Muratov} A.~L., {Kere{\v s}} D., {Faucher-Gigu{\`e}re} C.-A., {Hopkins} P.~F.,
  {Quataert} E., {Murray} N., 2015, \mnras, 454, 2691

\bibitem[{{Murray} \& {Rahman}(2010)}]{murray2010}
{Murray} N., {Rahman} M., 2010, \apj, 709, 424

\bibitem[{{Nan} {et~al}\mbox{.}(2011){Nan}, {Li}, {Jin}, {Wang}, {Zhu}, {Zhu},
  {Zhang}, {Yue}, \& {Qian}}]{nan2011}
{Nan} R. {et~al.}, 2011, International Journal of Modern Physics D, 20, 989

\bibitem[{{Noeske} {et~al}\mbox{.}(2007){Noeske}, {Weiner}, {Faber},
  {Papovich}, {Koo}, {Somerville}, {Bundy}, {Conselice}, {Newman},
  {Schiminovich}, {Le Floc'h}, {Coil}, {Rieke}, {Lotz}, {Primack}, {Barmby},
  {Cooper}, {Davis}, {Ellis}, {Fazio}, {Guhathakurta}, {Huang}, {Kassin},
  {Martin}, {Phillips}, {Rich}, {Small}, {Willmer}, \& {Wilson}}]{noeske2007}
{Noeske} K.~G. {et~al.}, 2007, \apjl, 660, L43

\bibitem[{{Noterdaeme} {et~al}\mbox{.}(2012){Noterdaeme}, {Petitjean},
  {Carithers}, {P{\^a}ris}, {Font-Ribera}, {Bailey}, {Aubourg}, {Bizyaev},
  {Ebelke}, {Finley}, {Ge}, {Malanushenko}, {Malanushenko},
  {Miralda-Escud{\'e}}, {Myers}, {Oravetz}, {Pan}, {Pieri}, {Ross},
  {Schneider}, {Simmons}, \& {York}}]{noterdaeme2012}
{Noterdaeme} P. {et~al.}, 2012, \aap, 547, L1

\bibitem[{{Obreschkow} \& {Rawlings}(2009{\natexlab{a}})}]{obreschkow2009b}
{Obreschkow} D., {Rawlings} S., 2009{\natexlab{a}}, \apjl, 696, L129

\bibitem[{{Obreschkow} \& {Rawlings}(2009{\natexlab{b}})}]{obreschkow2009a}
{Obreschkow} D., {Rawlings} S., 2009{\natexlab{b}}, \mnras, 394, 1857

\bibitem[{{Ostriker}, {McKee} \& {Leroy}(2010){Ostriker}, {McKee}, \&
  {Leroy}}]{ostriker2010}
{Ostriker} E.~C., {McKee} C.~F., {Leroy} A.~K., 2010, \apj, 721, 975

\bibitem[{{P{\'e}rez-Gonz{\'a}lez}
  {et~al}\mbox{.}(2008){P{\'e}rez-Gonz{\'a}lez}, {Rieke}, {Villar}, {Barro},
  {Blaylock}, {Egami}, {Gallego}, {Gil de Paz}, {Pascual}, {Zamorano}, \&
  {Donley}}]{perezgonz2008}
{P{\'e}rez-Gonz{\'a}lez} P.~G. {et~al.}, 2008, \apj, 675, 234

\bibitem[{{P{\'e}roux} {et~al}\mbox{.}(2005){P{\'e}roux}, {Dessauges-Zavadsky},
  {D'Odorico}, {Sun Kim}, \& {McMahon}}]{peroux2005}
{P{\'e}roux} C., {Dessauges-Zavadsky} M., {D'Odorico} S., {Sun Kim} T.,
  {McMahon} R.~G., 2005, \mnras, 363, 479

\bibitem[{{Planck Collaboration} {et~al}\mbox{.}(2015){Planck Collaboration},
  {Ade}, {Aghanim}, {Arnaud}, {Ashdown}, {Aumont}, {Baccigalupi}, {Banday},
  {Barreiro}, {Bartlett}, \& et~al.}]{planck2015}
{Planck Collaboration} {et~al.}, 2015, ArXiv 1502.01589

\bibitem[{{Popping} {et~al}\mbox{.}(2015){Popping}, {Caputi}, {Trager},
  {Somerville}, {Dekel}, {Kassin}, {Kocevski}, {Koekemoer}, {Faber},
  {Ferguson}, {Galametz}, {Grogin}, {Guo}, {Lu}, {Wel}, \&
  {Weiner}}]{popping2015}
{Popping} G. {et~al.}, 2015, \mnras, 454, 2258

\bibitem[{{Popping}, {Somerville} \& {Trager}(2014){Popping}, {Somerville}, \&
  {Trager}}]{popping2014}
{Popping} G., {Somerville} R.~S., {Trager} S.~C., 2014, \mnras, 442, 2398

\bibitem[{{Prochaska} \& {Wolfe}(2009)}]{prochaska2009}
{Prochaska} J.~X., {Wolfe} A.~M., 2009, \apj, 696, 1543

\bibitem[{{Rao}, {Turnshek} \& {Nestor}(2006){Rao}, {Turnshek}, \&
  {Nestor}}]{rao2006}
{Rao} S.~M., {Turnshek} D.~A., {Nestor} D.~B., 2006, \apj, 636, 610

\bibitem[{{Robitaille} \& {Whitney}(2010)}]{robitaille2010}
{Robitaille} T.~P., {Whitney} B.~A., 2010, \apjl, 710, L11

\bibitem[{{Sabra} {et~al}\mbox{.}(2015){Sabra}, {Saliba}, {Abi Akl}, \&
  {Chahine}}]{sabra2015}
{Sabra} B.~M., {Saliba} C., {Abi Akl} M., {Chahine} G., 2015, \apj, 803, 5

\bibitem[{{Saintonge} {et~al}\mbox{.}(2011){Saintonge}, {Kauffmann}, {Kramer},
  {Tacconi}, {Buchbender}, {Catinella}, {Fabello}, {Graci{\'a}-Carpio}, {Wang},
  {Cortese}, {Fu}, {Genzel}, {Giovanelli}, {Guo}, {Haynes}, {Heckman},
  {Krumholz}, {Lemonias}, {Li}, {Moran}, {Rodriguez-Fernandez}, {Schiminovich},
  {Schuster}, \& {Sievers}}]{saintonge2011}
{Saintonge} A. {et~al.}, 2011, \mnras, 415, 32

\bibitem[{{Saintonge} {et~al}\mbox{.}(2013){Saintonge}, {Lutz}, {Genzel},
  {Magnelli}, {Nordon}, {Tacconi}, {Baker}, {Bandara}, {Berta}, {F{\"o}rster
  Schreiber}, {Poglitsch}, {Sturm}, {Wuyts}, \& {Wuyts}}]{saintonge2013}
{Saintonge} A. {et~al.}, 2013, \apj, 778, 2

\bibitem[{{Salim} {et~al}\mbox{.}(2007){Salim}, {Rich}, {Charlot},
  {Brinchmann}, {Johnson}, {Schiminovich}, {Seibert}, {Mallery}, {Heckman},
  {Forster}, {Friedman}, {Martin}, {Morrissey}, {Neff}, {Small}, {Wyder},
  {Bianchi}, {Donas}, {Lee}, {Madore}, {Milliard}, {Szalay}, {Welsh}, \&
  {Yi}}]{salim2007}
{Salim} S. {et~al.}, 2007, \apjs, 173, 267

\bibitem[{{Salmi} {et~al}\mbox{.}(2012){Salmi}, {Daddi}, {Elbaz}, {Sargent},
  {Dickinson}, {Renzini}, {Bethermin}, \& {Le Borgne}}]{salmi2012}
{Salmi} F., {Daddi} E., {Elbaz} D., {Sargent} M.~T., {Dickinson} M., {Renzini}
  A., {Bethermin} M., {Le Borgne} D., 2012, \apjl, 754, L14

\bibitem[{{Sanders} {et~al}\mbox{.}(2015){Sanders}, {Shapley}, {Kriek},
  {Reddy}, {Freeman}, {Coil}, {Siana}, {Mobasher}, {Shivaei}, {Price}, \& {de
  Groot}}]{sanders2015}
{Sanders} R.~L. {et~al.}, 2015, \apj, 799, 138

\bibitem[{{Santini} {et~al}\mbox{.}(2009){Santini}, {Fontana}, {Grazian},
  {Salimbeni}, {Fiore}, {Fontanot}, {Boutsia}, {Castellano}, {Cristiani}, {de
  Santis}, {Gallozzi}, {Giallongo}, {Menci}, {Nonino}, {Paris}, {Pentericci},
  \& {Vanzella}}]{santini2009}
{Santini} P. {et~al.}, 2009, \aap, 504, 751

\bibitem[{{Schaye} {et~al}\mbox{.}(2010){Schaye}, {Dalla Vecchia}, {Booth},
  {Wiersma}, {Theuns}, {Haas}, {Bertone}, {Duffy}, {McCarthy}, \& {van de
  Voort}}]{schaye2010}
{Schaye} J. {et~al.}, 2010, \mnras, 402, 1536

\bibitem[{{Schechter}(1976)}]{schechter1976}
{Schechter} P., 1976, \apj, 203, 297

\bibitem[{{Schmidt}(1959)}]{schmidt1959}
{Schmidt} M., 1959, \apj, 129, 243

\bibitem[{{Shen} {et~al}\mbox{.}(2003){Shen}, {Mo}, {White}, {Blanton},
  {Kauffmann}, {Voges}, {Brinkmann}, \& {Csabai}}]{shen2003}
{Shen} S., {Mo} H.~J., {White} S.~D.~M., {Blanton} M.~R., {Kauffmann} G.,
  {Voges} W., {Brinkmann} J., {Csabai} I., 2003, \mnras, 343, 978

\bibitem[{{Sobral} {et~al}\mbox{.}(2014){Sobral}, {Best}, {Smail}, {Mobasher},
  {Stott}, \& {Nisbet}}]{sobral2014}
{Sobral} D., {Best} P.~N., {Smail} I., {Mobasher} B., {Stott} J., {Nisbet} D.,
  2014, \mnras, 437, 3516

\bibitem[{{Somerville}, {Popping} \& {Trager}(2015){Somerville}, {Popping}, \&
  {Trager}}]{somerville2015}
{Somerville} R.~S., {Popping} G., {Trager} S.~C., 2015, \mnras, 453, 4337

\bibitem[{{Speagle} {et~al}\mbox{.}(2014){Speagle}, {Steinhardt}, {Capak}, \&
  {Silverman}}]{speagle2014}
{Speagle} J.~S., {Steinhardt} C.~L., {Capak} P.~L., {Silverman} J.~D., 2014,
  \apjs, 214, 15

\bibitem[{{Springel} {et~al}\mbox{.}(2005){Springel}, {White}, {Jenkins},
  {Frenk}, {Yoshida}, {Gao}, {Navarro}, {Thacker}, {Croton}, {Helly},
  {Peacock}, {Cole}, {Thomas}, {Couchman}, {Evrard}, {Colberg}, \&
  {Pearce}}]{springel2005}
{Springel} V. {et~al.}, 2005, \nat, 435, 629

\bibitem[{{Steidel} {et~al}\mbox{.}(2014){Steidel}, {Rudie}, {Strom},
  {Pettini}, {Reddy}, {Shapley}, {Trainor}, {Erb}, {Turner}, {Konidaris},
  {Kulas}, {Mace}, {Matthews}, \& {McLean}}]{steidel2014}
{Steidel} C.~C. {et~al.}, 2014, \apj, 795, 165

\bibitem[{{Stevens} {et~al}\mbox{.}(2017){Stevens}, {del P.~Lagos},
  {Contreras}, {Croton}, {Padilla}, {Schaller}, {Schaye}, \&
  {Theuns}}]{stevens2016}
{Stevens} A.~R.~H., {del P.~Lagos} C., {Contreras} S., {Croton} D.~J.,
  {Padilla} N.~D., {Schaller} M., {Schaye} J., {Theuns} T., 2017, \mnras

\bibitem[{{Tacconi} {et~al}\mbox{.}(2013){Tacconi}, {Neri}, {Genzel}, {Combes},
  {Bolatto}, {Cooper}, {Wuyts}, {Bournaud}, {Burkert}, {Comerford}, {Cox},
  {Davis}, {F{\"o}rster Schreiber}, {Garc{\'{\i}}a-Burillo}, {Gracia-Carpio},
  {Lutz}, {Naab}, {Newman}, {Omont}, {Saintonge}, {Shapiro Griffin}, {Shapley},
  {Sternberg}, \& {Weiner}}]{tacconi2013}
{Tacconi} L.~J. {et~al.}, 2013, \apj, 768, 74

\bibitem[{{Toomre}(1964)}]{toomre1964}
{Toomre} A., 1964, \apj, 139, 1217

\bibitem[{{Tremonti} {et~al}\mbox{.}(2004){Tremonti}, {Heckman}, {Kauffmann},
  {Brinchmann}, {Charlot}, {White}, {Seibert}, {Peng}, {Schlegel}, {Uomoto},
  {Fukugita}, \& {Brinkmann}}]{tremonti2004}
{Tremonti} C.~A. {et~al.}, 2004, \apj, 613, 898

\bibitem[{{van der Wel} {et~al}\mbox{.}(2014){van der Wel}, {Franx}, {van
  Dokkum}, {Skelton}, {Momcheva}, {Whitaker}, {Brammer}, {Bell}, {Rix},
  {Wuyts}, {Ferguson}, {Holden}, {Barro}, {Koekemoer}, {Chang}, {McGrath},
  {H{\"a}ussler}, {Dekel}, {Behroozi}, {Fumagalli}, {Leja}, {Lundgren},
  {Maseda}, {Nelson}, {Wake}, {Patel}, {Labb{\'e}}, {Faber}, {Grogin}, \&
  {Kocevski}}]{vanderwel2014}
{van der Wel} A. {et~al.}, 2014, \apj, 788, 28

\bibitem[{{Volonteri}, {Natarajan} \& {G{\"u}ltekin}(2011){Volonteri},
  {Natarajan}, \& {G{\"u}ltekin}}]{volonteri2011}
{Volonteri} M., {Natarajan} P., {G{\"u}ltekin} K., 2011, \apj, 737, 50

\bibitem[{{Walter} {et~al}\mbox{.}(2008){Walter}, {Brinks}, {de Blok},
  {Bigiel}, {Kennicutt}, {Thornley}, \& {Leroy}}]{walter2008}
{Walter} F., {Brinks} E., {de Blok} W.~J.~G., {Bigiel} F., {Kennicutt}, Jr.
  R.~C., {Thornley} M.~D., {Leroy} A., 2008, \aj, 136, 2563

\bibitem[{{Walter} {et~al}\mbox{.}(2014){Walter}, {Decarli}, {Sargent},
  {Carilli}, {Dickinson}, {Riechers}, {Ellis}, {Stark}, {Weiner}, {Aravena},
  {Bell}, {Bertoldi}, {Cox}, {Da Cunha}, {Daddi}, {Downes}, {Lentati},
  {Maiolino}, {Menten}, {Neri}, {Rix}, \& {Weiss}}]{walter2014}
{Walter} F. {et~al.}, 2014, \apj, 782, 79

\bibitem[{{Wang} {et~al}\mbox{.}(2008){Wang}, {De Lucia}, {Kitzbichler}, \&
  {White}}]{wang2008}
{Wang} J., {De Lucia} G., {Kitzbichler} M.~G., {White} S.~D.~M., 2008, \mnras,
  384, 1301

\bibitem[{{Weinmann} {et~al}\mbox{.}(2012){Weinmann}, {Pasquali},
  {Oppenheimer}, {Finlator}, {Mendel}, {Crain}, \& {Macci{\`o}}}]{weinmann2012}
{Weinmann} S.~M., {Pasquali} A., {Oppenheimer} B.~D., {Finlator} K., {Mendel}
  J.~T., {Crain} R.~A., {Macci{\`o}} A.~V., 2012, \mnras, 426, 2797

\bibitem[{{White}, {Somerville} \& {Ferguson}(2015){White}, {Somerville}, \&
  {Ferguson}}]{white2015}
{White} C.~E., {Somerville} R.~S., {Ferguson} H.~C., 2015, \apj, 799, 201

\bibitem[{{Wolfire} {et~al}\mbox{.}(2003){Wolfire}, {McKee}, {Hollenbach}, \&
  {Tielens}}]{wolfire2003}
{Wolfire} M.~G., {McKee} C.~F., {Hollenbach} D., {Tielens} A.~G.~G.~M., 2003,
  \apj, 587, 278

\bibitem[{{Wong} \& {Blitz}(2002)}]{wong2002}
{Wong} T., {Blitz} L., 2002, \apj, 569, 157

\bibitem[{{Wootten} \& {Thompson}(2009)}]{wootten2009}
{Wootten} A., {Thompson} A.~R., 2009, IEEE Proceedings, 97, 1463

\bibitem[{{Zafar} {et~al}\mbox{.}(2013){Zafar}, {P{\'e}roux}, {Popping},
  {Milliard}, {Deharveng}, \& {Frank}}]{zafar2013}
{Zafar} T., {P{\'e}roux} C., {Popping} A., {Milliard} B., {Deharveng} J.-M.,
  {Frank} S., 2013, \aap, 556, A141

\bibitem[{{Zhao} {et~al}\mbox{.}(2009){Zhao}, {Jing}, {Mo}, \&
  {B{\"o}rner}}]{zhao2009}
{Zhao} D.~H., {Jing} Y.~P., {Mo} H.~J., {B{\"o}rner} G., 2009, \apj, 707, 354

\bibitem[{{Zoldan} {et~al}\mbox{.}(2017){Zoldan}, {De Lucia}, {Xie},
  {Fontanot}, \& {Hirschmann}}]{zoldan2016}
{Zoldan} A., {De Lucia} G., {Xie} L., {Fontanot} F., {Hirschmann} M., 2017,
  \mnras, 465, 2236

\bibitem[{{Zwaan} {et~al}\mbox{.}(2005){Zwaan}, {Meyer}, {Staveley-Smith}, \&
  {Webster}}]{zwaan2005}
{Zwaan} M.~A., {Meyer} M.~J., {Staveley-Smith} L., {Webster} R.~L., 2005,
  \mnras, 359, L30

\end{thebibliography}

\appendix

\section{Comparison between different models for disk sizes}
\label{app:disk}

In Section~\ref{subsec:diskmodel}, we introduce our updated model for disk
sizes, based on accumulation of angular momentum through different physical
processes.  In this section, we compare model predictions obtained using the
semi-analytic model described in HDLF16 with its original
prescriptions to model disk sizes, and our updated modelling.

\begin{figure*}
\includegraphics[width = 1.0\textwidth]{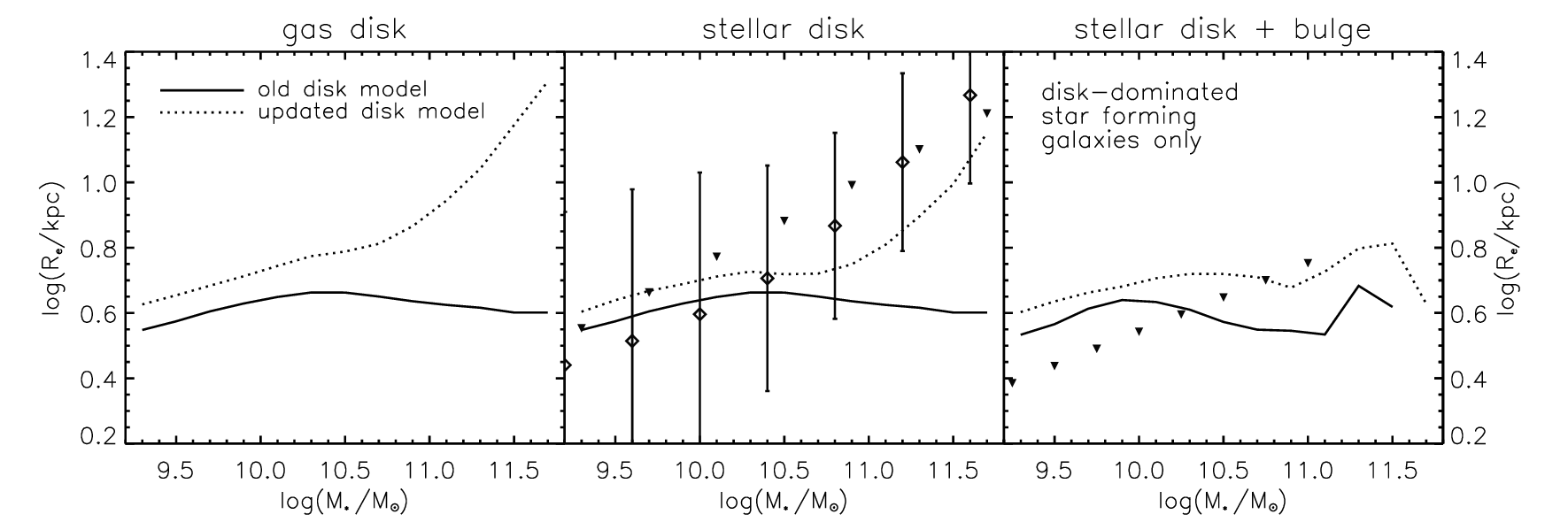}
\caption{The size-mass relation predicted by the model described
in HDLF16 using its original prescriptions for disk sizes (solid lines), and
our updated model (dotted lines). The left and middle panels show the half-mass
radius for the gaseous and stellar disks, respectively. The right panel shows
the half-mass radius of stars, considering both the bulge and disk components,
for disk-dominated, star forming galaxies only ($sSFR>0.3/t_{\rm H}$,
$M_{bulge}/M_{\star}<0.5$). The diamonds with error bars and triangles in the
middle panel are observed stellar disk sizes based on SDSS  and
GAMA\citep[][half-light radius for disk only]{dutton2011,lange2016}. The
triangles are observed sizes based on GAMA data \citep[][half-light
radius for disk and bulge]{lange2015}.}
\label{fig:diskmethodcompare}
\end{figure*}

The left and middle panels of Fig.~\ref{fig:diskmethodcompare} show the
half-mass radius of the gaseous and stellar disks as a function of galaxy
stellar mass.  Using our updated model for disk size, both radii are larger
than using the disk model adopted in HDLF16, particularly for
stellar masses larger than log$({\rm M}_{*}/{\rm M}_\odot)\sim 10.5$. The most
massive galaxies are bulge dominated and acquired their stellar
mass primarily through mergers and accretions of lower mass systems. In the
original model used in HDLF16, the size of the disk was not updated
during galaxy mergers, while we now trace sizes of both components adopting a
physically motivated scheme. If we consider only disk-dominated 
($M_{bulge}/M_{\star}<0.5$), star forming galaxies with specific star formation 
rate sSFR$>0.3/t_{\rm H}$, the two models predict a very similar 
size-mass relation. Results are shown in the right panel of 
Fig.~\ref{fig:diskmethodcompare} where we consider the half-mass radius of 
the composed system disk+bulge.

It is worth stressing that, although there are significant differences between
the disk sizes of massive galaxies predicted by the HDLF16
model and by the same physical model adopting our updated prescriptions for
disk size, this does not introduce significant differences for other galaxy
properties or statistics such as e.g. the galaxy stellar mass function and
other scaling relations. 

\section{Resolution tests}
\label{app:resolution}

We use the MS and MSII to quantify the resolution limits in our model.
These two simulations are based on the same cosmological model and are 
run using the same simulation code, but the mass resolution of MSII is 125
times higher than that of the MS.

\begin{figure}
\includegraphics[width = 0.4\textwidth]{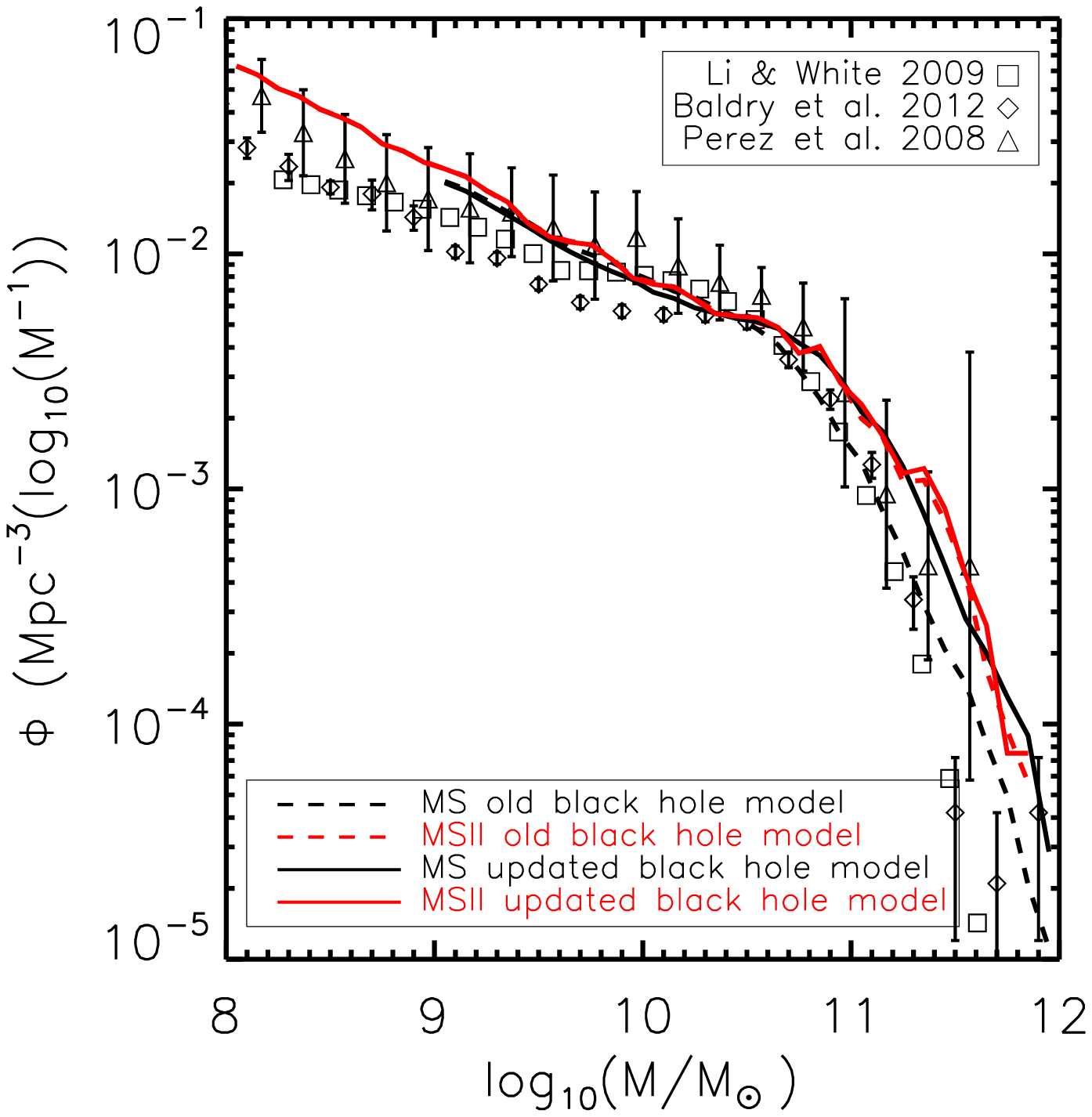}
\caption{Stellar mass functions based on the MS (black lines) and the MSII (red
lines). Dashed lines correspond to the model introduced in \citet[][the FIRE
  feedback scheme]{hirschmann2016}, while solid lines correspond to the same
  physical model including the updates described in
  Sections~\ref{subsec:diskmodel} and \ref{subsec:bhmodel} for the disk size and
  black hole model.}
\label{fig:smf_MSII}
\end{figure}

\begin{figure}
\includegraphics[width = 0.4\textwidth]{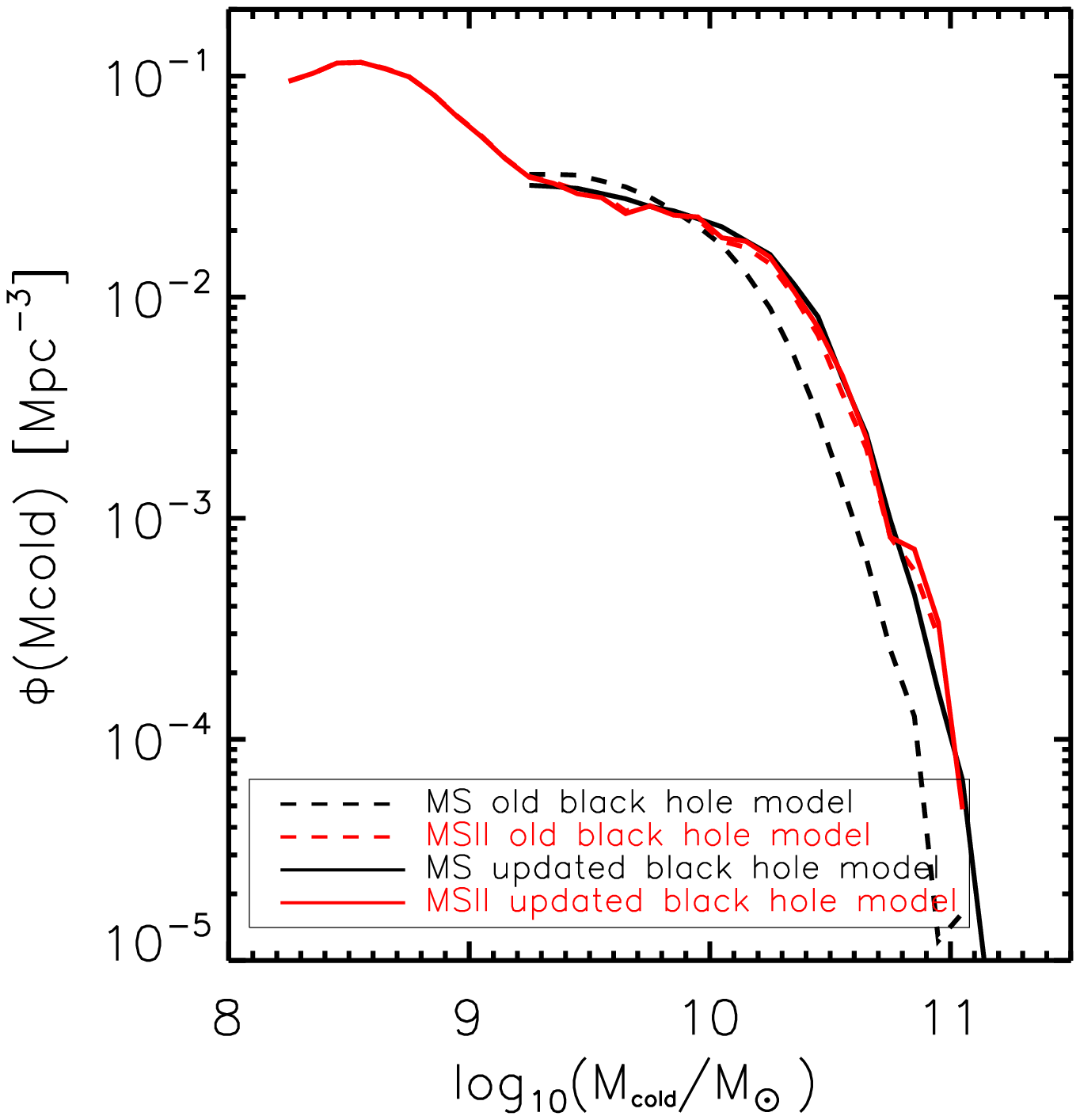}
\caption{Same as in Fig.~\ref{fig:smf_MSII} but for the cold gas mass function.}
\label{fig:cg_MSII}
\end{figure}

Fig.~\ref{fig:smf_MSII} shows the galaxy stellar mass function predicted by the
model presented in \citet[][their FIRE feedback scheme]{hirschmann2016}, both
based on the MS (dashed black line) and the MSII (dashed red line). The figure
shows that the convergence is good over the mass range log$({\rm M}_{*}/{\rm
M}_\odot) = 9 - 10.5$, while the model based on the MS tends to under-predict
the number densities of most massive galaxies with respect to the model based on the
MSII. Fig.~\ref{fig:cg_MSII} shows the corresponding results for the cold gas
mass function. Also in this case, there is a discrepancy at the massive end,
with the MS corresponding to lower number densities of gas rich galaxies with
respect to the MSII.

\begin{figure}

\includegraphics[width = 0.4\textwidth]{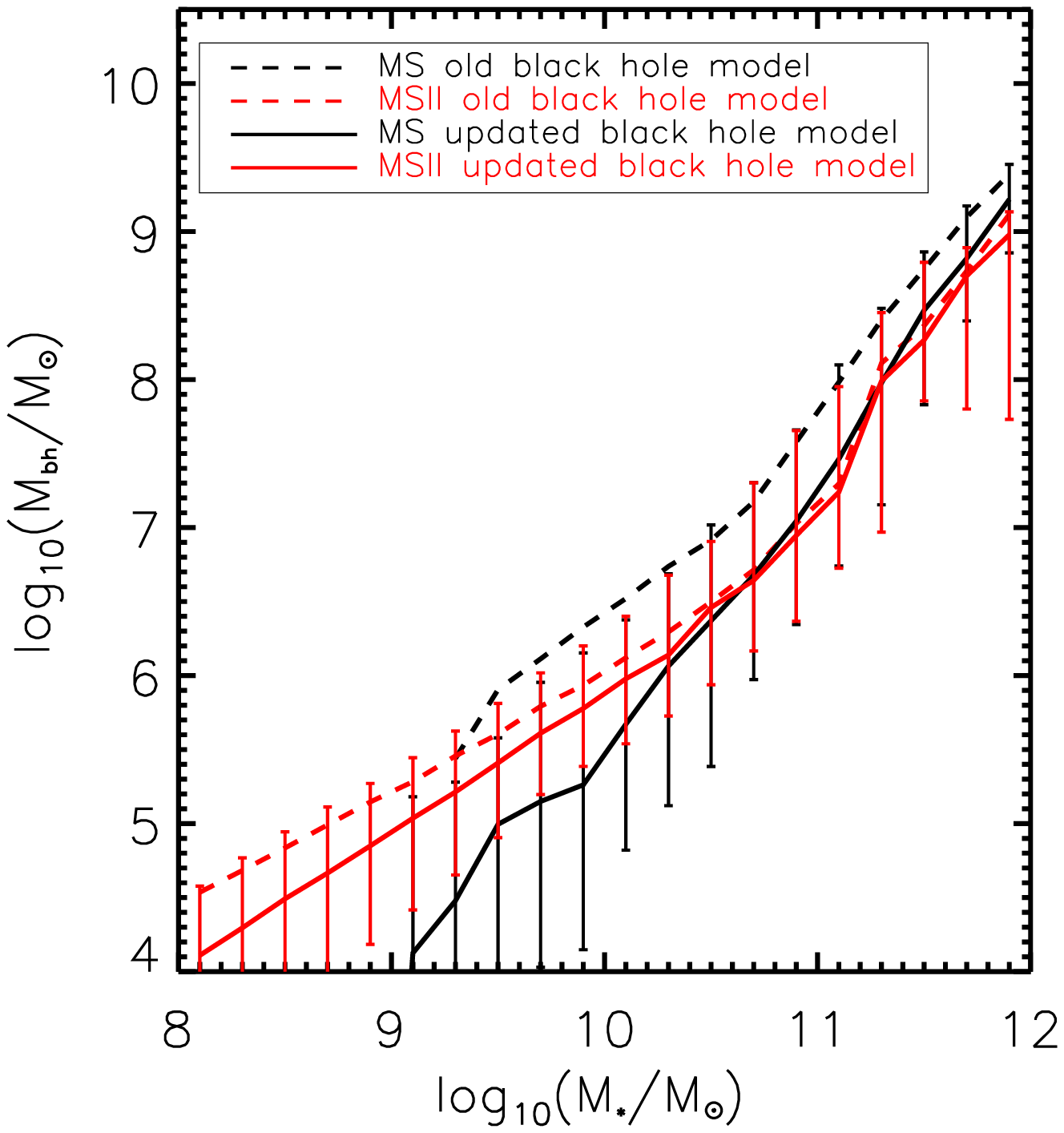}
\caption{Same as in Fig.~\ref{fig:smf_MSII}, but for black hole mass-galaxy 
stellar mass relation.}
\label{fig:bh_star_MSII}
\end{figure}

\begin{figure}
\includegraphics[width = 0.4\textwidth]{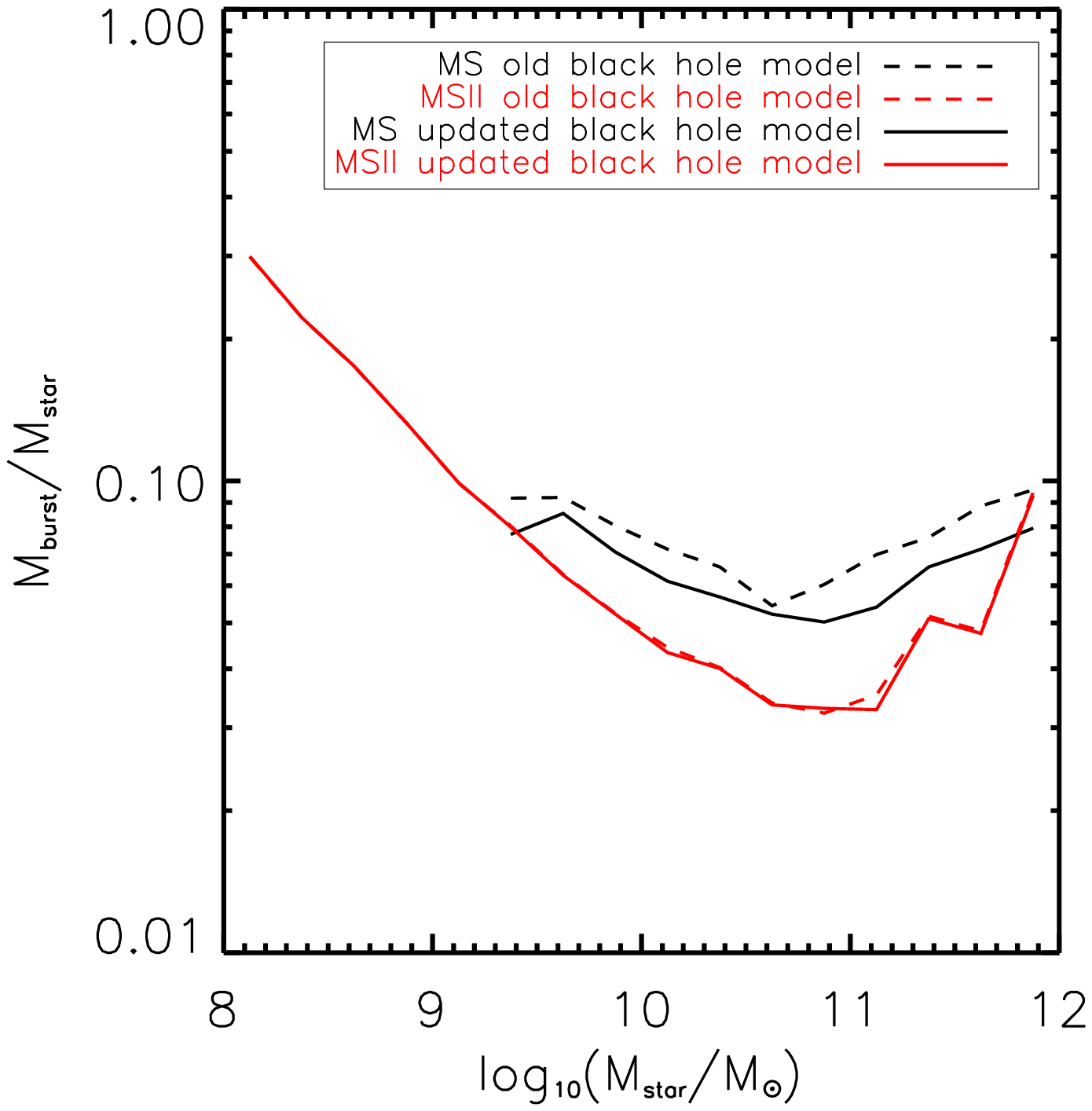}
\caption{Same as in Fig.~\ref{fig:smf_MSII}, but for the fraction of stars 
formed during merger driven star bursts. }
\label{fig:burstfrac}
\end{figure}

We find that this is largely due to a difference in the black hole
masses: specifically, if we switch off the accretion onto black holes during
galaxy mergers, predictions based on the MS and MSII for the galaxy stellar
mass function are consistent at the massive end. Fig.~\ref{fig:bh_star_MSII} 
shows the relation between the black hole mass and the galaxy stellar mass.
When considering the HDLF16 model, the relation based on the MSII is shifted 
down with respect to that based on the MS by about $\sim 0.5$~dex. The more 
massive black holes in the MS cause a more efficient suppression of galaxy 
formation (via AGN feedback) at the massive end. Thus, the galaxy stellar 
mass function based on the MS is below that based on the MSII.  

The results described above can be understood as follows: at early redshift,
star formation driven by cooling flows dominates the evolution of galaxies.  In
the MSII, dark matter halos are resolved earlier than that in the MS, and so
star formation starts earlier, locking a fraction of the gas available in
stars. Therefore, when the first mergers take place, less gas is available to
fuel the black hole growth. In the MS, the first resolved haloes are identified
at later times with respect to the MSII. As these haloes are in the rapid
cooling regime, larger amount of cold gas are dumped into the galaxies and
become available for black hole growth during the first galaxy mergers.  In our
model, gas rich mergers also result in a larger fraction of stars formed during
mergers. To quantify the importance of this channel, we record the mass
of stars formed in mergers $M_{\rm burst}$ for runs based on both simulations. 
Fig.~\ref{fig:burstfrac} shows the median fraction of $M_{\rm burst}$ 
with respect to the galaxy stellar mass as a function of the
latter. For the MS, the fraction is about $1.4$ times larger than that obtained
for the MSII. In both simulations, however, bursts contribute for less than
$\sim 10$ per cent of the total stellar mass
for galaxies with $M_{\star} > 10^9 \solarmass$. Thus we argue that the main
reason for the differences seen in the galaxy stellar mass function and cold
gas mass function is due to the systematic differences in the black hole growth. 

As explained in Section~\ref{subsec:bhmodel}, we update the black hole
model used in HDLF16 by `planting a black hole seed' in each
galaxy sitting at the centre of a halo with virial temperatures above $10^4$~K.
We rerun our resolution tests using the same physical model adopted
in HDLF16 but including our updated black hole model. The solid
lines shown in Figs.~\ref{fig:smf_MSII}, \ref{fig:cg_MSII},
and \ref{fig:bh_star_MSII} show results from these tests. Both the galaxy
stellar mass function and the cold gas mass function now converge well at the
most massive end. Our updated black hole growth model also predicts consistent
results for the relation between the black hole mass and the galaxy stellar
mass (see solid lines in Fig.~\ref{fig:bh_star_MSII}), although black holes 
tend to be more massive in the MSII for galaxies with $M_{\star} <
10^{10} \solarmass$.

\section{Different choices of $G'_0$ and $\rho_{sd}$}
\label{app:test_g0rhosd}

This section presents results of different tests related to the definition of
the interstellar radiation field ($G'_0$) within the GK11 model, and of the
density of dark matter and stars ($\rho_{sd}$) within the K13 model.

\begin{figure*}
\centering
\begin{subfigure}{.33\textwidth}
  \centering
  \includegraphics[width=1.0\linewidth]{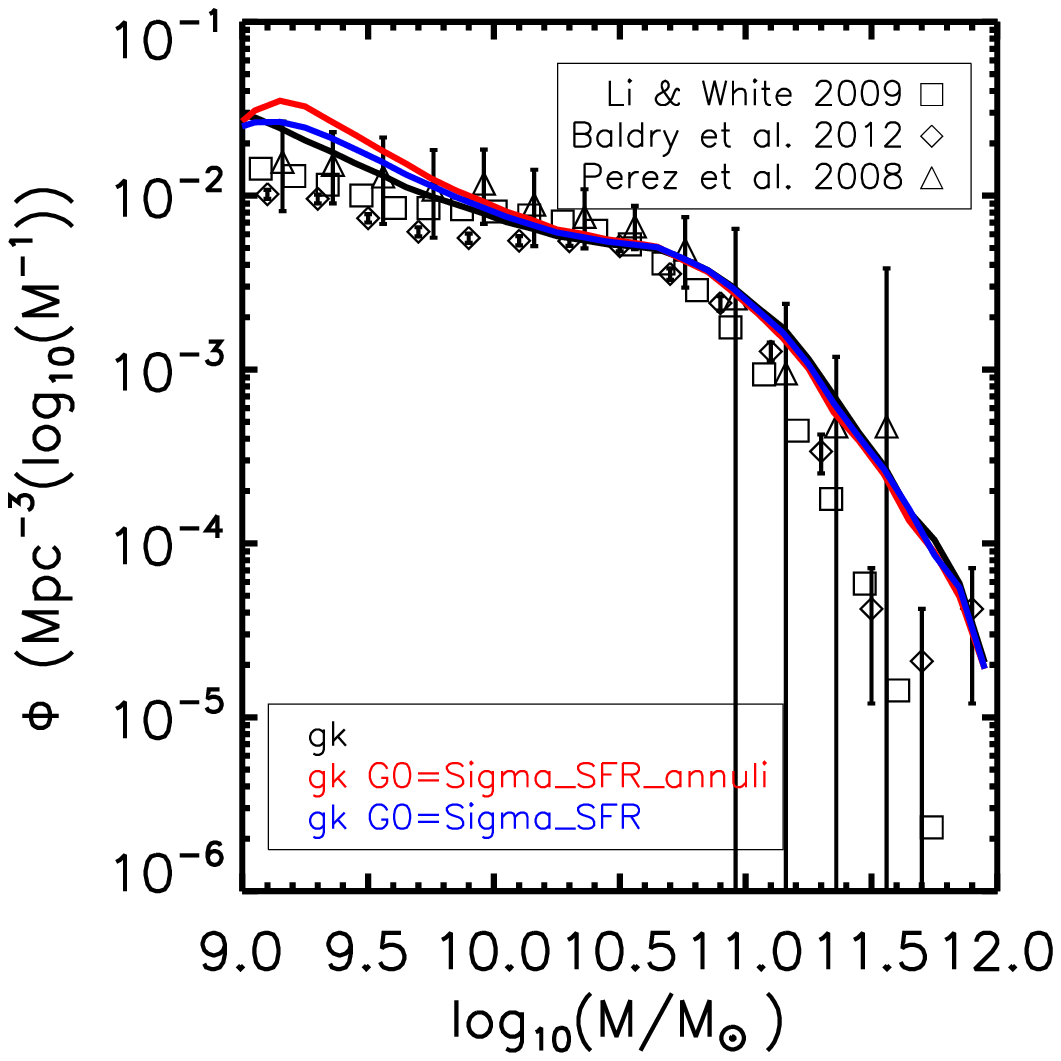}
\end{subfigure}%
\begin{subfigure}{.33\textwidth}
  \centering
  \includegraphics[width=1.0\linewidth]{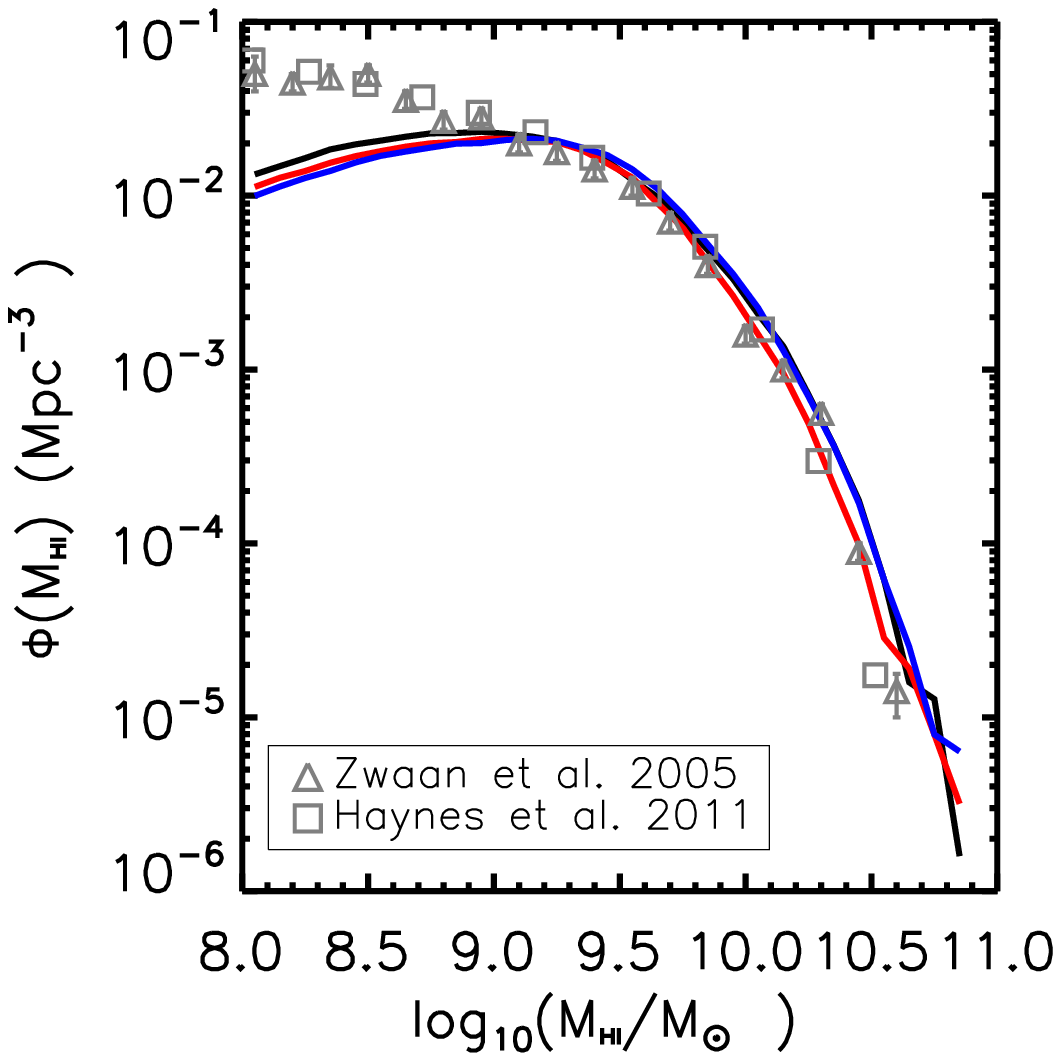}
\end{subfigure}
\begin{subfigure}{.33\textwidth}
  \centering
  \includegraphics[width=1.0\linewidth]{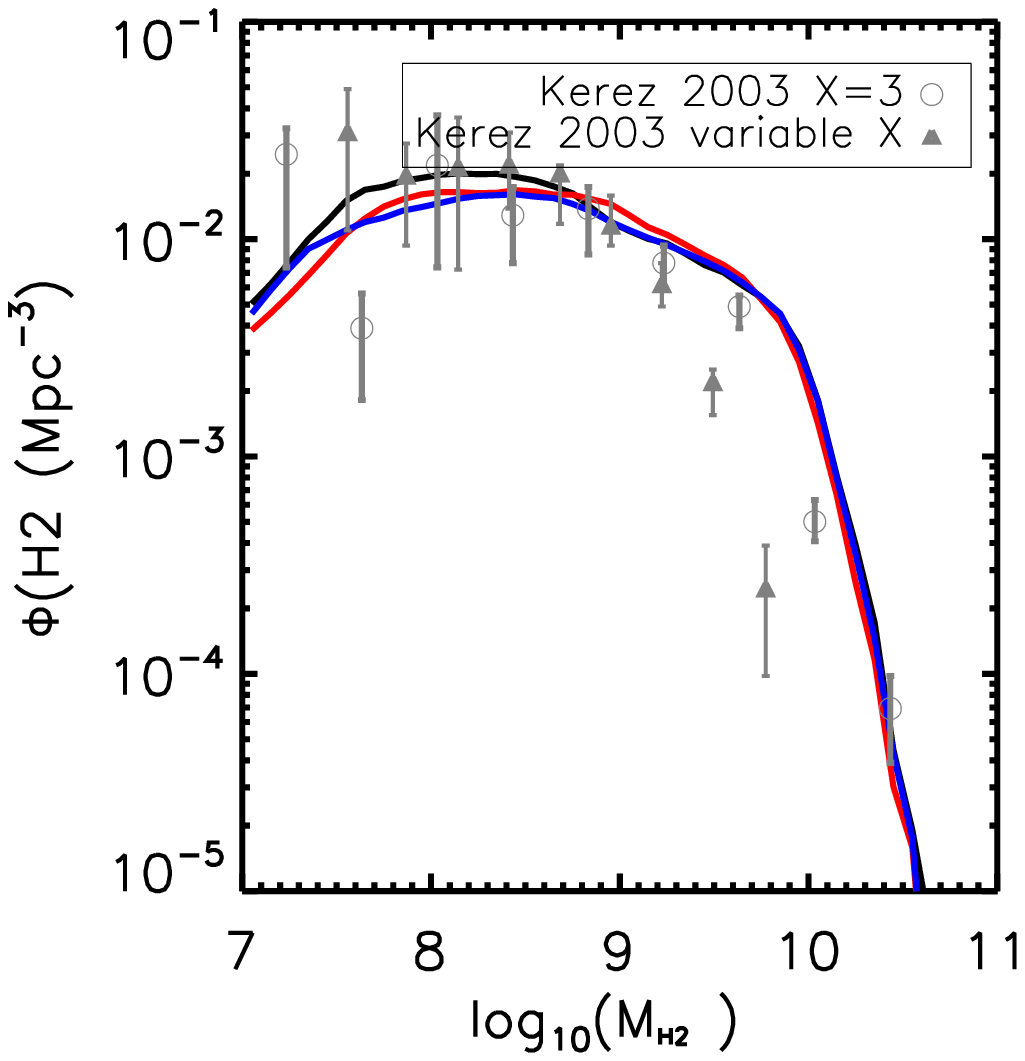}
\end{subfigure}
\caption{Results of tests for different assumptions to approximate the
  interstellar radiation field ($G'_0$) within the GK11 model. The black lines
correspond to our default model where we assume $G'_0$ is proportional to the
total star formation rate within the galaxy disk and normalized to the star
formation rate estimated for our galaxy. The blue lines correspond to the same
physical model but assuming $G'_0$ is proportional to the surface density of
SFR averaged over the entire disk. Finally, red lines show results based on the
same assumption but applied to each disk annulus. From left to right, the
different panels show the galaxy stellar mass function, the HI mass function,
and the H$_2$ mass function at $z=0$.}
\label{fig:g0comparison}
\end{figure*}

Our default assumption for $G'_0$ is given by
the star formation rate integrated over the entire gaseous disk, averaged over
the time interval between two subsequent snapshots and normalized to the
current rate of star formation estimated for our Galaxy. As discussed in
Section~\ref{sec:modelk13}, however, it would be more physical to express the
interstellar radiation field in terms of the surface density of star formation
rate. In Fig.~\ref{fig:g0comparison}, we compare the predicted galaxy stellar
mass function (left panel), HI mass function (middle panel) and H$_2$ mass
function (right panel) from our GK11 model with results obtained using
two alternative prescriptions for $G'_0$. In particular, blue lines correspond
to a model where $G'_0$ is assumed to be proportional to the surface density of
star formation averaged over the entire disk. In this case, we assume the
normalization factor to be $\Sigma_{\rm SFR, MW} = 5\times
10^{-4} \solarmass/yr/pc^2$. Red lines correspond to a model using the same
assumption but within each disk annulus. The figure shows that differences
between these different assumptions are very small (less than $\sim 0.1$~dex at
the low mass end in all three panels).

\begin{figure*}
\centering
\begin{subfigure}{.33\textwidth}
  \centering
  \includegraphics[width=1.0\linewidth]{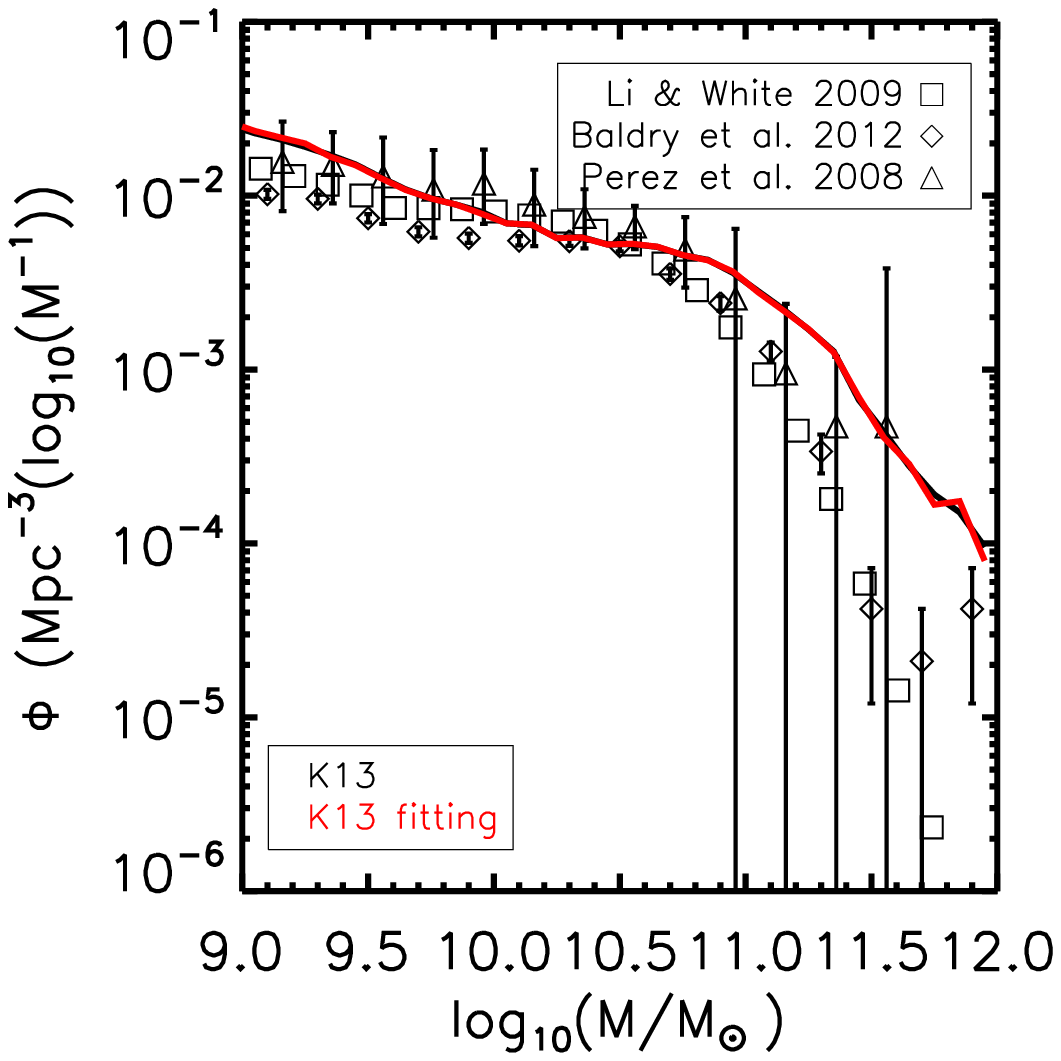}
\end{subfigure}%
\begin{subfigure}{.33\textwidth}
  \centering
  \includegraphics[width=1.0\linewidth]{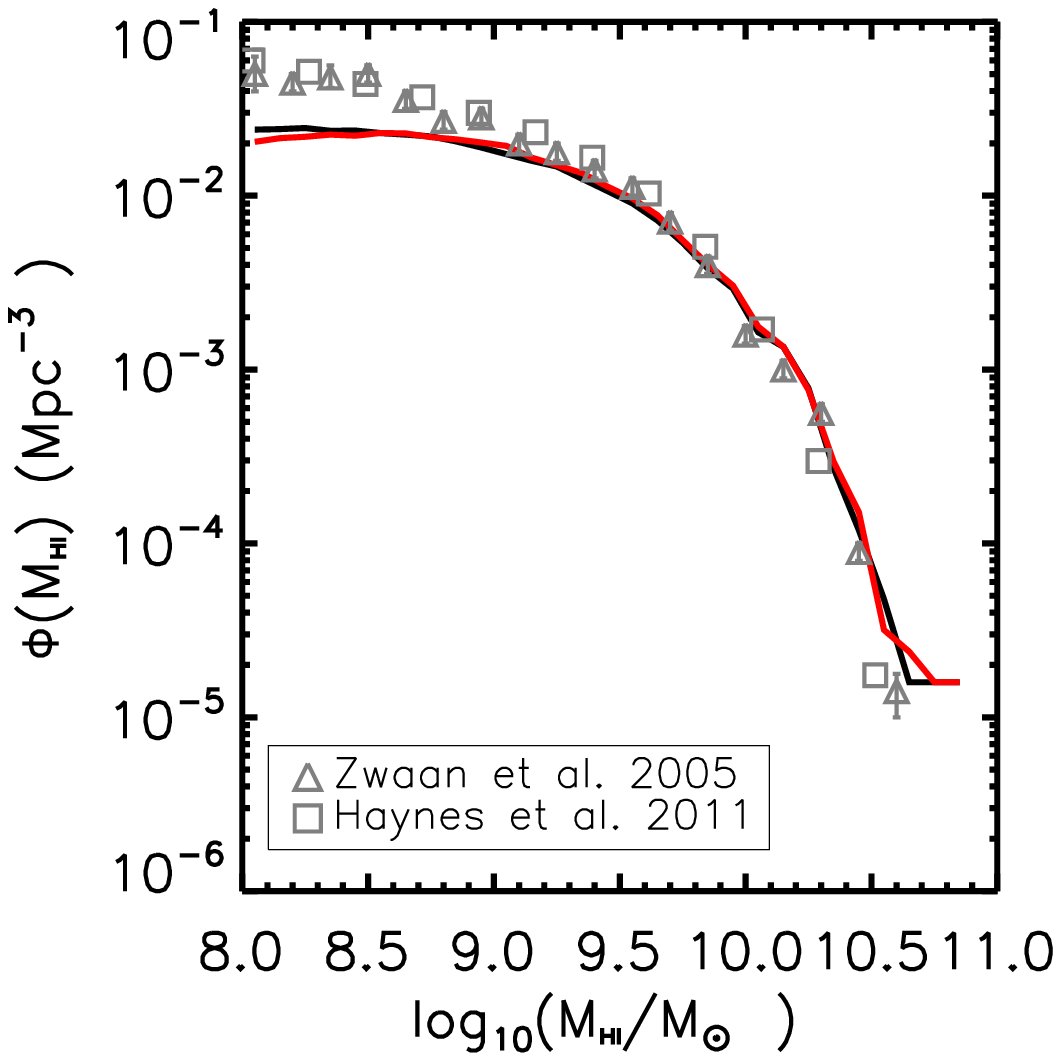}
\end{subfigure}
\begin{subfigure}{.33\textwidth}
  \centering
  \includegraphics[width=1.0\linewidth]{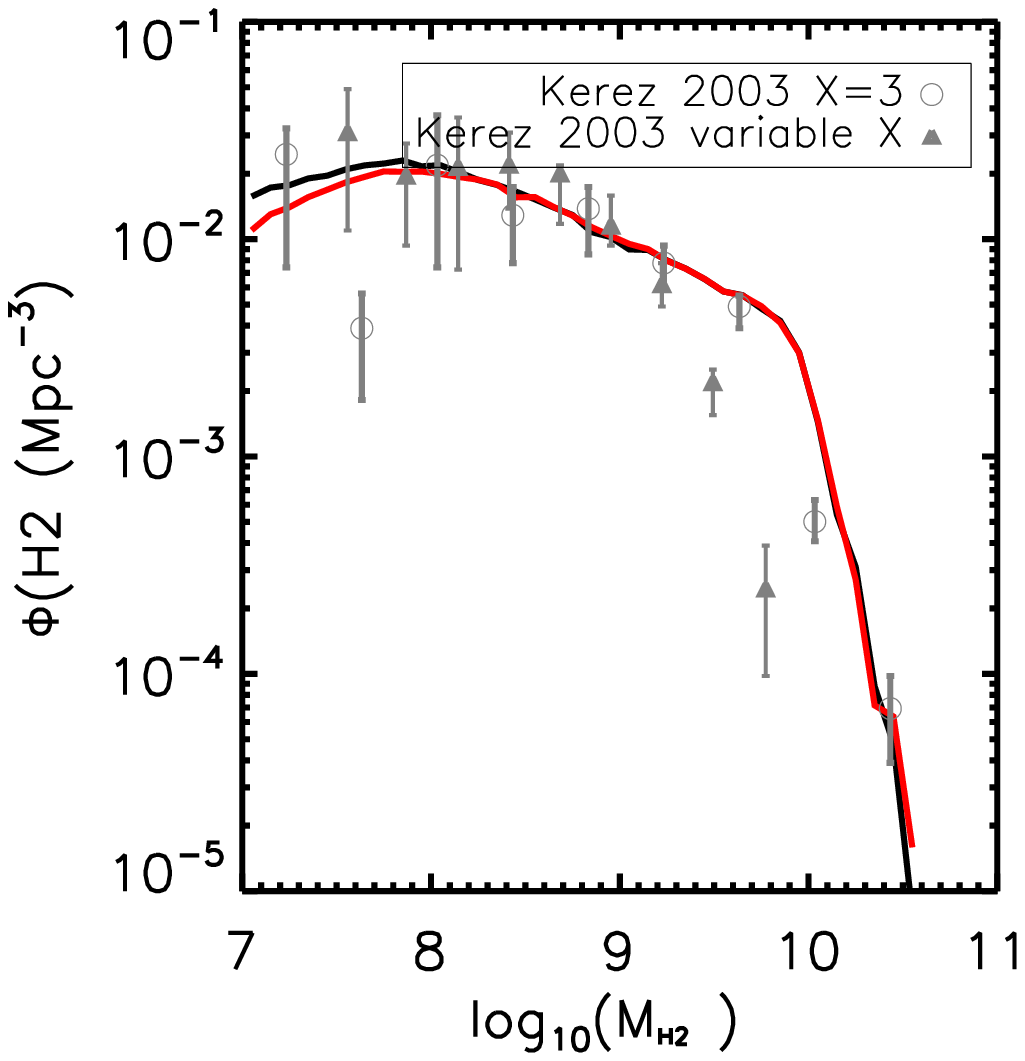}
\end{subfigure}
\caption{As for Fig.~\ref{fig:g0comparison} but this time for different
  assumptions for the density of dark matter and stars ($\rho_{sd}$) within the
  K13 model. The black lines correspond to our default model described in
  Section~\ref{sec:modelk13}. Red lines correspond to results based on the same
  physical model but using the lower limit for $\rho_{sd}$ resulting from the
  fitting function provided by \citet{krumholz2013}.}
\label{fig:nrsdcomparison}
\end{figure*}

Fig.~\ref{fig:nrsdcomparison} shows a similar comparison but this time for
tests made using different assumptions to compute $\rho_{sd}$ within the K13
model. As explained in Section~\ref{sec:modelk13}, our default model uses the
calculator provided by \citet{zhao2009} to assign a concentration to any halo
in the simulation. Assuming a NFW profile, this allows us to compute the
density of dark matter. Red lines shown in Fig.~\ref{fig:nrsdcomparison} 
correspond to a model adopting the lower limit given by the fitting formula 
provided by \citet{krumholz2013}.  We find that this parameter has little 
influence on the final model results and so significant amounts of computational 
time can be saved using a simpler approximation. 

\label{lastpage}
\end{document}